\documentclass[oneside,12pt]{article}

\usepackage[english]{babel}
\usepackage{jheppub} 

\usepackage[T1]{fontenc} 

\usetikzlibrary{positioning}
\usetikzlibrary{patterns,patterns.meta}
\usepackage[compat=1.1.0]{tikz-feynman}
\usepackage{amsmath,amssymb}
\usepackage{tikz}
\usepackage{graphicx}
\usepackage{cases}
\usepackage{setspace}
\usepackage{comment}
\usepackage{bm}
\usepackage{booktabs}
\usepackage{subcaption}


\usepackage{parskip}
\title{\boldmath Analyticity of the Black Hole S-Matrix}
\author[a]{Miguel Correia,}
\author[b]{Tushar Gopalka,}
\author[b]{Giulia Isabella}
\author[b]{and Anna M. Wolz}
\affiliation[a]{Department of Physics, McGill University, 3600 Rue University, Montr\'eal, H3A 2T8, QC Canada}
\affiliation[b]{Mani L. Bhaumik Institute for Theoretical Physics, Department of Physics and Astronomy, University of California Los Angeles, Los Angeles, CA 90095, USA}

\emailAdd{miguel.ribeirocorreia@mcgill.ca, tushar221997@g.ucla.edu, giuliaisabella@physics.ucla.edu, awolz@g.ucla.edu}

\abstract{
We establish the analytic structure of the S-matrix in the complex-frequency plane for classical wave scattering on a Schwarzschild background in four space-time dimensions. Our argument relies on the analytic continuation of the gravitational potential, with the singularity behind the horizon playing a crucial role.  We find that in the lower half-plane the partial-wave amplitudes are analytic except for the quasinormal-mode poles and the branch cut associated with late-time tails. 
As a direct consequence of causality, the retarded Green's function and absorption amplitude are analytic in the upper-half plane. We show, however, that Stokes phenomena can obstruct this analyticity domain from carrying over to the elastic amplitude, which instead develops a branch-cut in the upper-half plane. We also determine the effect of infrared (IR) regulators on the analytic structure, showing that polynomial boundedness requires a sharp lower bound on the IR cutoff in terms of the Schwarzschild radius.

}

\begin{document}

\maketitle

\newpage
\section{Introduction}

The analytic structure of the S-matrix encodes a great deal of information about the underlying dynamics of a quantum field theory (QFT). Basic principles such as causality, unitarity, and crossing symmetry impose stringent constraints on the location and nature of its singularities. For instance, poles identify the spectrum of intermediate states and reflect factorization properties, while discontinuities across branch cuts capture particle production and are controlled by unitarity \cite{Cutkosky:1960,Froissart:1961,Martin:1963UnitarityHE,Martin:1969ScatteringTheory,Eden:1966AnalyticSMatrix,Correia:2020xtr}. Conversely, the existence of extended regions of analyticity can often be traced back to micro-causality \cite{StreaterWightman:1989PCT,Lehmann:1958Ellipse}, the principle that operators at space-like separation must commute. 

This understanding, which extends beyond perturbation theory, lies at the heart of the modern S-matrix bootstrap, which has been applied to
a wide range of problems: from strongly-coupled gapped QFTs \cite{Paulos:2017fhb,Guerrieri:2018uew,Hebbar:2020ukp,Guerrieri:2023qbg,He:2023lyy,Cordova:2023wjp,Guerrieri:2024jkn,Guerrieri:2024ckc,Gumus:2024lmj,Correia:2025uvc,Cordova:2025bah,EliasMiro:2025rqo} to mapping out the space of quantum UV completions of gravity \cite{ Caron-Huot:2021rmr,Bellazzini:2015cra, Bern:2021ppb,  Bellazzini:2021shn,Caron-Huot:2022ugt,Serra:2022pzl, Chiang:2022jep,Haring:2022cyf,Beadle:2025cdx,Chang:2025cxc} and other effective field theories (EFTs) \cite{Bellazzini:2016xrt,Cheung:2016yqr,Bellazzini:2023nqj,Albert:2022oes,Serra:2022pzl}.  
In this context, fundamental properties of an unknown ultraviolet (UV) theory can be translated into constraints on its low-energy EFT through dispersion relations \cite{Adams:2006sv,Arkani-Hamed:2020blm,deRham:2017avq,Bellazzini:2020cot,Sinha:2020win,Tolley:2020gtv,Caron-Huot:2020cmc,Beadle:2024hqg,Paulos:2017fhb,EliasMiro:2022xaa}, allowing one to shape the space of Wilson coefficients that admit a consistent UV completion.

Despite the success of this program, most of the powerful non-perturbative results underlying it are firmly established only within a narrow class of theories: Lorentz invariant and gapped QFTs on flat space-time. This limit naturally excludes a large space of theories where many interesting physical systems live. Examples include systems with long-range interactions, condensed phases such as fluids and solids, cosmological backgrounds, open systems, and scattering on curved backgrounds. Recently, there has been growing interest in understanding the analytic structure of amplitudes and two-point correlators in such settings (see e.g. \cite{Hui:2023pxc,Hui:2025aja,Creminelli:2024lhd,Creminelli:2023kze,Chowdhury:2025dlx,Chowdhury:2025qyc,DiPietro:2021sjt}), although a general picture of these properties remains an open question.

In this work, we focus our attention on one such example: the scattering of classical waves on black hole backgrounds. This setup is particularly interesting because it not only breaks Lorentz symmetry and features long-range interactions, but also constitutes an open system, as the wave can be absorbed by the event horizon. The associated S-matrix captures these features in its analytic structure, encoding the dynamical response of the black hole to external perturbations. 

Beyond this theoretical motivation, determining the analytic structure of the black hole S-matrix provides a new arena in which to construct dispersion relations and to relate fundamental properties of the ultraviolet theory to effective field theories in a classical setting, an area where positivity bounds and S-matrix bootstrap techniques have not yet been systematically explored. A particularly relevant application is the study of compact objects in the strong-gravity regime. In this context, bounding their Love numbers, the Wilson coefficients of the worldline effective action \cite{Goldberger:2004jt,Goldberger:2005cd,Goldberger:2009qd,Porto:2016pyg}, would be especially interesting given their measurable imprint on gravitational-wave signals and their potential detectability in future experiments \cite{Punturo:2010zz,Amaro-Seoane:2017vxj,Reitze:2019iox}. Morover, the leading-order (static) Love number for black holes has been shown to vanish \cite{Binnington:2009bb,Damour:2009vw,Kol:2011vg,Porto:2016zng,Charalambous:2021kcz,Hui:2021vcv,Berens:2025okm,Parra-Martinez:2025bcu}, raising intriguing questions about its naturalness, which could also be addressed from a first principle perspective. Finally, in the case of neutron stars, such bounds could translate into nontrivial consistency conditions on their equation of state \cite{Damour:2009vw,Postnikov:2010yn,Hinderer:2007mb}.

The dynamics of classical waves on a Schwarzschild background are governed by the Regge-Wheeler and Zerilli equations \cite{Regge:1957td,Zerilli:1970se}, which at fixed angular momentum reduce to a one-dimensional wave equation with an effective potential. Its solution at spatial infinity can be expressed as a superposition of incoming and outgoing plane waves, while the S-matrix is identified with the reflection amplitude, or elastic amplitude, $R(\omega)$, namely the ratio of the outgoing and incoming connection coefficients. 

The basic expectation of analyticity in this classical setup comes from the simplest manifestation of causality: no signal can be detected before it is emitted. Imposing this condition on the retarded Green’s function
\begin{equation}
G_R(t-t',x,x') = 0\,, \qquad t-t' < 0\,,
\end{equation}
where $x$ and $x'$ denote the spatial coordinates of observer and source. This relation implies analyticity in the upper half of the complex frequency plane. However, this argument, often referred to as the \textit{signal model} \cite{Camanho:2014apa}, does not immediately extend to the S-matrix (and indeed we show in this work that it is violated). 

Concerning the lower half plane, it is well established that the retarded Green's function showcases two types of singularities. The first are poles located symmetrically about the imaginary axis: the quasinormal modes \cite{Regge:1957td,Vishveshwara:1970cc,Chandrasekhar:1975zza,Berti:2009kk,Press:1971wr,Vishveshwara:1970zz,KonoplyaZhidenko2011, Ferrari:1984zz}. These modes govern the exponential damping of the signal observed during the ringdown phase of gravitational-wave emission \cite{Berti:2009kk}. The second set of singularities corresponds to a branch cut extending along the negative imaginary axis, known to control the late-time tail behavior of the signal \cite{Leaver:1986gd,price_nonspherical_1972_2439,price_nonspherical_1972_2419,Casals:2012ng,Casals:2015nja}.

In contrast to the retarded Green’s function, the analytic structure of the transmission and reflection coefficients of the black hole S-matrix has not yet been established (see \cite{Castro:2013lba,Casals:2015nja, Motl:2003cd,Neitzke:2003mz,Arnaudo:2025uos} for work in this direction). Naively, one might expect that 
the elastic amplitude $R(\omega)$ would inherit the analytic properties of the retarded Green’s function, as suggested by the asymptotic relation \cite{Andersson:1996cm}
\begin{align}\label{eq:classicalLSZintro}
G_R(\omega,x,x') \xrightarrow[x,x'\to\infty]{}
\frac{1}{2i\omega}\left[e^{i\omega(x'-x)} - R(\omega)\,e^{i\omega(x'+x)}\right]\,.
\end{align}
Yet, as we point out in this work, this expectation does not necessarily hold, due to the possible emergence of \textit{Stokes phenomena} \cite{Stokes:1864,Barry:1989zz,Dingle:1973}, which can arise when the source and the observer are taken to infinity, 
$x,x'\to\infty$.\footnote{The Stokes phenomenon \cite{Stokes:1864, Dingle:1973, Barry:1989zz} can introduce singularities in the asymptotic limit of the Green’s function. In particular, because the wave equation has an irregular singular point at $x=\infty$, exponentially suppressed and enhanced terms can interchange dominance across the imaginary $\omega$ axis, leading to possible singularities of the reflection coefficient there. Stokes phenomena typically appear in asymptotic perturbative expansions and play a central role in the theory of resurgence \cite{Marino:2015yie,Aniceto:2018bis,Serone:2024uwz}.}
In this work we fill this gap by establishing the analyticity properties of the transmission and reflection coefficients.

\subsection*{Summary of the results}

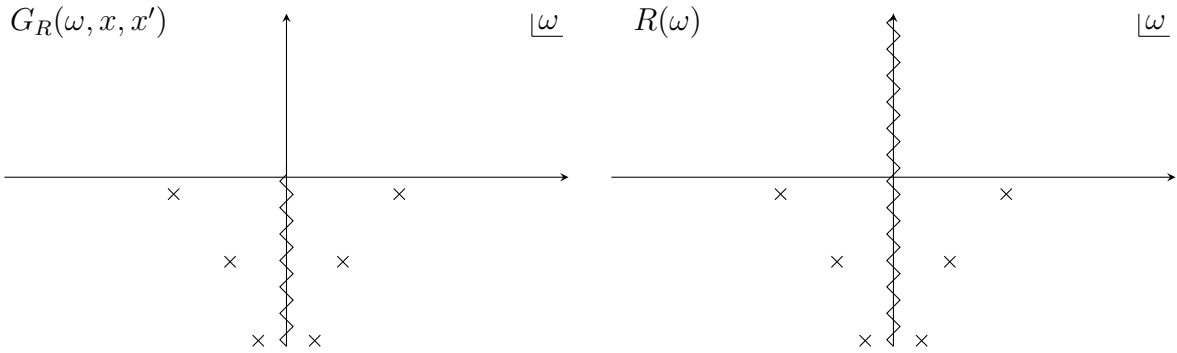
\begin{figure}[t]
  \centering
  \begin{tikzpicture}[>=stealth, scale=.75]
    \node at (-3.5,5.25) {$G_R(\omega,x,x')$};

    \draw[black, ->] (-5,2.5) -- (5,2.5);

    \draw[black, decorate, decoration=zigzag] 
           (0,-.5) -- (0,2.75);

    \draw[->] (0,-0.5) -- (0,5.4);

    \node at (4.90-.25,5.25) {$\omega$};
    \draw (4.6-.25,5.4) -- (4.6-.25,5.0);
    \draw (4.6-.25,5.0) -- (5.15-.25,5.0);
    
    \draw (0.1-2,0.1+2.2) -- (-0.1-2,-0.1+2.2);
    \draw (-0.1-2,0.1+2.2) -- (0.1-2,-0.1+2.2);
    \draw (0.1+2,0.1+2.2) -- (-0.1+2,-0.1+2.2);
    \draw (-0.1+2,0.1+2.2) -- (0.1+2,-0.1+2.2);
    \draw (0.1-1,0.1+1.) -- (-0.1-1,-0.1+1.);
    \draw (-0.1-1,0.1+1.) -- (0.1-1,-0.1+1.);
    \draw (0.1+1,0.1+1.) -- (-0.1+1,-0.1+1.);
    \draw (-0.1+1,0.1+1.) -- (0.1+1,-0.1+1.);
    \draw (0.1-.5,0.1-.4) -- (-0.1-.5,-0.1-.4);
    \draw (-0.1-.5,0.1-.4) -- (0.1-.5,-0.1-.4);
    \draw (0.1+.5,0.1-.4) -- (-0.1+.5,-0.1-.4);
    \draw (-0.1+.5,0.1-.4) -- (0.1+.5,-0.1-.4);

  \end{tikzpicture}
  \quad
  \begin{tikzpicture}[>=stealth, scale=.75]
    \node at (-4,5.25) {$R(\omega)$};
    
    \draw[black, ->] (-5,2.5) -- (5,2.5);

    \draw[black, decorate, decoration=zigzag] 
           (0,-.5) -- (0,5.4);

    \draw[->] (0,-0.5) -- (0,5.4);

    \node at (4.90-.25,5.25) {$\omega$};
    \draw (4.6-.25,5.4) -- (4.6-.25,5.0);
    \draw (4.6-.25,5.0) -- (5.15-.25,5.0);
    
    \draw (0.1-2,0.1+2.2) -- (-0.1-2,-0.1+2.2);
    \draw (-0.1-2,0.1+2.2) -- (0.1-2,-0.1+2.2);
    \draw (0.1+2,0.1+2.2) -- (-0.1+2,-0.1+2.2);
    \draw (-0.1+2,0.1+2.2) -- (0.1+2,-0.1+2.2);
    \draw (0.1-1,0.1+1.) -- (-0.1-1,-0.1+1.);
    \draw (-0.1-1,0.1+1.) -- (0.1-1,-0.1+1.);
    \draw (0.1+1,0.1+1.) -- (-0.1+1,-0.1+1.);
    \draw (-0.1+1,0.1+1.) -- (0.1+1,-0.1+1.);
    \draw (0.1-.5,0.1-.4) -- (-0.1-.5,-0.1-.4);
    \draw (-0.1-.5,0.1-.4) -- (0.1-.5,-0.1-.4);
    \draw (0.1+.5,0.1-.4) -- (-0.1+.5,-0.1-.4);
    \draw (-0.1+.5,0.1-.4) -- (0.1+.5,-0.1-.4);

  \end{tikzpicture}
  \caption{Analytic structure of \textbf{Left:} $G_R(\omega,x,x')$ and \textbf{Right:} $R(\omega)$ of the black hole. There is a branch cut along the negative imaginary axis for $G_R$ and along the entire imaginary axis for $R$, as well as quasinormal mode poles in the LHP (crosses, locations are approximate).}
  \label{fig:GBHsketch}
\end{figure}

The analytic structure for wave scattering on Schwarzschild potential obtained are summarized in Figure \ref{fig:GBHsketch} for both the retarded Green's function $G_R(\omega,x,x')$ and the reflection coefficient $R(\omega)$. 

In the lower half-plane (LHP), we show that the retarded Green’s function and the reflection coefficient share the same analytic structure: a branch cut along the lower part of the imaginary axis and a sequence of poles symmetric about it, corresponding respectively to late-time tails and quasinormal modes. However, in the upper half-plane (UHP), while $G_R(\omega,x,x')$ remains analytic, as required by causality, we find that $R(\omega)$ develops a branch cut along the positive imaginary axis. This discontinuity arises from logarithmic terms induced by the long-range nature of gravity and by finite-size effects of the black hole, associated with the running of dynamical Love numbers.

Our derivation of the regions of analyticity for $G_R(\omega, x,x')$ and $R(\omega)$ employs ideas from inverse-scattering theory and non-relativistic quantum mechanics \cite{Taylor:1972,https://doi.org/10.1002/cpa.3160320202,Yafaev:1992, Reed:1979MMPIII,Newton:2013ScatteringReprint}, and applies broad\-ly to classical wave scattering on one-dimensional potentials. Crucially, it generalizes standard techniques originally developed for short-range interactions to incorporate long-range gravitational effects and the dissipative, open nature of the black hole system. 

We find that the analyticity guaranteed by causality for the retarded Green’s function in the UHP is not, by itself, sufficient to ensure analyticity of $R(\omega)$. The missing ingredient is a ``locality'' assumption on the potential: that it admits an analytic continuation in the complex $x$-plane. This is typically verified for potentials that arise from a field theory, as is the case for the Schwarzschild gravitational potential that we study here. 

In particular, singularities of the potential in the complex $x$-plane manifest as non-analy\-ticities in frequency space for the S-matrix.
Perhaps surprisingly, when applying our proof to the Schwarzschild potential, the obstruction to the analytic continuation in position space is due to the black hole singularity. Although hidden behind the event horizon, this singularity becomes accessible upon continuation into the complex plane and determines the proven maximal domain of analyticity of the black hole S-matrix.

\subsection*{Outline of the paper}

Section~\ref{sec:asymG} presents the main derivation of the analyticity properties of the retarded Green’s function and transmission and reflection coefficients for generic potentials. The results are then applied to the Pöschl–Teller potential in Section~\ref{sec:poschlteller} and to the Schwarzschild potential in Section~\ref{sec:BH}. In both cases, we test the predicted regions of analyticity against explicit solutions obtained in various limits. We conclude in Section~\ref{sec:discussion} with a summary and outlook.

Appendix~\ref{app:potentialsproof} complements the analyticity argument of Section~\ref{sec:asymG} by imposing additional fall-off conditions on the potential. 
 Appendix~\ref{app:boundstates} uses the Laplace transform to show that bound state poles in the upper half-plane of the retarded Green’s function remain consistent with causality. 
 Appendix~\ref{app:SchwHighFreqCorr} shows how to systematically add high-frequency corrections to the Schwarzschild reflection coefficient.
Appendix~\ref{app:bounds} applies S-matrix bootstrap methods to bound Wilson coefficients in a point-particle EFT without long-range forces.
 Appendix~\ref{app:QFT} derives the retarded Green’s function and partial-wave S-matrix in QFT via the LSZ reduction formula and illustrates the Stokes phenomenon for the Yukawa potential.

\section{Scattering theory and domains of analyticity}\label{sec:asymG}

We consider the one-dimensional scattering of classical waves described by the wave equation
\begin{equation}
\label{eq:waveeq}
    \left[{d^2 \over dx^2} + \omega^2 - V(x) \right] \phi(\omega,x) = 0
\end{equation}
where $x \in \mathbb{R}$, $\omega$ is the frequency of the scattered wave, $V(x)$ is a time-independent potential, and $\phi(\omega,x)$ is the wavefunction. 

Here, we explore the analytic properties in complex $\omega$ of the wavefunction, retarded Green's function, and S-matrix for a general potential $V(x)$ that decays sufficiently fast at infinity $V(x \to \pm \infty) \lesssim {1 \over |x|}$ and remains bounded for finite $x$. Potentials with this behavior describe asymptotically flat backgrounds, such as the Schwarzschild potential, where at spatial infinity the solution is a linear combination of plane waves. 

We describe how general properties of 
$V(x)$ translate into analytic domains of the retarded Green’s function and the S-matrix, showing in particular that analyticity of the potential in $x$ leads to an enlargement of the analyticity domain of the S-matrix. In Chapters \ref{sec:poschlteller} and \ref{sec:BH} we test these conclusions on a simple toy model and later on the physically relevant case of black hole scattering.

Our analysis brings together ideas from inverse-scattering theory and scattering in non-relativistic quantum mechanics \cite{Taylor:1972,Yafaev:1992,https://doi.org/10.1002/cpa.3160320202,Reed:1979MMPIII,Newton:2013ScatteringReprint}.

\subsection{Boundary conditions and the S-matrix} 
As already mentioned above, if the potential decays sufficiently fast, the solution  at infinity can be written as a linear combination of plane waves
\begin{equation}
\label{eq:BCplus}
 \phi(\omega,x\to +\infty) = e^{- i \omega x} - R(\omega) \, e^{i \omega x},
\end{equation}
where the time dependent solution is given by the Fourier transform
\begin{equation}
\label{eq:FT}
\phi(t,x) = \int_{- \infty}^{+ \infty}e^{- i \omega t} \phi(\omega,x) \, d\omega \,.
\end{equation}
We see that the first piece in \eqref{eq:BCplus} amounts to the ingoing wave (traveling to the left), while the second piece corresponds to the outgoing wave (traveling to the right), whose coefficient $R(\omega)$ is defined as the \emph{reflection} amplitude in the S-matrix.\footnote{We normalize the coefficient of the ingoing $e^{-i\omega x}$ solution to $1$.} Note that we are suppressing any quantum numbers, such as the orbital angular momentum $\ell$ in the case of rotational symmetry, where $R(\omega)=S_\ell(\omega)$ is also known as the \emph{partial wave} S-matrix \cite{Taylor:1972}.

We consider open systems, so we impose a purely outgoing boundary condition at $x\to-\infty$
\begin{equation}
\label{eq:BCminus}
    \phi(\omega,\;x \to - \infty) = T(\omega)\, e^{- i \omega x}
\end{equation}
where $T(\omega)$ is the \emph{transmission} amplitude. Note that for scattering off black holes, the presence of a horizon requires $T(\omega) \neq 0$, while a Dirichlet boundary condition could be recovered by setting $T(\omega) = 0$. 

Let us list some basic properties of the S-matrix. If $\phi(t,x)$ corresponds to a classical field then it is a real quantity, which implies the relation $\phi^*(\omega,x) = \phi(-\omega^*,x)$ via \eqref{eq:FT}. Applying this to the boundary conditions \eqref{eq:BCplus} and \eqref{eq:BCminus} leads to the symmetries
\begin{equation}\label{eq:crossing}
  \text{Reflection symmetry:} \qquad  R^*(\omega) = R(-\omega^*), \;\;\; T^*(\omega) = T(-\omega^*).
\end{equation}
 Furthermore, the wave equation \eqref{eq:waveeq} has a conserved current due to the time independence of $V(x)$,
\begin{equation}
    {d \over d x} \left[\phi^* {d \phi \over d x} - \phi {d \phi^* \over d x} \right] = 0
\end{equation}
whose integrated form along with the chosen asymptotics gives a unitarity relation between the reflection and transmission amplitudes \cite{Taylor:1972,Yafaev:1992,Reed:1979MMPIII}
\begin{equation}
\label{eq:unitarity}
  \text{Unitarity:} \qquad  |R(\omega)|^2 + |T(\omega)|^2 = 1.
\end{equation}
In the black hole setting this means that the total incoming flux is conserved between the reflected part and the component absorbed by the horizon \cite{Futterman:1988book,Berti:2009kk}.
We now turn to discussing the analyticity properties of the S-matrix and its relation to causality, which leads us to introduce the retarded Green's function.

\subsection{Retarded Green's function} 

The retarded Green's function $G_R(\omega,x,x')$ satisfies the inhomogeneous equation
\begin{equation}
\label{eq:waveeqgreen}
    \left[{d^2 \over dx^2} + \omega^2 - V(x) \right] G_R(\omega,x,x') = \delta(x-x')
\end{equation}
and its retarded causal nature is fixed by the condition that it vanishes for times prior to the source activation at $t'$
\begin{equation}\label{eq:causality}
G_R(t-t',x,x')=0\,,\,\,\,\, \text{for any } t-t'<0\, ,
\end{equation}
where we have used again the time-independence of the potential to express the Green's function in terms of the time difference $t-t'$. Without loss of generality, we set $t'=0$
 and therefore
\begin{equation}\label{eq:FTGR}
G_R(\omega,x,x') = \int_{- \infty}^{+\infty} dt \, G_R(t,x,x')\, e^{i\omega t}\,,
\end{equation}
which is analytic in the upper-half of the complex-$\omega$ plane. Using the Wronskian representation, we build the retarded Green's function as:
\begin{equation}
\label{eq:Green}
    G_R(\omega,x,x') = {\phi_L(\omega,x) \,\phi_R(\omega,x')\, \Theta(x'-x) + \phi_L(\omega,x')\, \phi_R(\omega,x) \,\Theta(x-x') \over W[\phi_L,\phi_R](\omega)}
\end{equation}
where $\phi_L(\omega,x)$ and $\phi_R(\omega,x)$ are linearly independent solutions to the equation \eqref{eq:waveeq} and
\begin{equation}\label{eq:wronskian}
    W[\phi_L,\phi_R](\omega)= \phi_L(\omega,x) {d \phi_R(\omega,x) \over d x} - {d \phi_L(\omega,x) \over d x} \phi_R(\omega,x)
\end{equation}
is the Wronskian between the two solutions. The Wronskian is independent of $x$ because there is no ${d \phi \over d x}$ term in the differential equation \eqref{eq:waveeq}, and it is straightforward to check that \eqref{eq:Green} satisfies the differential equation~\eqref{eq:waveeqgreen}.

The retarded causal condition \eqref{eq:causality} is established if $G_R(\omega,x,x')$ is analytic for $\mathrm{Im} \,\omega > 0$, which is in turn guaranteed if $\phi_L(\omega,x)$ and $\phi_R(\omega,x)$ are analytic for $\mathrm{Im} \,\omega > 0$, and the Wronskian $W[\phi_L,\phi_R](\omega)$ has no zeros for $\mathrm{Im} \, \omega > 0$.\footnote{These Wronskian zeros would correspond to poles in the UHP of the retarded Green's function, typically associated with bound-states. While their presence would naively be incompatible with causality, it is important to note that if the potential $V(x)$ is bounded from below only a finite number of bound-states exists \cite{Yafaev:1992, Newton:1983JOT, Chadan:2002bound1D}. In this case, usage of the Laplace transform instead of Fourier transform amounts to a deformation of the frequency plane contour above all the bound states which preserves the requirement of causality; refer to Appendix \ref{app:boundstates} for more details.} 

The boundary conditions of the two independent solutions are 
\begin{equation}\boxed{
\label{eq:bcLR}
    \phi_L(\omega,x \to -\infty) = e^{- i \omega x}, \qquad     \phi_R(\omega,x \to +\infty) = e^{ i \omega x}.
    }
\end{equation}
These translate to causal boundary conditions for the retarded Green's function, namely that an asymptotic observer at $x\to\pm\infty$ only sees waves propagating outwards from a source located at finite $x'$. For example, 
\begin{equation}
\label{eq:GRBC}
    G_R(\omega,x \to + \infty,x' \text{ fixed}) \sim e^{+ i \omega x}.
\end{equation}
Since the left outgoing solution $\phi_L(\omega,x)$ satisfies the same boundary condition as the physical solution $\phi(\omega,x)$ \eqref{eq:BCminus}, they are related up to some normalization. This leads to the asymptotic relation \eqref{eq:classicalLSZintro} between the reflection coefficient $R(\omega)$ and the retarded Green's function $G_R(\omega,x,x')$ when both the source and the observer are taken to infinity, $x, x' \to \infty$.

In the following section, we will show that these boundary conditions, together with very general assumptions on the potential $V(x)$ (bounded, integrable and positive for all $x$),

\begin{equation}
\label{eq:Vintbound}
  V(x) > 0 \; \text{ for } x \in \mathbb{R}, \quad \text{ and } \quad \int^{+\infty}_{-\infty} |V(x)| dx < \infty\,,
\end{equation}

rigorously imply that $\phi_L(\omega,x)$, $\phi_R(\omega,x)$, and consequently $G_R(\omega,x,x')$ are analytic for $\mathrm{Im}\,\omega>0$, thus establishing that $G_R(\omega,x,x')$ given by \eqref{eq:Green} is indeed the retarded Green's function.

However, we will see in Sections \ref{sec:analyticpotentials} and \ref{sec:analyticpotentials2}  that establishing analyticity for the reflection amplitude $R(\omega)$ requires stronger assumptions on the potential $V(x)$, in particular analyticity in $x$. 
Finally, in Section \ref{sec:smatrixpoles} we investigate the zeros of the Wronskian
which correspond to poles of the S-matrix. We find that such poles exist only on the positive imaginary axis (bound states) and in the lower half-plane (quasinormal modes). 
These results are summarized in Figure~\ref{fig:allResults}.

\subsection{Analyticity from causality}
\label{sec:jost}

The homogeneous solutions $\phi_L(\omega,x)$ and $\phi_R(\omega,x)$ satisfy the boundary conditions~\eqref{eq:bcLR}, which have been normalized to a constant.
This normalization will be useful to establish analyticity in the asymptotic limit $x\to\infty$, which is an idea that can be traced back to Jost \cite{JostPais1951}. In the context of quantum mechanics these are known as \emph{Jost solutions} \cite{Taylor:1972,Mizera:2023tfe}.
It suffices to consider the left-outgoing solution $\phi_L(\omega,x)$, as the analysis for the right-outgoing solution $\phi_R(\omega,x)$ is entirely analogous. 

We start by writing an integral equation for $\phi_L(\omega,x)$.
From the method of variation of constants (see Chapter 3 in \cite{Teschl2012}), we obtain the Volterra integral relation \cite{Lalescu1908_Volterra, Talukdar1991} satisfying the boundary condition \eqref{eq:bcLR} and the wave equation \eqref{eq:waveeq}
\begin{equation}
\label{eq:psig}
    \phi_L(\omega, x) = e^{- i \omega x} + \int_{- \infty}^x {(e^{i \omega (x- x')} - e^{i \omega (x'- x)}) V(x') \phi_L(\omega, x') \over 2 i \omega} dx'\,.
\end{equation}
Introducing $\chi_L(\omega,x)= e^{i \omega x}\phi_L(\omega,x) $ we find the compact form
\begin{equation}
\label{eq:chiL}
    \chi_L(\omega, x) = 1 + \int_{- \infty}^x G_0(\omega,x-x')\,  V(x') \,\chi_L(\omega,x')\,  dx'
\end{equation}
with
\begin{equation}
\label{eq:G0}
    G_0(\omega,x-x') = {e^{2i \omega (x- x')} - 1 \over 2 i \omega}\,.
\end{equation}
Iterating \eqref{eq:chiL} we can write the formal series
\begin{equation}
\label{eq:serieschi}
    \chi_L(\omega,x) = \sum_{n=0}^\infty  \chi_L^{(n)}(\omega,x)
\end{equation}
where $\chi_L^{(0)}(\omega,x) = 1$ and each ``Born'' iteration is given by
\begin{align}
\label{eq:chin}
    \chi_L^{(n)}(\omega,x) &= \int_{- \infty}^x  G_0(\omega,x-x') \,V(x')\, \chi^{(n-1)}_L(\omega,x')\,dx'  \\
    &=  \int_{x>x_1>\dots>x_n>-\infty} G_0(\omega,x-x_1) \cdots G_0(\omega,x_{n-1}-x_n) \, V(x_1) \cdots V(x_n) \, dx_1 \cdots dx_n\,. \notag
\end{align}
We will now show that the series \eqref{eq:serieschi} converges for $\mathrm{Im}\, \omega > 0$. 

The first observation is that the free Green's function \eqref{eq:G0} obeys the bound
\begin{equation}
\label{eq:G0bound}
    |G_0(\omega,x-x')| \leq {1 \over |\omega|} \qquad \text{ for }\;\; x - x' > 0 \;\; \text{ and } \;\; \mathrm{Im \; \omega > 0}.
\end{equation}
We now see the utility of using the Jost functions: due to the integration region $x>x_1>\dots>x_n>-\infty$, each combination $x_{j-1}-x_j$ appearing in the $G_0$'s is always positive.
This leads to the following bound on every term in the Born series 
\begin{align}
    |\chi^{(n)}_L(\omega,x)| &\leq {1 \over |\omega|^n} \int_{x>x_1>\dots>x_n} |V(x_1)| \cdots |V(x_n)| \, dx_1 \dots dx_n \notag \\
    &= {1 \over n!\, |\omega|^n} \left[\int_{- \infty}^x |V(x')| \,dx' \right]^n\,.\qquad \mathrm{Im} \,\omega \geq 0
\end{align}
The boundedness of the potential \eqref{eq:Vintbound} guarantees that this bound remains finite. 
This result implies that the series \eqref{eq:serieschi} is uniformly convergent\footnote{For a given domain $\Omega$ in the complex $\omega$-plane, we take the supremum of the bound~\eqref{eq:chibound} over $\Omega$. The resulting $\omega$-independent bound allows us to apply the Weierstrass $M$-test, which ensures uniform convergence of the series~\eqref{eq:serieschi}.}
\begin{equation}
\label{eq:chibound}
    |\chi_L(\omega,x)| \leq \sum_{n=0}^\infty  |\chi_L^{(n)}(\omega,x)| \leq \exp \left[ {1 \over |\omega|} \int_{- \infty}^x |V(x')| \,dx' \right].
\end{equation}
In addition to uniform convergence, each term in the Born series \eqref{eq:serieschi} is analytic for $\text{Im}\,\omega>0$. These two properties guarantee that 
 $\phi_L(\omega,x) = e^{- i \omega x} \chi_L(\omega,x)$ is analytic for $\mathrm{Im} \, \omega > 0$.  Analogous arguments show that the right outgoing solution $\phi_R(\omega,x)$ is analytic for $\mathrm{Im}\, \omega > 0$. 

It follows that the Wronskian \eqref{eq:wronskian} is equally analytic, and if it has no zeros we conclude that the retarded Green's function \eqref{eq:Green} is analytic in the UHP (zeros of the Wronskian are discussed further in Section~\ref{sec:smatrixpoles} and Appendix \ref{app:boundstates}).

We now turn to determining the analyticity of the reflection $R(\omega)$ and transmission $T(\omega)$ coefficients.
We first define the connection coefficients $A_\text{in}(\omega)$ and $A_\text{out}(\omega)$ of the Jost solution
\begin{equation}
\label{eq:phidecomposition}
    \phi_L(\omega,x\to+\infty) = A_{\mathrm{in}}(\omega) \, e^{-i \omega x} + A_{\mathrm{out}}(\omega)\, e^{i \omega x}\,.
\end{equation}
 Using the boundary condition for $\phi_R(\omega,x\to\infty)$ in \eqref{eq:bcLR}, the Wronskian  \eqref{eq:wronskian} is given by
\begin{equation}
    W[\phi_L,\phi_R](\omega) = 2 i \omega A_{\mathrm{in}}(\omega)
\end{equation}
and via \eqref{eq:BCplus} and \eqref{eq:BCminus}, the S-matrix elements read
\begin{equation}
\label{eq:RT}
    R(\omega) = - {A_{\mathrm{out}}(\omega) \over A_{\mathrm{in}}(\omega)}\,, \qquad T(\omega) = {1 \over A_{\mathrm{in}}(\omega)}\,.
\end{equation}
The connection coefficients obey the integral representations 
\begin{equation}\boxed{
\label{eq:Ainchi}
    A_{\mathrm{in}}(\omega) = 1 - {1 \over 2 i \omega} \int_{- \infty}^\infty V(x') \,\chi_L(\omega,x') dx'}
\end{equation}
and
\begin{equation}
\label{eq:Aoutchi}\boxed{
    A_{\mathrm{out}}(\omega) = {1 \over 2 i \omega} \int_{- \infty}^\infty e^{- 2 i \omega x'} V(x') \,\chi_L(\omega,x') dx'}\,,
\end{equation}
derived from the Volterra equation \eqref{eq:psig}.

\textbf{Analyticity in UHP.} Using these integral relations and the proven analyticty for $\chi_L(\omega,x)$, some analyticity domain for the connection coefficients follows. In particular, the integral representation for $A_\text{in}(\omega)$ \eqref{eq:Ainchi} remains convergent given the bound \eqref{eq:chibound}, and moreover the proven analyticity of $\chi_L(\omega, x')$ allows $A_\text{in}(\omega)$ to be analytically continued for $\text{Im}\, \omega >0$.\footnote{This result also follows from the basic observation that $A_\text{in}(\omega)$ is given by the Wronskian \eqref{eq:wronskian} of two analytic functions.}

Unfortunately, at this stage, we cannot say anything about analyticity of $A_\text{out}(\omega)$ in the upper-half-plane. For $\mathrm{Im} \,\omega > 0$, the  factor $e^{- 2 i \omega x'}$ in the integral relation \eqref{eq:Aoutchi} grows exponentially $\sim e^{2\, \mathrm{Im}\,\omega \,x'}$ for positive $x'$, making the integral diverge.

Given that the transmission amplitude $T(\omega)$ only depends on $A_\text{in}(\omega)$ via \eqref{eq:RT} we find that
\begin{equation}
    T(\omega) \text{ is meromorphic for } \mathrm{Im}\, \omega > 0.
\end{equation}
In particular, zeros of $A_\text{in}(\omega)$ lead to poles of $T(\omega)$, the same poles present in the retarded Green's function $G_R(\omega,x,x')$ given in equation \eqref{eq:Green}. 

Analyticity in the lower-half-plane of $T(\omega)$ requires additional assumptions on the potential, which we will now discuss.

\subsection{Extended analyticity for the transmission amplitude}\label{sec:analyticpotentials}

Additional properties of the potential will allow us to extend the regions of analyticity of the Jost solutions, Green's function, and S-matrix elements. In particular, we consider potentials that can be analytically continued into a region of the complex-$x$ plane, examples of which include the P\"oschl-Teller potential (see Section \ref{sec:poschlteller}) and the Schwarzschild gravitational potential (see Section \ref{sec:BH}). Additional relevant classes of potentials include those that decay exponentially (such as the Yukawa potential \cite{Taylor:1972}, explored in Appendix~\ref{sec:yukawa}) and those that are compactly supported. These assumptions translate into extended  regions of analyticity, as established in Appendix \ref{app:potentialsproof}.

Here, we assume the potential $V(x)$ can be analytically continued in the complex-$x$ plane. In particular, we assume that the potential $V(x)$ is analytic in the domain:
\begin{equation}
\label{eq:Van}
    \begin{split}
    - \theta_\text{max}< \arg x  < \theta_\text{max} \;\cup\; \pi- \theta_\text{max}&< \arg x  < \pi \;\cup\; -\pi < \arg x  < -\pi+\theta_\text{max}\,,
    \end{split}
\end{equation}
as shown in Figure~\ref{fig:Vxanalytic}.\footnote{Here, we take the region of analyticity to be symmetric about the imaginary axis for simplicity. One can extend our arguments to different angles $(\theta_{1},\theta_{2})$.}
\begin{figure}[t]
  \centering
  \begin{minipage}[t]{0.45\textwidth}
    \centering
    \begin{tikzpicture}[>=stealth,scale=.75]
      \fill[blue!15] (0,0) -- (1.8,3) -- (5,3) -- (5,-3) -- (1.8,-3) -- cycle;
      \fill[blue!15] (-1.8,-3) -- (-5,-3) -- (-5,3) -- (-1.8,3) -- (0,0) -- cycle;
      \draw [thick,domain=-58:58] plot ({cos(\x)}, {sin(\x)});
      \node at (1.5,0.7) {$\theta_{\textrm{max}}$};
      \node at (1.6,-0.7) {$-\theta_{\textrm{max}}$};
      \draw[->,thick] (0,-3) -- (0,3);
      \draw[->,thick] (-5,0) -- (5,0);
      \node at (4.825,2.75) {$x$};
      \draw[thick] (4.60,3) -- (4.6,2.55);
      \draw[thick] (4.6,2.55) -- (5,2.55);
    \end{tikzpicture}
    \caption{$V(x)$ is analytic and decays sufficiently fast in the shaded blue region.}
    \label{fig:Vxanalytic}
  \end{minipage}
  \hfill
  \begin{minipage}[t]{0.45\textwidth}
    \centering
    \begin{tikzpicture}[>=stealth,scale=.75]
      \draw[thick,blue] (-3,-3) -- (1.5,1.5);
      \draw[->,>=angle 45,thick,blue] (-3,-3) -- (-1.5,-1.5);
      \fill[blue] (1.5,1.5) circle (2pt) node[above] {$x$};
      \draw [thick,domain=0:45] plot ({cos(\x)}, {sin(\x)});
      \node at (1.2,0.6) {$\theta$};
      \draw[->,thick] (0,-3) -- (0,3);
      \draw[->,thick] (-5,0) -- (5,0);
      \node at (4.825,2.8) {$x'$};
      \draw[thick] (4.55,3) -- (4.55,2.5);
      \draw[thick] (4.55,2.5) -- (5,2.5);
    \end{tikzpicture}
    \caption{Integration contour for the integral in Eq.~\eqref{eq:complexxint1}, with $x=e^{i\theta}y$.}
    \label{fig:xcontour}
  \end{minipage}
\end{figure}
Moreover, we also assume that the potential decays sufficiently fast in this wedge:
\begin{equation}\label{eq:Vbound}
    V(|x| \to \infty) \lesssim {1 \over |x|}, \;\; \text{ for } \;\;\arg x \in \eqref{eq:Van}
\end{equation}

Let us first show that this implies that the Jost solutions $\chi_L(\omega,x)$ and $\chi_R(\omega,x)$ are analytic in a slated half of the complex-$\omega$ plane. We move $x$ into the complex plane by an angle $\theta$,
\begin{equation}
    x=e^{i\theta}y\,,\qquad y\in\mathbb{R}\,,\qquad\theta\in\left[-\theta_\text{max},\theta_\text{max}\right]
\end{equation}
and deform the contour of integration of each Born iteration \eqref{eq:chin}, such that the new contour of integration is inside the domain \eqref{eq:Van}, as shown in Figure \ref{fig:xcontour}.

In particular, for the first iteration, we have 
\begin{align}\label{eq:complexxint1}
\chi^{(1)}_L(\omega,x) &=  \int_{- e^{i\theta}\infty}^x dx'\, G_0(\omega,x-x') \,V(x') \\
&= \int_{- \infty}^y dy'\, e^{i \theta} G_0\big(\omega, e^{i \theta}(y-y')\big) \,V(e^{i \theta} y')\,,
\end{align}
where
\begin{equation}
\label{eq:G0y}
     G_0\big(\omega, e^{i \theta}(y-y')\big) = {e^{2i \omega e^{i \theta}  (y- y')} - 1 \over 2 i \omega}\,,
\end{equation}
and $y'$ parameterizes the contour, with $y'<0$ being integration in the left half plane of $x'$ and $y'>0$ being integration in the right half plane of $x'$ up to the point $x=e^{i\theta}y$.

We can write the series expansion in \eqref{eq:chin} along the deformed contour as
\begin{align}
\label{eq:chiny}
    \chi_L^{(n)}(\omega,e^{i \theta} y) &= \int_{- \infty}^y  e^{i \theta} G_0\big(\omega, e^{i \theta}(y-y')\big) \,V(e^{i \theta} y') \, \chi^{(n-1)}_L(\omega,e^{i\theta}y') \, dy' \\
    &= e^{in \theta} \int_{y>y_1>\dots>y_n>-\infty}  G_0\big(\omega, e^{i \theta}(y-y_1)\big) \cdots G_0\big(\omega, e^{i \theta}(y_{n-1}\!-y_n)\big) \notag \\
    &\qquad \qquad \qquad\qquad \qquad \qquad \;\;\;\; \times  \;V(e^{i \theta} y_1) \dots V(e^{i \theta} y_n) \; dy_1 \cdots dy_n \,.\notag
\end{align}
The bound \eqref{eq:G0bound} now reads
\begin{equation}
\label{eq:G0boundy}
    | G_0\big(\omega, e^{i \theta}(y-y')\big)| \leq {1 \over |\omega|} \qquad \text{ for }\;\; y - y' > 0 \;\; \text{ and } \;\; \mathrm{Im} \, (e^{i \theta} \omega) > 0.
\end{equation}
We therefore find the bound on the $n$th term in the series
\begin{align}
    |\chi^{(n)}_L(\omega,e^{i \theta} y)| &\leq {1 \over |\omega|^n} \int_{y>y_1>\dots>y_n>-\infty} |V(e^{i \theta} y_1)| \cdots |V(e^{i \theta} y_n)| \, dy_1 \dots dy_n \notag \\
    &= {1 \over n!\, |\omega|^n} \left[\int_{- \infty}^y |V(e^{i \theta}y')| \,dy' \right]^n, \qquad  \mathrm{Im} \, (e^{i \theta} \omega) > 0.
\end{align}
This implies that the series \eqref{eq:serieschi} is uniformly convergent,
\begin{equation}
\label{eq:chiboundy1}
    |\chi_L(\omega,e^{i \theta} y)| \leq \sum_{n=0}^\infty  |\chi_L^{(n)}(\omega,e^{i \theta} y)| \leq \exp \left[ {1 \over |\omega|} \int_{- \infty}^y |V(e^{i \theta}y')| \,dy' \right], \qquad  \mathrm{Im} \, (e^{i \theta} \omega) > 0.
\end{equation}

Since each term in the Born series in analytic, this bound establishes that the Jost solution $\phi_L(\omega,e^{i \theta} y) = e^{- i \omega e^{i \theta}y} \chi_L(\omega,e^{i \theta} y)$ is analytic for $\mathrm{Im} \, (e^{i \theta} \omega) > 0$.
Since the quantity $\chi_L(\omega,e^{i\theta}y)$ is analytic and reduces to the original definition of the Jost solution for real $\omega$, in equation \eqref{eq:chiL}, it is a valid analytic continuation of $\chi_L(\omega,x)$ in the domain \eqref{eq:Van}.

\textbf{Analyticity of $A_\text{in}(\omega)$.} With this result we can show that the connection coefficient $A_\text{in}(\omega)$ given in \eqref{eq:Ainchi} is analytic in a slice of the lower half-plane. This follows from deforming the contour of the integral at an angle $\theta$:
\begin{equation}
\label{eq:Ainchi2}
    A_{\mathrm{in}}(\omega) = 1 - {e^{i \theta} \over 2 i \omega} \int_{- \infty}^\infty V(e^{i \theta} y) \,\chi_L(\omega,e^{i \theta} y) \,dy.
\end{equation}
This expression is analytic for $\mathrm{Im} \, (e^{i \theta} \omega) > 0$, and given the range of analyticity \eqref{eq:Van} of the potential, we find that
\begin{equation}
    A_\text{in}(\omega) \text{ is analytic in the range }  - \theta_\text{max} < \arg \omega < \pi \cup -\pi < \arg \omega< -\pi+ \theta_\text{max}
\end{equation}
as depicted in the dotted region of Figure~\ref{fig:AinAoutRanalytic}.

\textbf{Analyticity of $T(\omega)$.} Given the relation \eqref{eq:RT} we establish that $T(\omega)$ is meromorphic in the domain
\begin{equation}
 - \theta_\text{max} < \arg \omega < \pi \cup -\pi < \arg \omega< -\pi+ \theta_\text{max}
\end{equation}
shown as the dotted region in Figure~\ref{fig:AinAoutRanalytic}.

\subsection{Extended analyticity for the reflection amplitude}

\label{sec:analyticpotentials2} 

We now show that a similar argument involving contour deformation into the complex $x$-plane allows to prove an extended domain of analyticity for the reflection coefficient $R(\omega)$. Given the relation \eqref{eq:RT} and having established analyticity for $A_\text{in}(\omega)$ it remains to show analyticity for $A_\text{out}(\omega)$.

\begin{figure}[t]
  \centering
  \begin{tikzpicture}[>=stealth]
    \draw[->,thick] (0,-3) -- (0,3);
    \draw[black, ->,thick] (-5,0) -- (5,0);
    \draw[blue,thick] (-5,-3) -- (0,0);
    \draw[blue,thick] (0,0) -- (3,.9);
    \draw[->,>=angle 45,thick,blue] (-5,-3) -- (-3,-9/5);
    \draw [thick,domain=0:17] plot ({1.5*cos(\x)}, {1.5*sin(\x)});
    \node at (1.9,0.225) {$\theta_+$};
    \draw [thick,domain=180:211] plot ({1.5*cos(\x)}, {1.5*sin(\x)});
    \node at (-1.8,-0.5) {$\theta_-$};

    \fill[blue] (3,.9) circle (2pt) node[above] {$y$};
    \fill[blue] (2.5,.75) circle (2pt) node[above] {$y_1$};
    \fill[blue] (1,.3) circle (2pt) node[above] {$y_j$};
    \node[rotate=15,blue] at (1.8,0.83) {\Large $\cdots$};

    \fill[blue] (-.9,-0.55) circle (2pt) node[below right] {$y_{j+1}$};
    \fill[blue] (-2.15,-1.3) circle (2pt) node[below right] {$y_n$};
    \node[rotate=32,blue] at (-1.25,-1.12) {\Large $\cdots$};

    \node at (4.85,2.85) {$x'$};
    \draw[thick] (4.60,3) -- (4.6,2.6);
    \draw[thick] (4.6,2.6) -- (5,2.6);

    \node at (-4.85,-2.4) {\color{blue}$\gamma(x)$};
  \end{tikzpicture}
  \caption{Integration contour $\gamma(x)$ for the iterated integral in Eq.~\eqref{eq:chi1angles} in the complex-$x'$ plane with $x=e^{i\theta_-}y\,\Theta(-y)+e^{i\theta_+}y\,\Theta(y)$. The contour shown has $y>0$ and chooses $y_j$ to be the first integral where the integration variable is positive. $y_n,...,y_{j+1}$ are negative and rotated by $\theta_-$, while $y_j,...,y_1$ and the fixed variable $y$ are positive and rotated by $\theta_+$. }
  \label{fig:contour}
\end{figure}
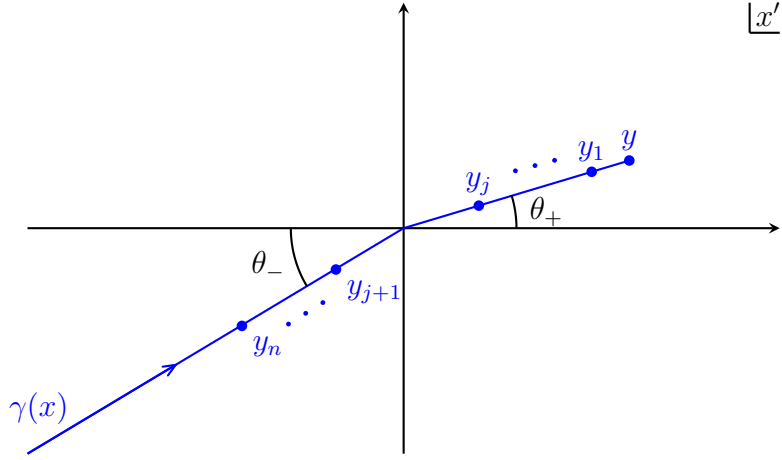

\textbf{Analyticity of $A_\text{out}(\omega)$.} We start again with extending the analyticity domain of $\chi_L(\omega,x)$, but instead of slating the full contour over $x$ at a single angle $\theta$, we deform the integration from $- \infty \to 0$ by an angle $\theta_-$ and from $0 \to + \infty$ by a different angle $\theta_+$. We write this new contour $\gamma(x)$, shown in Figure~\ref{fig:contour}, where
\begin{equation}
    x=e^{i\theta_-}y\,\Theta(-y)+e^{i\theta_+}y\,\Theta(y)\,.
\end{equation}

At leading order we write the analytic continuation of $\chi_L$ to this contour
\begin{align}\label{eq:chi1angles}
\chi^{(1)}_L(\omega,x) 
&=  \int_{\gamma(x)}  G_0(\omega,x-x') \,V(x')\, dx' \\
&= \Theta(-y)\int_{- \infty}^y  e^{i \theta_-} G_0\big(\omega, e^{i \theta_-}(y-y')\big) \,V(e^{i \theta_-} y')\, dy'\nonumber\\
&\quad+\Theta(y)\left[\int_{- \infty}^0  e^{i \theta_-} G_0\big(\omega, e^{i \theta_-}(y-y')\big) \,V(e^{i \theta_-} y')\, dy'\right.\nonumber \\
    &\qquad\qquad\quad\left.+ \int_{0}^{y}  e^{i \theta_+} G_0\big(\omega, e^{i \theta_+}(y-y')\big) \,V(e^{i \theta_+} y')\, dy'\right] \nonumber\,.
\end{align}
For the general iteration, we can rewrite \eqref{eq:chin} across the contour depicted in Figure \ref{fig:contour} in the compact form
\begin{align}
\label{eq:chiny2}
    \tilde{\chi}_L^{(n)}(\omega,y) &= \int_{- \infty}^y \tilde{G}_0(\omega,y,y') \,\tilde{V}(y')\, \tilde{\chi}^{(n-1)}_L(\omega,y')\, dy'  \\
    &=  \int_{y>y_1>\dots>y_n>-\infty} \tilde{G}_0(\omega,y,y_1) \cdots \tilde{G}_0(\omega,y_{n-1},y_n) \, \tilde{V}(y_1) \dots \tilde{V}(y_n) \, dy_1 \cdots dy_n\,, \notag
\end{align}
where
\begin{equation}
\label{eq:chisplit}
    \tilde{\chi}_L^{(n)}(\omega,y) \equiv \chi_L^{(n)}(\omega,e^{i \theta_-}y)\, \Theta(-y) \;+ \;\chi_L^{(n)}(\omega,e^{i \theta_+}y)\, \Theta(y)\,, 
\end{equation}
and similarly for the potential
\begin{equation}
\label{eq:Vsplit}
    \tilde{V}(y) \equiv e^{i \theta_-} V(e^{i \theta_-} y) \,\Theta(-y)\; + \;e^{i \theta_+} V(e^{i \theta_+} y) \,\Theta(y).
\end{equation}
The Green's function, depending on both and $y$ and $y'$, is given in terms of three disjoint contributions
\begin{align}
\label{eq:Gshit}
    \tilde{G}_0(\omega,y,y') &\equiv \;G_0\big(\omega, e^{i \theta_-}(y-y')\big) \,\Theta(-y) \,\Theta(-y') \\
     &+ \;G_0\big(\omega, e^{i \theta_+}y-e^{i \theta_-}y'\big) \,\Theta(y) \,\Theta(-y') \notag \\
      &+\;G_0\big(\omega, e^{i \theta_+}(y-y')\big) \,\Theta(y) \,\Theta(y') \,,\notag
\end{align} 
depending on which branch $y$ and $y'$ are on. Note that $y > y'$ implies that a possible fourth piece proportional to $\Theta(-y) \Theta(y')$ does not have support.

Notice also that for $y < 0$, the ordering of the integral \eqref{eq:chiny2} implies that all integration variables must be negative, $y_j < 0$, meaning that there is no dependence on $\theta_+$. In this case we just recycle what we concluded in the previous analysis, namely the bound \eqref{eq:chiboundy1} replacing $\theta = \theta_-$.

On the other hand, for $y > 0$ both $\theta_+$ and $\theta_-$ are relevant. We can separate the iterated integrals in \eqref{eq:chiny2} in a way that makes explicit the split between $y_j > 0$ and $y_j < 0$ regions. In particular,
\begin{equation}
    \int_{y>y_1>\dots>y_n>-\infty} \!\!\!\! dy_1 \cdots dy_n = \sum_{j=0}^{n} \int_{y>y_1> \cdots>y_{j}>0}  \!\!\!\! dy_1 \cdots dy_{j} 
   \int_{0>y_{j+1}> \cdots>y_{n}>-\infty}  \!\!\!\! dy_{j+1} \cdots dy_n
\end{equation}
where $y_0 \equiv y$ is assumed positive and not integrated over, see Figure \ref{fig:contour}.

Using this split integral representation we may now plug the explicit expressions \eqref{eq:chisplit}, \eqref{eq:Vsplit} and \eqref{eq:Gshit} into \eqref{eq:chiny2}. So we find
\begin{align}
\label{eq:chiny3}
 &\tilde{\chi}_L^{(n)}(\omega,y) = \notag \\
 &\sum_{j=0}^n \Bigg[\int_{y>y_1> \cdots>y_j>0}   dy_1 \cdots dy_{j}\, e^{i \theta_+j} \; G_0\big(\omega, e^{i \theta_+}(y-y_1)\big)\cdots \,G_0\big(\omega, e^{i \theta_+}(y_{j-1}-y_{j})\big)\notag \\
 &\;\;\;\;\;\;\;\times V(e^{i \theta_+} y_1) \cdots V(e^{i \theta_+} y_{j})\Bigg] \\
   &\;\;\times\Bigg[\int_{0>y_{j+1}> \cdots>y_{n}>-\infty} dy_{j+1} \cdots dy_n\, e^{i \theta_-(n-j)} \; G_0\big(\omega, e^{i \theta_+}y_{j}-e^{i \theta_-}y_{j+1}\big)  \;\notag \\
   &\;\;\;\;\;\;\; \times \; G_0\big(\omega, e^{i \theta_-}(y_{j+1}-y_{j+2})\big) \cdots \,G_0\big(\omega, e^{i \theta_-}(y_{n-1}-y_n)\big) \; V(e^{i \theta_-} y_{j+1}) \cdots V(e^{i \theta_-} y_n)\Bigg]\notag \,.
\end{align}
Now we will bound this integral in the (shaded) region\footnote{The figure shows $\theta_->\theta_+>0$, but the derivation holds for any $\theta_-$ and $\theta_+$.}
\begin{equation}\label{eq:domainy}
\mathrm{Im}(e^{i \theta_-} \omega) > 0 \; \cap \; \mathrm{Im}(e^{i \theta_+} \omega) < 0 :\qquad\raisebox{-3.5em}{
\begin{tikzpicture}[line width=1.1, scale=0.5,>=stealth]
    \fill[gray!25] (0,0) -- (5,-3) -- (5,-1.5) --cycle;
    \draw[thick] (0,0) -- (5,-3);
    \draw[thick] (0,0) -- (5,-1.5);
    \draw [thick,domain=-17:0] plot ({1.5*cos(\x)}, {1.5*sin(\x)});
    \node at (2.5,-0.45) {\footnotesize $-\theta_+$};
    \draw [thick,domain=-31:0] plot ({3.5*cos(\x)}, {3.5*sin(\x)});
    \node at (4,-1.8) {\footnotesize $-\theta_-$};

    \draw[->,thick] (0,-3) -- (0,3);
    \draw[ ->,thick] (-5,0) -- (5,0);
    
    \node at (4.85,2.85) {\footnotesize $\omega$};
    \draw[thin] (4.5,3) -- (4.5,2.6);
    \draw[thin] (4.5,2.6) -- (5,2.6);
\end{tikzpicture}
}
\end{equation}
The motivation for choosing this region with $\mathrm{Im}(e^{i \theta_+} \omega) < 0$ is so that the factor $e^{-2i\omega x'}$ in the integral \eqref{eq:Aoutchi} is suppressed for $\text{Re}(x')>0$. The other condition $\mathrm{Im}(e^{i \theta_-} \omega) > 0$ provides exponential suppression in the opposite, $\text{Re}(x')<0$ direction.
If $\theta_+ = \theta_-$ the overlapping region is empty, hence we take these two angles to be different. 

To proceed we note the following bounds on the Green's function $G_0$ valid in this domain,
\begin{align}
|G_0\big(\omega, e^{i \theta_+}(y-y')\big)| &< {e^{2 |\mathrm{Im}(e^{i \theta_+}\omega )|(y-y')} \over |\omega|},  \\
|G_0\big(\omega, e^{i \theta_+}y-e^{i \theta_-}y'\big) |  &< {e^{2 |\mathrm{Im}(e^{i \theta_+}\omega )|y} \over |\omega|}, \\
 |G_0\big(\omega, e^{i \theta_-}(y-y')\big)| & < {1 \over |\omega|}.
\end{align}
Applying these bounds to \eqref{eq:chiny3} we find
\begin{align}
\label{eq:chinboundy3}
 |\tilde{\chi}_L^{(n)}(\omega,y)| &\leq \sum_{j=0}^n \int_{y>y_1> \cdots>y_j>0}   {e^{2 |\mathrm{Im}(e^{i \theta_+}\omega )|(y-y_j)} \over |\omega|^j}\; |V(e^{i \theta_+} y_1)| \cdots |V(e^{i \theta_+} y_{j})| \\
   &\;\;\;\;\; \times\int_{0>y_{j+1}> \cdots>y_{n}>-\infty}   \; {e^{2 |\mathrm{Im}(e^{i \theta_+}\omega )|y_j} \over |\omega|} \times {1 \over |\omega|^{n-j-1}}  \times |V(e^{i \theta_-} y_{j+1})| \cdots |V(e^{i \theta_-} y_n)| \notag  \\
&= { e^{2 |\mathrm{Im}(e^{i \theta_+}\omega )|y} \over |\omega|^n} \sum_{j=0}^n \int_{y>y_1> \cdots>y_j>0} |V(e^{i \theta_+} y_1)| \cdots |V(e^{i \theta_+} y_{j})| \; dy_1 \cdots dy_j \notag \\
& \qquad \qquad \qquad \qquad  \times \int_{0>y_{j+1}> \cdots>y_{n}>-\infty} |V(e^{i \theta_-} y_{j+1})| \cdots |V(e^{i \theta_-} y_n)| \; dy_{j+1} \cdots dy_n \notag 
\end{align}
Since both integrands are fully symmetric under permutation of the $y_j$, we find the bound
\begin{equation}
    |\tilde{\chi}_L^{(n)}(\omega,y)| \leq { e^{2 |\mathrm{Im}(e^{i \theta_+}\omega )|y} \over |\omega|^n} \sum_{j=0}^n {1 \over j!(n-j)!} \bigg[ \int_0^y dy' \,|V(e^{i \theta_+} y')| \bigg]^j \bigg[ \int^0_{-\infty} dy' \,|V(e^{i \theta_-} y')| \bigg]^{n-j}.
\end{equation}
Finally, we use the binomial theorem to arrive at
\begin{equation}
    |\tilde{\chi}_L^{(n)}(\omega,y)| \leq { e^{2 |\mathrm{Im}(e^{i \theta_+}\omega )|y} \over n! \, |\omega|^n} \bigg[\int_0^y dy' \,|V(e^{i \theta_+} y')| +  \int^0_{-\infty} dy' \,|V(e^{i \theta_-} y')|\bigg]^n.
\end{equation}
This bound implies that the series for $\tilde{\chi}_L(\omega,y)$ is uniformly convergent,
\begin{align}
    |\chi_L(\omega,e^{i \theta_+}y)|& \leq \sum_{n=0}^\infty |\tilde{\chi}_L^{(n)}(\omega,y)| \notag \\
    &\leq \exp \left[2 |\mathrm{Im}(e^{i \theta_+}\omega )|y + \frac{1}{|\omega|}\int_0^y dy' \,|V(e^{i \theta_+} y')| + \frac{1}{|\omega|} \int^0_{-\infty} dy' \,|V(e^{i \theta_-} y')| \right],
\end{align} 
and thus $\chi_L(\omega,e^{i \theta_+}y)$ is analytic in the domain \eqref{eq:domainy}.

We are now ready to prove analyticity of $A_\text{out}(\omega)$ in this domain. We deform the integration contour of \eqref{eq:Aoutchi} along the same rays slated by $\theta_-$ and $\theta_+$, 
\begin{align}
\label{eq:Aoutchi2}
    A_{\mathrm{out}}(\omega) &= {e^{i \theta_-} \over 2 i \omega} \int_{- \infty}^0 e^{- 2 i \omega \, e^{i \theta_-}y} \; V(e^{i \theta_-}y) \;\chi_L(\omega,e^{i \theta_-}y) \,dy  \\
    &\;\;\;+ \;{e^{i \theta_+} \over 2 i \omega} \int^{+ \infty}_0 e^{- 2 i \omega \, e^{i \theta_+}y} \; V(e^{i \theta_+}y) \;\chi_L(\omega,e^{i \theta_+}y) \,dy \,\notag.
\end{align}
The integral in the first line is convergent for $\mathrm{Im}(e^{i \theta_-} \omega) > 0$. This follows from: (i) the factor $e^{- 2 i \omega \, e^{i \theta_-}y} \lesssim e^{-2 |\mathrm{Im}(\omega \, e^{i \theta_-})| |y|}$ providing exponential damping, (ii) the bound on $\chi_L(\omega,e^{i \theta_-} y)$ given in \eqref{eq:chiboundy1}, and (iii) the assumptions \eqref{eq:Van} that the potential decays sufficiently fast in the complex plane and has no singularities along this ray.

Likewise, the integral in the second line is convergent for $\mathrm{Im}(e^{i \theta_+} \omega) < 0$. Again the factor $e^{- 2 i \omega \, e^{i \theta_+}y} \lesssim e^{-2 |\mathrm{Im}(\omega \, e^{i \theta_+})| |y|}$ provides exponential damping which in this case is useful to cancel  the exponential growth of the bound \eqref{eq:chinboundy3} on the Jost solution, $\chi_L(\omega,e^{i \theta_+} y) \lesssim e^{2 |\mathrm{Im}(\omega \, e^{i \theta_+})| |y|}$. Finally, the assumptions \eqref{eq:Van} guarantee that the potential decays sufficiently fast in the complex plane and has no singularities along this ray.

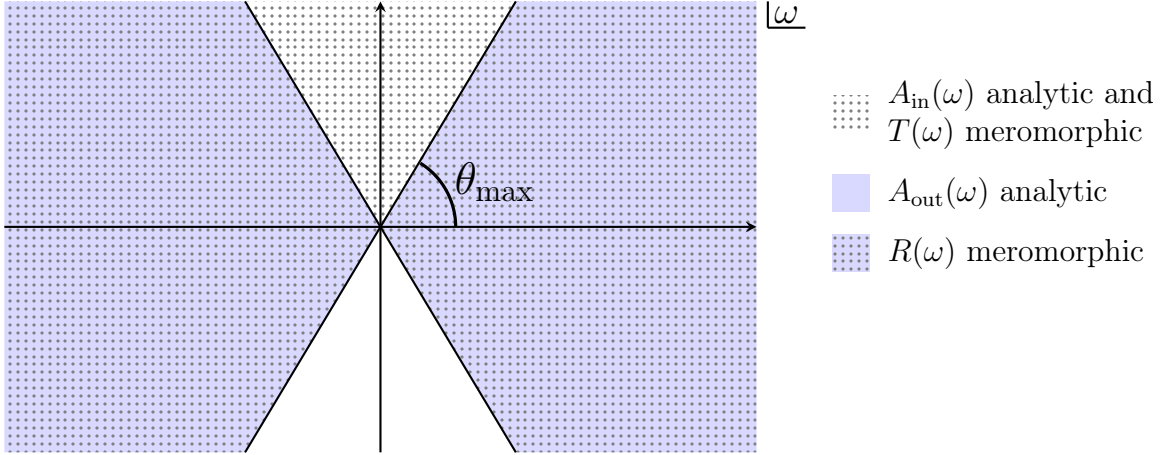
\begin{figure}[t]
  \centering
  \begin{tikzpicture}[>=stealth]
    \fill[blue!15] (0,0) -- (1.8,3) -- (5,3) -- (5,-3)-- (1.8,-3) --cycle;
    \fill[blue!15] (0,0) -- (-1.8,-3) -- (-5,-3) -- (-5,3)-- (-1.8,3) --cycle;
    \fill[pattern={Dots[distance=3pt,radius=.6pt]},pattern color=gray] (0,0) -- (-1.8,-3) -- (-5,-3) -- (-5,3)-- (5,3) -- (5,-3) -- (1.8,-3) -- cycle;
    
    \draw[thick] (1.8,3) -- (-1.8,-3);
    \draw[thick] (-1.8,3) -- (1.8,-3);
    \draw [line width=1.2pt,domain=0:58] plot ({cos(\x)}, {sin(\x)});
    \node at (1.45,0.6) {{ \Large $\theta_{\textrm{max}}$}};

    \draw[->,thick] (0,-3) -- (0,3);
    \draw[ ->,thick] (-5,0) -- (5,0);
    
    \node at (5.4,2.825) {{\large $\omega$}};
    \draw[thick] (5.15,3) -- (5.15,2.65);
    \draw[thick] (5.15,2.65) -- (5.65,2.65);

  \begin{scope}[shift={(6,1.5)}] 
    \fill[pattern={Dots[distance=3pt,radius=.6pt]},pattern color=gray] (0,-.25) rectangle (0.5,0.25);
    \node[right] at (0.6,0.25) {$A_\text{in}(\omega)$ analytic and};
    \node[right] at (0.6,-.25) {$T(\omega)$ meromorphic};

    \fill[blue!15] (0,-0.8-.5) rectangle (0.5,-0.3-.5);
    \node[right] at (0.6,-0.55-.5) {$A_\text{out}(\omega)$ analytic};

    \fill[blue!15] (0,-1.6-.5) rectangle (0.5,-1.1-.5);
    \fill[pattern={Dots[distance=3pt,radius=.6pt]},pattern color=gray] (0,-1.6-.5) rectangle (0.5,-1.1-.5);
    \node[right] at (0.6,-1.35-.5) {$R(\omega)$ meromorphic};
  \end{scope}
  \end{tikzpicture}
  \caption{Summary of the established analyticity domains derived for a potential $V(x)$ satisfying \eqref{eq:Van} and \eqref{eq:Vbound}. $A_\text{in}(\omega)$ is analytic in the dotted region and $A_\text{out}(\omega)$ is analytic in the shaded blue region. Therefore $T(\omega)$ is meromorphic in the dotted region and  $R(\omega)$ is meromorphic in the overlap of the dotted and blue shaded regions. The regions are defined by diagonal lines at an angle $\pm\theta_\text{max}$ with respect to the real axis.}
  \label{fig:AinAoutRanalytic}
\end{figure}

This proves that $A_\text{out}(\omega)$ is analytic in the domain \eqref{eq:domainy}. 
Since $V(x)$ is analytic in the domain \eqref{eq:Van},
we conclude that $A_\text{out}(\omega)$ is analytic in the range
\begin{equation}
- \theta_\text{max} < \arg \omega < \theta_\text{max}\;\cup\; \pi- \theta_\text{max} < \arg \omega <\pi\;\cup\; -\pi < \arg \omega <-\pi+ \theta_\text{max}\,.
\end{equation}
This region is shown as the blue shaded region in Figure~\ref{fig:AinAoutRanalytic}.

\textbf{Analyticity of $R(\omega)$.} We are poised to translate this result into analyticity of the reflection coefficient $R(\omega)$. Given formula \eqref{eq:RT}, the analyticity domain of $R(\omega)$ is the intersection of the analyticity domains of $A_\text{in}(\omega)$ and $A_\text{out}(\omega)$, excluding zeroes of $A_\text{in}(\omega)$ which lead to poles. In summary, $R(\omega)$ is meromorphic for
\begin{equation}
- \theta_\text{max} < \arg \omega < \theta_\text{max}\;\cup\; \pi- \theta_\text{max} < \arg \omega <\pi\;\cup\; -\pi < \arg \omega <-\pi+ \theta_\text{max}\,,
\end{equation}
shown as the overlapping dotted and blue shaded region in Figure~\ref{fig:AinAoutRanalytic}.\footnote{The proof can be generalized for a potential that is analytic and integrable in a region different to the one defined in Eq.~\eqref{eq:Van}. Loosely speaking, if the potential is analytic through an angle in the left (right) half plane of $x$, then $R(\omega)$ is meromorphic through that angle in the lower (upper) half plane of $\omega$. Similarly, if the potential is analytic through an angle in the lower (upper) half plane of $x$, then $R(\omega)$ is meromorphic through that angle in the right (left) half plane of $\omega$.}

Note that while we have proved regions of analyticity and meromorphicity of the different functions, we have \textit{not} proven that the complements of these regions \textit{will} have singularities. We are ignorant about the analytic structure outside of the determined regions until the exact solution for a specific potential is obtained. 

We will now show that the location of the poles in the S-matrix, coming from $A_\text{in}(\omega) = 0$, can also be constrained based on general arguments.

\subsection{Poles of the S-matrix: $A_\text{in}(\omega)=0$}\label{sec:smatrixpoles}

In the above section, we have seen that the reflection amplitude is meromorphic in the blue and dotted region of the complex-$\omega$ plane shown in Figure~\ref{fig:AinAoutRanalytic}. Within this region, there can be isolated poles arising from zeros of $A_{\text{in}}$. We now show that they can only occur either everywhere in the LHP or along the imaginary axis of the UHP. 

We start with the differential equation~\eqref{eq:waveeq} written as
\begin{equation}
\label{ode}
D\phi(\omega,x)=\omega^2\,\phi(\omega,x)
\end{equation}
with
\begin{equation}
    D\equiv-\frac{d^2}{dx^2}+V(x)\,.
\end{equation}
We multiply equation (\ref{ode}) by $\phi^*$ and subtract the complex-conjugate equation after multiplying by $\phi$, resulting in the following expressions:
\begin{align}
\phi^*D\phi-(D\phi)^*\phi
&= \big(\omega^2-(\omega^2)^*\big)\,|\phi|^2,\\
\phi^*D\phi-(D\phi)^*\phi
&= -\,\frac{d}{dx}\!\big(\phi^*\phi'-(\phi^*)'\phi\big),
\end{align}
where we temporarily suppress the dependence on the variables $(\omega,x)$.
Integrating from $x=a$ to $x=b$ yields
\begin{equation}
\label{eq:greensturm}
\int_a^b \! dx \,\big(\phi^* D\phi - (D\phi)^* \phi\big)
= -\,\Big[\phi^* \phi' - (\phi^*)' \phi\Big]_{x=a}^{x=b}
= -\,\big((\omega^2)^* - \omega^2\big)\!\int_a^b |\phi|^2\,dx.
\end{equation}

Importantly, when $\omega_0$ solves $A_\text{in}(\omega_0)=0$, the Jost solution $\phi_L(\omega=\omega_0,x)$ obeys  outgoing boundary conditions at \emph{both} $x\to\pm\infty$, following (\ref{eq:bcLR}) and (\ref{eq:phidecomposition}). 

Therefore, in the limit \(b\to+\infty\), \(a\to-\infty\), equation \eqref{eq:greensturm} with $\phi=\phi_L$ becomes
\begin{equation}
\label{eq:Bplusminus}
\Big[\phi_L^*\phi_L-(\phi_L^*)'\phi_L\Big]_{a}^{b}
=2i\,\mathrm{Re}\,\omega_0\Big(|A_\text{out}|^2 e^{-2\,\mathrm{Im}\,\omega_0\,b}+e^{+2\,\mathrm{Im}\,\omega_0\,a}\Big).
\end{equation}
Substituting (\ref{eq:Bplusminus}) into equation (\ref{eq:greensturm}) we find
\begin{equation}
\label{eq:master}
\mathrm{Re}\,\omega_0 \,\Big( |A_\text{out}|^2 e^{-2\,\mathrm{Im}\,\omega_0\,b}+e^{+2\,\mathrm{Im}\,\omega_0\,a}\Big)
=-2 \,\mathrm{Re}\,\omega_0\,\mathrm{Im}\,\omega_0 \int_a^b |\phi_L|^2 dx.
\end{equation}

The relation \eqref{eq:master} constrains the location of $\omega_0$ to one of two possibilities. Either $\omega_0$ is purely imaginary ($\mathrm{Re}\, \omega_0 = 0$), with both of sides of the equation vanishing.
In this case, it is possible to have $\omega_0 = i \kappa$, with $\kappa > 0$, corresponding to a bound-state \cite{Mizera:2023tfe}.\footnote{While these bound-state poles are located in the UHP they do not contradict causality. In particular, using a Laplace transform instead of a Fourier transform in the definition of the retarded Green’s function \eqref{eq:FTGR} explicitly shows the preservation of causality. See Appendix~\ref{app:boundstates} for details.}

The other scenario is $\mathrm{Re}\, \omega_0 \neq 0$, with Eq.~\eqref{eq:master} implying that $\mathrm{Im} \, \omega_0$ is negative
\begin{equation}
\label{eq:ImQNM}
    \mathrm{Im}\, \omega_0 = -{|A_\text{out}|^2 e^{-2\,\mathrm{Im}\,\omega_0\,b}+e^{+2\,\mathrm{Im}\,\omega_0\,a} \over 2 \int_a^b |\phi_L|^2 dx} < 0\,,
\end{equation}
corresponding to a quasinormal mode (QNM). A QNM is a non-normalizable solution that diverges at spatial infinity and is exponentially damped in time.\footnote{Note that in this case the limit $b \to + \infty$, $a \to - \infty$ must be taken with care as both numerator and denominator in \eqref{eq:ImQNM} become divergent. Their ratio should however give a finite result given that the left-hand-side of \eqref{eq:ImQNM} is finite, even though we have not shown this explicitly here. See Chapter 12 of \cite{Taylor:1972} for a different argument that does not suffer from this limitation.} They are expected to play a role in the ringdown phase of black hole mergers \cite{Regge:1957td,Vishveshwara:1970cc,Chandrasekhar:1975zza,Berti:2009kk,Press:1971wr,Vishveshwara:1970zz,KonoplyaZhidenko2011, Ferrari:1984zz}.

In summary, we have the two possibilities:
\[
\boxed{
\begin{cases}
 \text{Re}\,\omega_0 \in \mathbb{R} \text{ and } \text{Im}\,\omega_0<0 \quad (\text{QNM}),\\[4pt]
 \text{Re}\,\omega_0 = 0 \text{ and } \text{Im}\,\omega_0>0  \quad (\text{Bound-state}).
\end{cases}
}
\]

Let us now show that the bound-state poles, located in the UHP, are not present for scattering potentials that are always positive, as is the case for the Schwarzschild potential.

Taking $\omega_0=i\kappa$, with $\kappa>0$ and substituting into \eqref{ode} gives
\begin{equation}
-\phi_L''+V\phi_L=-\kappa^2\phi_L.
\end{equation}
which, after multiplying by $\phi_L^*$ and integrating on the real axis results in
\begin{equation}
\label{eq:eqboundstate}
\int_{-\infty}^\infty\!\Big(|\phi_L'|^2+V|\phi_L|^2+\kappa^2|\phi_L|^2\Big)\,dx
=\Big[\phi_L^*\phi_L'\Big]_{-\infty}^{+\infty}.
\end{equation}
Since $A_{\text{in}}(i \kappa)=0$, the Jost solution $\phi_L(i \kappa,x)$ is exponentially suppressed in both limits $x\to\pm\infty$, given the boundary conditions \eqref{eq:bcLR} and \eqref{eq:phidecomposition}, setting the RHS  of \eqref{eq:eqboundstate} to zero. If \(V(x)>0\), each term on the LHS is nonnegative, enforcing $\phi_L(i\kappa,x)=0$. Therefore, all bound-state solutions are trivial and no bound-state poles are present for positive potentials.

\begin{figure}[t]
    \centering
    \begin{tikzpicture}[>=stealth]
        \draw[thick] (-8,5) -- (8,5);
        \draw[thick] (-8,1) -- (8,1);
        \draw[thick] (-8,-3) -- (8,-3);
        \draw[thick] (-8,-7) -- (8,-7);
        \draw[thick] (-8/3,5) -- (-8/3,-10.5);
        \draw[thick] (8/3,5) -- (8/3,-10.5);
        \node[above] at (-8+8/3,5+.28) {$V(x)$ property};
        \node[above] at (-8/3,5+.35) {$\Longrightarrow$};
        \node[align=center,above] at (0,5) {$G_R,T$ meromorphic \\ in shaded region};
        \node[align=center,above] at (16/3,5) {$R$ meromorphic \\ in shaded region};
        \node[align=center] at (-8+8/3,2.95) {Bounded:\\ $\int^{+\infty}_{-\infty} |V(x)| dx < \infty$};
        \begin{scope}[shift={(-8+8/3,-1.)},scale=0.5] 
            \node[align=center] at (0,0) {$V(x)$ analytic for \\ complex $x$ in the \\wedge \eqref{eq:Van} and bounded \\
            asymptotically $\eqref{eq:Vbound}$};
        \end{scope}
        \node[align=center] at (-8+8/3,-5) {Exponential decay: \\
        $V(x \to \pm \infty) \sim e^{-\mu_\pm |x|}$};
        \node[align=center] at (-8+8/3,-9) {Compactly supported:\\$V(x \to \pm \infty) \lesssim e^{-\mu_\pm |x|}$};
        \begin{scope}[shift={(0,3)},scale=0.5] 
            \fill[black!15] (-4,0) rectangle (4,3);
            \draw[->,thick] (0,-3) -- (0,3);
            \draw[ ->,thick] (-4,0) -- (4,0);
            \node at (3.8,2.8) {\footnotesize $\omega$};
            \draw[thin] (3.45,3) -- (3.45,2.5);
            \draw[thin] (3.45,2.5) -- (4,2.5);
        \end{scope}
        \begin{scope}[shift={(16/3,3)},scale=0.5] 
            \draw[->,thick] (0,-3) -- (0,3);
            \draw[ ->,thick] (-4,0) -- (4,0);
            \node at (3.8,2.8) {\footnotesize $\omega$};
            \draw[thin] (3.45,3) -- (3.45,2.5);
            \draw[thin] (3.45,2.5) -- (4,2.5);
        \end{scope}
        \begin{scope}[shift={(0,-1)},scale=0.5] 
            \fill[black!15] (0,0) -- (-1.8,-3) -- (-4,-3) -- (-4,3)-- (4,3) -- (4,-3) -- (1.8,-3) -- cycle;
            \draw[->,thick] (0,-3) -- (0,3);
            \draw[ ->,thick] (-4,0) -- (4,0);
            \node at (3.8,2.8) {\footnotesize $\omega$};
            \draw[thin] (3.45,3) -- (3.45,2.5);
            \draw[thin] (3.45,2.5) -- (4,2.5);
            \draw [thick,domain=-58:0] plot ({cos(\x)}, {sin(\x)});
            \node at (2,-0.7) {\footnotesize $-\theta_\text{max}$};
        \end{scope}
        \begin{scope}[shift={(16/3,-1)},scale=0.5] 
            \fill[black!15] (0,0) -- (1.8,3) -- (4,3) -- (4,-3)-- (1.8,-3) --cycle;
            \fill[black!15] (0,0) -- (-1.8,-3) -- (-4,-3) -- (-4,3)-- (-1.8,3) --cycle;
            \draw[->,thick] (0,-3) -- (0,3);
            \draw[ ->,thick] (-4,0) -- (4,0);
            \node at (3.8,2.8) {\footnotesize $\omega$};
            \draw[thin] (3.45,3) -- (3.45,2.5);
            \draw[thin] (3.45,2.5) -- (4,2.5);
            \draw [thick,domain=-58:58] plot ({cos(\x)}, {sin(\x)});
            \node at (1.8,.7) {\footnotesize $\theta_\text{max}$};
            \node at (2,-0.7) {\footnotesize $-\theta_\text{max}$};
        \end{scope}
        \begin{scope}[shift={(0,-5)},scale=0.5] 
            \fill[black!15] (-4,-1) rectangle (4,3);
            \draw[->,thick] (0,-3) -- (0,3);
            \draw[ ->,thick] (-4,0) -- (4,0);
            \node at (3.8,2.8) {\footnotesize $\omega$};
            \draw[thin] (3.45,3) -- (3.45,2.5);
            \draw[thin] (3.45,2.5) -- (4,2.5);
            \draw[-,thick] (-.5,-1) -- (.5,-1);
            \node[right] at (0,-1.5) {$-\frac{1}{2}${\footnotesize$\text{min}(\mu_-,\mu_+)$}};
        \end{scope}
        \begin{scope}[shift={(0,-9)},scale=0.5] 
            \fill[black!15] (-4,-3) rectangle (4,3);
            \draw[->,thick] (0,-3) -- (0,3);
            \draw[ ->,thick] (-4,0) -- (4,0);
            \node at (3.8,2.8) {\footnotesize $\omega$};
            \draw[thin] (3.45,3) -- (3.45,2.5);
            \draw[thin] (3.45,2.5) -- (4,2.5);
        \end{scope}
        \begin{scope}[shift={(16/3,-5)},scale=0.5] 
            \fill[black!15] (-4,-1) rectangle (4,1);
            \draw[->,thick] (0,-3) -- (0,3);
            \draw[ ->,thick] (-4,0) -- (4,0);
            \node at (3.8,2.8) {\footnotesize $\omega$};
            \draw[thin] (3.45,3) -- (3.45,2.5);
            \draw[thin] (3.45,2.5) -- (4,2.5);
            \draw[-,thick] (-.5,-1) -- (.5,-1);
            \node[right] at (0,-1.5) {$-\frac{1}{2}${\footnotesize$\text{min}(\mu_-,\mu_+)$}};
            \draw[-,thick] (-.5,1) -- (.5,1);
            \node[right] at (0.5,1.5) {$\frac{1}{2}${\footnotesize$\mu_+$}};
        \end{scope}
        \begin{scope}[shift={(16/3,-9)},scale=0.5] 
            \fill[black!15] (-4,-3) rectangle (4,3);
            \draw[->,thick] (0,-3) -- (0,3);
            \draw[ ->,thick] (-4,0) -- (4,0);
            \node at (3.8,2.8) {\footnotesize $\omega$};
            \draw[thin] (3.45,3) -- (3.45,2.5);
            \draw[thin] (3.45,2.5) -- (4,2.5);
        \end{scope}
    \end{tikzpicture}
    \caption{Summary of analyticity domains established for different potentials. The property of $V(x)$ listed in the left column implies the corresponding shaded regions of meromorphicity of the retarded Green's function and transmission amplitude (middle column) and the reflection amplitude (right column). The first two rows are proven in Section~\ref{sec:asymG}, and the last two rows are proven in Appendix~\ref{app:potentialsproof}. In all these cases, positivity of the potential further implies that poles can only be located in the lower-half-plane. The gravitational black hole potentials and corresponding S-matrix fall in the category of the second row with $\theta_\text{max} = \pi/2 - \epsilon$ with $\epsilon >0$ arbitrarily small. }
    \label{fig:allResults}
\end{figure}

To conclude, analyticity and boundedness of the potential $V(x)$, as expressed in \eqref{eq:Van} and \eqref{eq:Vbound}, respectively, imply 
\begin{equation}\label{eq:GTsummary}
\boxed{
\begin{split}
    &G_R(\omega,x,x')\text{ and }T(\omega)\text{ are meromorphic for }\\
    &- \theta_\text{max} < \arg \omega < \pi \cup -\pi < \arg \omega< -\pi+ \theta_\text{max}
    \end{split}
    }
\end{equation}
\begin{equation}\label{eq:Rsummary}
\boxed{
\begin{split}
&R(\omega)\text{ is meromorphic for }\\
&- \theta_\text{max} < \arg \omega < \theta_\text{max}\;\cup\; \pi- \theta_\text{max} < \arg \omega <\pi\;\cup\; -\pi < \arg \omega <-\pi+ \theta_\text{max}\,.
\end{split}
}
\end{equation}
Moreover, in these regions of meromorphicity, the only poles arise from zeros of $A_\text{in}(\omega)$ which are necessarily in the LHP due to positivity of the potential. We summarize the analyticity domains derived in this Chapter and Appendix \ref{app:potentialsproof} in Figure \ref{fig:allResults}. We will now apply these results to the exactly solvable P\"oschl-Teller potential, and later in Chapter \ref{sec:BH} to the Schwarzschild gravitational potential.

\section{Example: the P\"oschl-Teller potential}\label{sec:poschlteller}

In this Chapter we illustrate the general analyticity results derived previously for the scattering against the P\"oschl-Teller potential barrier \cite{Poschl:1933zz}. This model has been studied extensively due to its rich behavior and its ability to be solved analytically~\cite{Cevik:2016mnr}. In particular, it has been used as a toy model for the Schwarzschild black hole potential at small orbital angular momentum $\ell$, allowing one to estimate the location of QNMs~\cite{Ferrari:1984zz,Berti:2009kk}.

The  P\"oschl-Teller potential is given by
\begin{equation}\label{eq:PTpot}
    V(x)=\frac{\alpha^2\lambda(1-\lambda)}{\cosh^2(\alpha x)}
\end{equation}
with parameters $\alpha>0$ and $\lambda=1/2+ib$ with $b>0$.

\subsection{Analyticity domains of the S-matrix}
\begin{figure}
    \centering
    \begin{tikzpicture}
        \node[anchor=south west,inner sep=0] (image) at (0,0) {\includegraphics[width=1\linewidth]{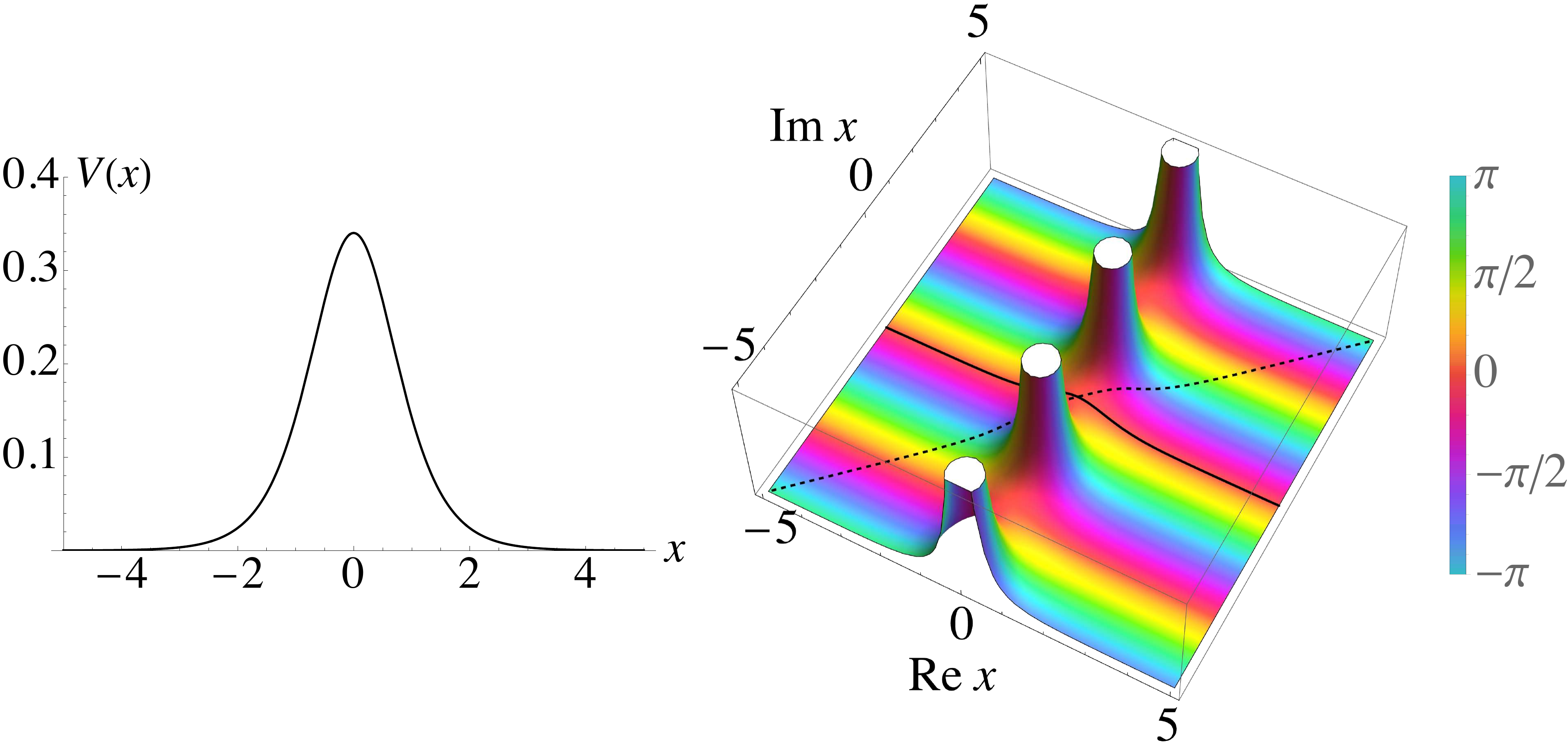}};
        \begin{scope}[x={(image.south east)},y={(image.north west)}]
            \draw [line width=1.pt,domain=-8.9:18.2,color=black] plot ({.5+.28*cos(\x)}, {.4+.28*sin(\x)});
            \node at (.795,0.43) {{\large  $\theta$}};
        \end{scope}
    \end{tikzpicture}
    \caption{The P\"oschl--Teller potential barrier $V(x)$ with $\alpha=1,\lambda=1/2+0.3\,i$ for \textbf{Left:} $x\in\mathbb{R}$ and \textbf{Right:} $x\in\mathbb{C}$. The height of the complex plot is the magnitude of the potential $|V(x)|$, while the color corresponds to the argument of the potential $\arg V(x)$. The potential has an infinite number of poles on the imaginary axis, located at $x=\tfrac{i}{\alpha}\left(\tfrac{\pi}{2}+2\pi  n\right)$, $n\in\mathbb{Z}$, and decays exponentially on the real axis as $x\rightarrow\pm\infty$. The solid line on the right plot shows the integration contour for real $x$, and the dashed line shows the integration contour for $x=e^{i\theta}y$. We can take $\theta \to \pi/2 - \epsilon$ with $\epsilon>0$ arbitrarily small without encountering singularities.}
    \label{fig:PTpotential}
\end{figure}

The P\"oschl-Teller potential \eqref{eq:PTpot} is bounded, satisfying \eqref{eq:Vintbound}, and decays exponentially
\begin{equation}\label{eq:PTfalloff}
    V(x\rightarrow \pm\infty)\sim e^{-2\alpha|x|}\,.
\end{equation}
Moreover, $V(x)$  can be analytically continued into the complex-$x$ plane, and has poles on the imaginary-$x$ axis. Therefore, it satisfies conditions \eqref{eq:Van} and \eqref{eq:Vbound} with $\theta_\text{max}=\pi/2- \epsilon$ with $\epsilon >0$ arbitrarily small. The potential is plotted in Figure~\ref{fig:PTpotential}, both for real $x$ and in the complex-$x$ plane. 

With these conditions met, following the analysis of Section~\ref{sec:analyticpotentials}, we conclude that the transmission and reflections coefficients, $T(\omega)$ and $R(\omega)$, have analyticity domains given in \eqref{eq:GTsummary} and \eqref{eq:Rsummary}, respectively. In particular, for $T(\omega)$ we show that it is analytic in the UHP and meromorphic in the LHP except on the negative imaginary axis. For the reflection amplitude $R(\omega)$ we show that it shares the same analytic structure of $T(\omega)$ in the LHP, and in the UHP it is analytic except along the imaginary axis, where it could have singularities. 

In Appendix~\ref{app:potentialsproof}, we also showed that for exponentially decaying potentials, the analytic domain in $\omega$ extends to an additional horizontal strip whose width is set by the exponential falloff (see Figure~\ref{fig:allResults}). For the Pöschl–Teller potential, this implies extra analyticity in the region $-\alpha <\text{Im}\, \omega<\alpha$.

We now analyze the exact solution for scattering off the P\"oschl-Teller potential barrier and verify these proven analyticity domains. 

\subsection{Inspection of the exact solution}

The exact Jost solutions for scattering against the P\"oschl-Teller potential with the desired boundary conditions 
are given by (see e.g. \cite{Berti:2009kk} for a derivation)
\begin{align}\label{eq:psisPT}
        \phi_L(\omega,x)&=\xi^{-\frac{i\omega}{2\alpha}}(1-\xi)^{-\frac{i\omega}{2\alpha}} {}_2F_1\left(\lambda-\frac{i\omega}{\alpha},1-\lambda-\frac{i\omega}{\alpha},1-\frac{i\omega}{\alpha},\xi\right)\,,\\
        \phi_R(\omega,x)&=\frac{\Gamma\left(1-\frac{i\omega}{\alpha}\right)\Gamma\left(-\frac{i\omega}{\alpha}\right)}{\Gamma\left(\lambda-\frac{i\omega}{\alpha}\right)\Gamma\left(1-\lambda-\frac{i\omega}{\alpha}\right)}\xi^{\frac{i\omega}{2\alpha}}(1-\xi)^{-\frac{i\omega}{2\alpha}} {}_2F_1\left(\lambda,1-\lambda,1+\frac{i\omega}{\alpha},\xi\right)\\
        &\quad-\frac{\Gamma\left(1+\frac{i\omega}{\alpha}\right)\Gamma\left(-\frac{i\omega}{\alpha}\right)}{\Gamma\left(\lambda\right)\Gamma\left(1-\lambda\right)}\xi^{-\frac{i\omega}{2\alpha}}(1-\xi)^{-\frac{i\omega}{2\alpha}} {}_2F_1\left(\lambda-\frac{i\omega}{\alpha},1-\lambda-\frac{i\omega}{\alpha},1-\frac{i\omega}{\alpha},\xi\right)\,,\notag
\end{align}
with
\begin{equation}
    \xi\equiv \frac{1+\tanh(\alpha x)}{2}\,,\qquad 0<\xi<1\,.
\end{equation}
From the asymptotic behavior of $\phi_L(\omega,x)$ in \eqref{eq:phidecomposition}, we extract the connection coefficients 
\begin{equation}\label{eq:PTAinAout}
    \begin{split}
        A_{\textrm{in}}(\omega)&= \frac{\Gamma\left(1-\frac{i\omega}{\alpha}\right)\Gamma\left(-\frac{i\omega}{\alpha}\right)}{\Gamma\left(1-\lambda-\frac{i\omega}{\alpha}\right)\Gamma\left(\lambda-\frac{i\omega}{\alpha}\right)}\,,\\
        A_{\textrm{out}}(\omega)&= \frac{\Gamma\left(1-\frac{i\omega}{\alpha}\right)\Gamma\left(\frac{i\omega}{\alpha}\right)}{\Gamma\left(1-\lambda\right)\Gamma\left(\lambda\right)}\,,
    \end{split}
\end{equation}
and build the retarded Green's function (for $x<x'$) via equation~\eqref{eq:Green}
\begin{equation}\label{eq:GPT}
    G_R(\omega,x,x')=-\frac{\Gamma\left(\lambda-\frac{i\omega}{\alpha}\right)\Gamma\left(1-\lambda-\frac{i\omega}{\alpha}\right)}{2i\omega\,\Gamma\left(1-\frac{i\omega}{\alpha}\right)\Gamma\left(-\frac{i\omega}{\alpha}\right)}\phi_L(\omega,x)\phi_R(\omega,x')\,,
\end{equation}
the transmission amplitude \eqref{eq:RT}
\begin{equation}
    T(\omega)=\frac{\Gamma\left(1-\lambda-\frac{i\omega}{\alpha}\right)\Gamma\left(\lambda-\frac{i\omega}{\alpha}\right)}{\Gamma\left(1-\frac{i\omega}{\alpha}\right)\Gamma\left(-\frac{i\omega}{\alpha}\right)}\,,
\end{equation}
and the reflection amplitude
\begin{equation}
    R(\omega)=-\frac{\Gamma\left(\frac{i\omega}{\alpha}\right)\Gamma\left(1-\lambda-\frac{i\omega}{\alpha}\right)\Gamma\left(\lambda-\frac{i\omega}{\alpha}\right)}{\Gamma\left(1-\lambda\right)\Gamma\left(\lambda\right)\Gamma\left(-\frac{i\omega}{\alpha}\right)}\,.
\end{equation}

\begin{figure}
    \centering
    \includegraphics[width=1\linewidth]{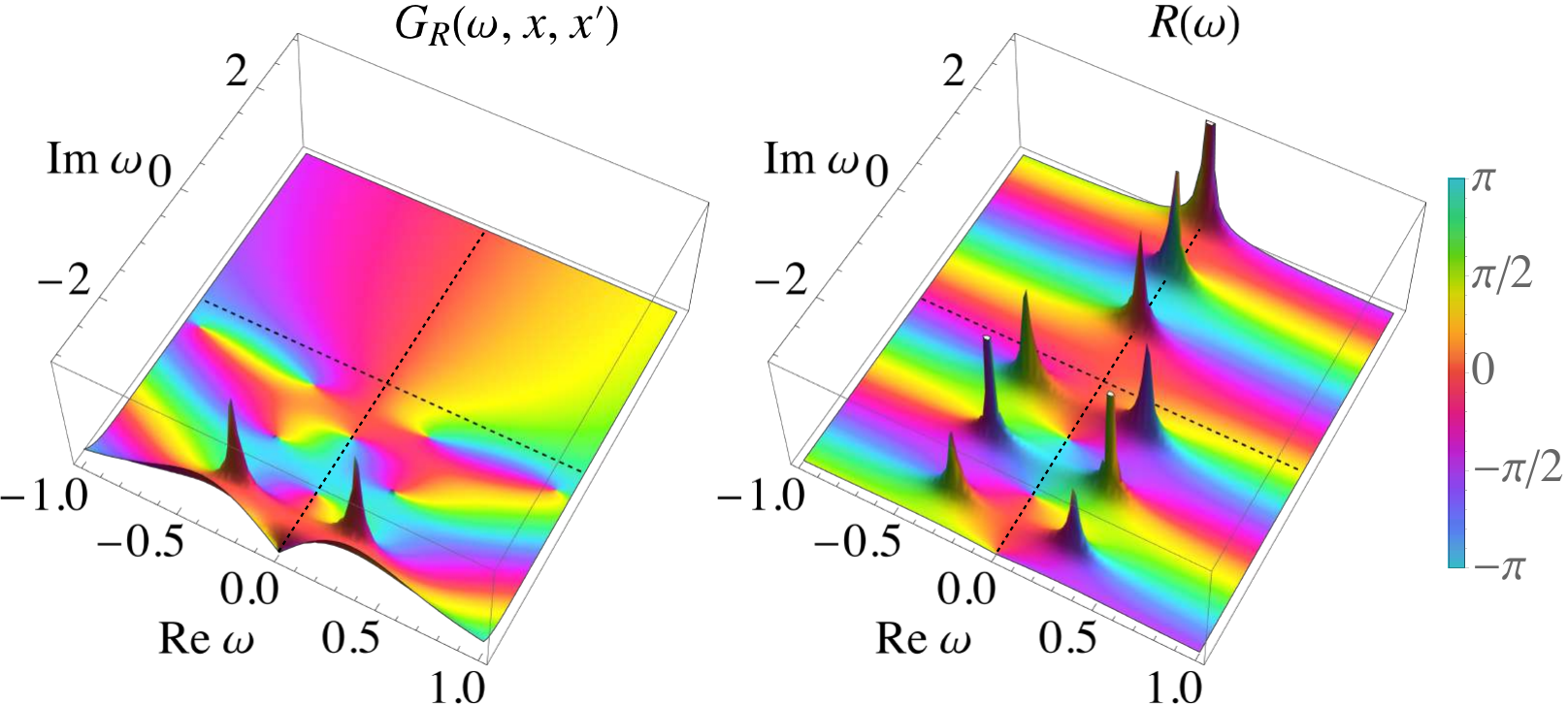}
    \caption{Solution to the P\"oschl Teller potential barrier with $\alpha=1,\lambda=1/2+0.3i$. Dotted lines mark the real and imaginary axes. \textbf{Left:} $G_R(\omega, x=2,x'=3)$. \textbf{Right:} $R(\omega)$. The poles in the LHP are QNMs. The poles in the UHP of $R(\omega)$ reflect the Stokes phenomenon.}
    \label{fig:PTGandS}
\end{figure}

\textbf{Analyticity domains.} Both functions, $G_R(\omega,x,x')$ and $R(\omega)$, are plotted in the complex-$\omega$ plane in Figure~\ref{fig:PTGandS}. The transmission amplitude $T(\omega)$ shares the same analytic structure as the retarded Green's function, so we do not discuss it further here. The retarded Green’s function and S-matrix both exhibit poles in the LHP with $\mathrm{Re}\, \omega \neq 0$, arising from zeros of the Wronskian. This condition defines the QNMs of the potential, characterized by $A_{\mathrm{in}}(\omega)=0$ and $\mathrm{Im}\, \omega <0$ (see Section \ref{sec:smatrixpoles}).

As required by causality, the Green's function is analytic in the UHP of $\omega$ for finite $x$ and $x'$. However, as anticipated by our general proof of analyticity, $R(\omega)$ has poles on the positive imaginary axis where $A_{\mathrm{out}}(\omega)$ diverges, due to the term $\Gamma\left(\frac{i\omega}{\alpha}\right)$. Finally, we confirm that both $G_R(\omega,x,x')$ and $R(\omega)$ are analytic within the strip $-\alpha< \text{Im}\, \omega < \alpha$ due to the exponential decay of the potential \eqref{eq:PTfalloff} (as shown in Appendix \ref{app:potentialsproof}).

\textbf{Stokes phenomena.} While $G_R(\omega,x,x')$ and $R(\omega)$ are related by the asymptotic relation \eqref{eq:classicalLSZintro}, we observe that they do not share the same analyticity domain. As we will now see, this is due to Stokes phenomena occurring in the limit $x,x' \to \infty$.

We first notice that $\phi_L(\omega,x)$ in \eqref{eq:psisPT} is manifestly analytic for $\mathrm{Im} \,\omega > 0$ because ${}_2F_1(a,b,c,\xi)$ only has singularities when $c$ is a non-positive integer. To see how the singularities of $R(\omega)$ appear, we rewrite the hypergeometric function using the identity
\begin{equation}
\begin{split}
    {}_2F_1(a,b,c,\xi)&=(1-\xi)^{c-a-b}\frac{\Gamma(c)\Gamma(a+b-c)}{\Gamma(a)\Gamma(b)}{}_2F_1(c-a,c-b,c-a-b+1,1-\xi)\\
    &\quad+\frac{\Gamma(c)\Gamma(c-a-b)}{\Gamma(c-a)\Gamma(c-b)}{}_2F_1(a,b,1-c+a+b,1-\xi)\,.
    \end{split}
\end{equation}
Therefore
\begin{equation}\label{eq:psi1PT1}
    \begin{split}
        \phi_L(\omega,x)=\xi^{-\frac{i\omega}{2\alpha}}(1-\xi)^{-\frac{i\omega}{2\alpha}} &\left[(1-\xi)^{\frac{i\omega}{\alpha}}A_{\textrm{in}}(\omega){}_2F_1\left(1-\lambda,\lambda,1+\frac{i\omega}{\alpha},1-\xi\right)\right.\\
        &\left.+A_{\textrm{out}}(\omega)\,{}_2F_1\left(\lambda-\frac{i\omega}{\alpha},1-\lambda-\frac{i\omega}{\alpha},1-\frac{i\omega}{\alpha},1-\xi\right)\right]
    \end{split}
\end{equation}
where we identify $A_{\text{in}}(\omega)$ and $A_{\text{out}}(\omega)$ as those obtained in \eqref{eq:PTAinAout}.

The first hypergeometric term in \eqref{eq:psi1PT1} and $A_{\textrm{out}}(\omega)$ have simple poles when $1+\frac{i\omega}{\alpha}$ and $\frac{i\omega}{\alpha}$ are non-positive integers, respectively. The residue of $\phi_L(\omega,x)$ at $\omega=i\alpha n$, $n=0,1,2,...$ is equal to zero, which is computed using 
\begin{equation}
   \!\!\!\!\!\lim_{c\rightarrow-n} \!\!\frac{{}_2F_1(a,b,c,z)}{\Gamma(c)}\!=\!\frac{\Gamma(a+n+1)\Gamma(b+n+1)}{\Gamma(a)\Gamma(b)\Gamma(n+2)}z^{n+1}{}_2F_1\left(a+n+1,b+n+1,n+2,z\right).
\end{equation}
Therefore $ \phi_L(\omega,x)$ is analytic in the UHP of $\omega$. 

However, if one evaluates $ \phi_L(\omega,x)$ in the limit $x\rightarrow\infty$, then the hypergeometric functions in~\eqref{eq:psi1PT1} asymptote to 1, and no longer cancel the poles coming from  $A_\text{out}(\omega)$ in Eq. \eqref{eq:PTAinAout}.  We conclude that the poles at $\omega=i\alpha n$, for $n\geq1$, emerge in the asymptotic limit $\phi_L(\omega,x\rightarrow\infty)$.

\textbf{Leading order approximation.} The poles of $R(\omega)$ in the UHP also appear in the leading order Born solution, which is given by the Fourier transform of the potential 
\begin{equation}
    R(\omega) \propto \int_{- \infty}^\infty e^{-2 i \omega x}\, V(x)\, dx \propto \int_{-\infty}^{\infty} \frac{e^{-2i\omega x}}{\cosh^2(\alpha x)}\, dx = \frac{2\pi\omega}{\alpha\sinh\left(\frac{\pi\omega}{\alpha}\right)}.
\end{equation}
Note that this expression has precisely the poles in the UHP of the full solution, located at $\omega = i \alpha \, n$, for $n$ integer. It does not have the QNMs, though, as they only appear nonperturbatively in the full answer.

Singularities in the UHP are present even in an approximated version of the P\"oschl--Teller potential using the following product identity \cite{melnikov2011}
\begin{equation}
    \cosh(x)\approx \prod^N_{n=1}\left(1+\frac{x^2}{\pi^2(n-1/2)^2}\right)\,.
\end{equation}
At finite $N$, the leading order solution for $R(\omega)$ has a square-root branch cut $\sim \sqrt{\omega^2}$ along the imaginary axis, and as $N\rightarrow\infty$ this branch-cut condenses into the poles of the exact solution.

\section{Analytic structure of the Schwarzschild black hole S-matrix}\label{sec:BH}

Here we study the scattering of classical waves off a fixed Schwarzschild black hole background, which is described by the \textit{Regge-Wheeler} and \textit{Zerilli} equations \cite{Regge:1957td,Zerilli:1970se}. These equations have been extensively studied (see \cite{Berti:2009kk} for a review), particularly in the low-frequency regime, where Black Hole Perturbation Theory~(BHPT)~\cite{Mano:1996mf,Mano:1996gn,Mano:1996vt} provides efficient techniques for computing the partial wave S-matrix. 

Despite the large body of literature on the subject  there has not been a rigorous study of the analytic structure of the S-matrix (see \cite{Castro:2013lba,Casals:2015nja, Motl:2003cd,Neitzke:2003mz} for work in this direction). In this section, we fill this gap by applying the results derived in Chapter \ref{sec:asymG}. Our results are fully general and apply to the scattering of scalar, vector, and tensor perturbations on a Schwarzschild background.

\subsection{Analyticity domains of the S-matrix}\label{sec:BHinverse}

Importantly, the spherical symmetry of the Schwarzschild background reduces the problem to one-dimensional scattering of waves in the radial direction $r$. However, since the gravitational potential behaves like $1/r$ at large distances, the asymptotic solutions are not given in terms of plane waves, and the S-matrix cannot be defined in terms of coefficients of outgoing/ingoing plane waves. Instead, the radial solutions reduce to Coulomb-type wavefunctions, which behave like
\begin{equation}
    \phi(\omega,r\rightarrow\infty)\sim e^{\pm i\omega  \left( r + R_s\log \left(\frac{r}{R_s}-1\right) \right)}\,,
\end{equation}
where $R_s=2GM$ is the Schwarzschild radius.

Crucially, with an appropriate change of coordinates $r\rightarrow x$, where $x$ is known as the tortoise coordinate,\footnote{The tortoise coordinate is also often denoted by $r^*$.} defined by
\begin{align}\label{eq:tortoiseSch}
\frac{dx}{dr}=\left( 1-\frac{R_s}{r}\right)^{-1}\quad \Rightarrow
\quad x=r+R_s\log\left( \frac{r}{R_s}-1\right)\, ,
\end{align}
the potential decays faster than $1/|x|$, as we show later, and the asymptotic states reduce to plane waves, allowing us to define an S-matrix as in \eqref{eq:BCplus}.\footnote{An analogous IR-safe definition exists for electromagnetic (Coulomb) scattering in relativistic quantum mechanics~\cite{Lippstreu:2025jit}.}

On the opposite end, the tortoise coordinate maps the event horizon $r=R_s$ to $x=-\infty$, where waves can escape, leaving the system. Wave scattering on the Schwarzschild background therefore amounts to scattering on the real line across a potential barrier with open boundary conditions at both ends. This map allows us to rigorously apply the results of Section \ref{sec:asymG}.

Concretely, in tortoise coordinates the Regge-Wheeler and Zerilli equations reduce to the one-dimensional form \cite{Berti:2009kk}
\begin{equation}\label{eq:WEBH}
\left[\frac{d^2}{dx^2}+\omega^2-V(r(x))\right]\phi(\omega,r(x))=0\,,
\end{equation}
where the Regge-Wheeler potential is given by
\begin{equation}\label{eq:Schpot}
   V(r) = V_{\mathrm{RW}}(r) \equiv \left(1-\frac{R_s}{r}\right)\left(\frac{\ell(\ell+1)}{r^2}+\frac{R_s}{r^3}(1-s^2)\right)   \,,
\end{equation}
with $\ell$ the orbital angular momentum, and $s=0,\pm1,\pm2$ is the spin of the massless wave. The radial variable $r(x)$ is implicitly given in terms of $x$ according to \eqref{eq:tortoiseSch}.

\begin{figure}
    \centering
    \begin{tikzpicture}
        \node[anchor=south west,inner sep=0] (image) at (0,0) {\includegraphics[width=1\linewidth]{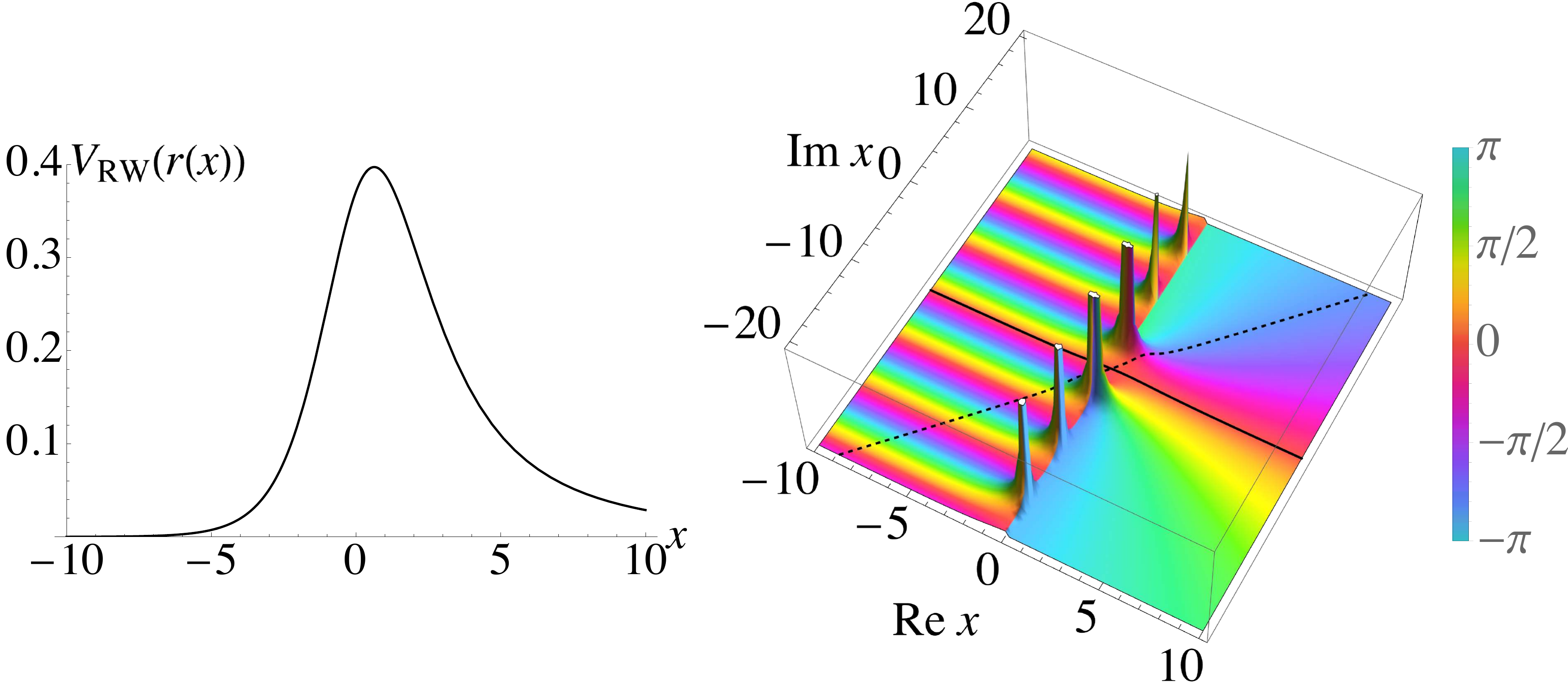}};
        \begin{scope}[x={(image.south east)},y={(image.north west)}]
            \draw [line width=1.pt,domain=-5.5:20.5,color=black] plot ({.5+.29*cos(\x)}, {.4+.29*sin(\x)});
            \node at (.8,0.45) {{\large $\theta$}};
        \end{scope}
    \end{tikzpicture}
    \caption{The Schwarzschild potential barrier $V_\text{RW}(r(x))$ in units $R_s=1$ with $\ell=1,s=0$ for \textbf{Left:} $x\in\mathbb{R}$ and \textbf{Right:} $x\in\mathbb{C}$. The potential has an infinite number of poles on the imaginary axis, located at $x=i (1+ 2n) \pi R_s, \;\; n \in \mathbb{Z}$. Each pole is a branch point, with the branch cuts chosen to run to $+i\infty$ ($-i\infty$) for branch points in the UHP (LHP). The potential decays exponentially as $x\rightarrow-\infty$ and decays as $x^{-2}$ as $x\to\infty$. The solid line on the right plot is the real axis, and the dashed line is slated by an angle $\theta$. We can take $\theta \in(-{\pi \over 2}, {\pi \over 2}) $ without encountering singularities.}
    \label{fig:SchwPotential}
\end{figure}

The Regge-Wheeler potential describes the parity odd sector for spin-$2$ perturbations, while the parity-even sector is determined by the Zerilli potential \cite{Zerilli:1970se}
\begin{equation}  \label{eq:Zpot}
V_Z(r)
= \left(1-\frac{R_s}{r} \right)\,
  \frac{
    8 \lambda^{2} (\lambda + 1) r^{3}
    + 12 \lambda^{2} R_s r^{2}
    + 18 \lambda R_s^{2} r
    + 9 R_s^{3}  
  }{
    r^{3}(2 \lambda r + 3 R_s)^{2}
  }, \quad  \lambda=\frac{(l-1)(l+2)}{2}.
  \end{equation}
In order to apply the results of Section \ref{sec:asymG} and determine analyticity properties of the black hole S-matrix, we analyze the potential's asymptotic fall-off behavior. We find
\begin{subequations}\label{eq:Vschasym}
\begin{align}
V_{\mathrm{RW}}(r(x))&=
\begin{cases}
\displaystyle
\frac{\ell(\ell+1)}{x^2}
    +\frac{R_s\!\left(1-\ell(\ell+1)\right)+2\ell(\ell+1)\log x}{x^3}
    +\mathcal{O}\!\left(\frac{1}{x^4}\right)\,, & \qquad \; x\to\infty\\[0.2cm]
\displaystyle
\frac{\ell(\ell+1)+1-s^2}{R_s^2}\, e^{x/R_s}
    \left(1+\mathcal{O}\!\left(e^{x/R_s}\right)\right)\,, & \qquad \; x\to-\infty\,,
\end{cases}\\[0.3cm]
V_{\mathrm{Z}}(r(x))&=
\begin{cases}
\displaystyle
\frac{2(1+\lambda)}{x^{2}}
    +\frac{-6 R_s(1 +\frac {5}{6} \lambda + \frac{1}{3} \lambda^{2})
           + 4 \lambda (1 + \lambda)\log x}
          {\lambda x^{3}}
    +\mathcal{O}\!\left(\frac{1}{x^4}\right)\,, & x\to\infty\\[0.2cm]
\displaystyle
\frac{3 + 4\lambda(1+\lambda)}{R_s^{2}\,(3 + 2\lambda)}\,
    e^{x/R_s}\left(1+\mathcal{O}\!\left(e^{x/R_s}\right)\right)\,, & x\to-\infty\,.
\end{cases}
\end{align}
\end{subequations}
The nature of the Regge-Wheeler potential for real values of $x$ can be seen explicitly from the graph on the left panel in Figure \ref{fig:SchwPotential}. The asymptotic falloff at both $x\to\pm\infty$ allows us to define an IR-finite S-matrix in terms of the connection coefficients \eqref{eq:phidecomposition}.\footnote{It is common to include an additional factor of $e^{\mp \frac{i\pi}{2}\ell}$ in the definition of the S-matrix \eqref{eq:RT} \cite{Taylor:1972}. Our analysis is not affected by this overall normalization.}

\textbf{Analyticity.} Both the Regge-Wheeler and Zerilli potentials are bounded on the real axis as required by the condition in \eqref{eq:Vintbound}. In the complex $x$-plane, the singularities of the potential $V(r(x))$ arise in two distinct ways: (i) from singularities of 
$V(r)$ itself as a function of $r$, and (ii) from singularities in the coordinate transformation $r(x)$.

The first class of singularities can be identified by direct inspection of the potentials \eqref{eq:Schpot} and \eqref{eq:Zpot}.
For the Regge–Wheeler potential, a pole appears at the black hole singularity $r=0$.
Using the tortoise coordinate relation \eqref{eq:tortoiseSch}, this corresponds to
\begin{equation}
r = 0 \,\, \Leftrightarrow \,\, x = i (1 + 2n)\pi R_s, \qquad n \in \mathbb{Z}.
\end{equation}
Thus, the black hole singularity maps to an infinite series of poles along the imaginary axis.

For the case of the Zerilli potential, in addition to the singularity at $r=0$, one finds further poles arising from the denominator $(2\lambda r+3R_s)^2$.
These occur at 
 $r=-\frac{3R_s}{2\lambda}$, but their images under the tortoise map lie outside the principal (first) Riemann sheet of the complex $x$-plane.
They become accessible only after analytic continuation across one of the branch cuts and therefore do not affect our analysis.

The second class of singularities arises from the tortoise coordinate map itself \eqref{eq:tortoiseSch}. 
In particular, the logarithm has a branch-point at $r = R_s$, which corresponds to $x \to \infty$. The inverse relation $r(x)$ can be expressed in terms of productLog functions, generating an infinite set of branch cuts extending from infinity to the points $x  = i (1+ 2n) \pi R_s$. We choose to orient the branch-cuts along the imaginary axis, in order to apply the results of Chapter \ref{sec:asymG}.

The Regge–Wheeler potential in the complex 
$x$-plane is shown in the right panel of Figure~\ref{fig:SchwPotential}, confirming the analytic structure discussed above.\footnote{Note that the region of the imaginary axis between $-i\pi \,R_s$ and $+i\pi \, R_s$ is free of branch cuts, so the potential remains analytic along this strip.}
The Zerilli potential exhibits the same analytic structure on the first sheet and is therefore not shown here.

Given the analyticity domain of the Regge–Wheeler and Zerilli potentials, together with their polynomial suppression at infinity \eqref{eq:Vschasym}, the derivation of Section~\ref{sec:analyticpotentials} applies.
This leads to the following properties of the reflection and transmission coefficients for scattering on a Schwarzschild background:
\begin{equation}\label{eq:BHanaly}\boxed{
\begin{aligned}
   &T(\omega) \text{\,\,and\,\,} G_R(\omega,x,x') \text{ are meromorphic on } \omega \in\mathbb{C} \setminus \{ i \kappa \mid \kappa \leq 0 \}\\[0.3cm]
   &R(\omega) \text{ is meromorphic on } \omega \in  \mathbb{C} \setminus \{ i \kappa \mid \kappa \in \mathbb{R} \}.
   \end{aligned}}
\end{equation}
Moreover, since the potentials are strictly positive, all poles lie in the lower half of the complex frequency plane, $\mathrm{Im} \, \omega < 0$ (see Section~\ref{sec:smatrixpoles}). These poles correspond to quasinormal modes and contribute to the retarded Green’s function, as well as to the transmission and reflection amplitudes.

\subsection{Low-frequency expansion}\label{sec:BHPT}

In the previous section, we used the results of Chapter \ref{sec:asymG} to determine the regions of analyticity of the reflection and transmission amplitudes, $R(\omega)$ and $T(\omega)$.
We will now verify the explicit analytic structure using the low-frequency expansion $|R_s\omega|\ll1$. For concreteness, we will study the reflection coefficient for a scalar wave ($s=0$), even though the features observed here will be general. 
We provide the low-frequency result with the correct analytic continuation into complex $\omega$, show that it is compatible with the previous section, and comment on the origin and meaning of the singularities.

The low-frequency expansion of the reflection coefficient $R(\omega)$ can be obtained with the MST method \cite{Mano:1996vt,Mano:1996gn,Mano:1996mf,Sasaki:2003xr} or alternatively from expressions given in terms of Nekrasov-Shatashvili (NS) functions in \cite{Bautista:2023sdf}.\footnote{It is well known that the Regge–Wheeler equation \eqref{eq:WEBH} can be brought to the normal form of a Heun equation (see e.g.~\cite{Fiziev:2005ki}), featuring a regular singularity at the horizon and an irregular singularity at infinity.
Remarkably, this equation can be mapped to the semiclassical limit of a Liouville conformal block, where its solutions are expressed in terms of Nekrasov–Shatashvili (NS) functions \cite{Bonelli:2022ten,Dodelson:2022yvn,Bautista:2023sdf,Aminov:2023jve}.} Importantly, the results are obtained in the physical region $\omega>0$, and their analytic continuation to the complex $\omega$-plane requires some care.\footnote{There are a variety of expressions in the literature; for instance, in \cite{Bautista:2023sdf}, the short-distance logarithms are written with a linear dependence on $\omega$, which is incompatible with reflection symmetry. Meanwhile, the long-distance exponentiated logarithm is expressed as 
$\log(|\omega|)$, which is non-holomorphic and therefore cannot be analytically continued. These expressions are, of course, valid in the physical region $\omega>0$.} 
In particular, enforcing consistency with the analyticity properties \eqref{eq:BHanaly} and the reflection symmetry \eqref{eq:crossing} unambiguously determines the analytic structure of the solution.

We express the reflection coefficient through an exponential representation of the S-matrix
\begin{equation}\label{eq:phaseshift}
    R(\omega) = (-1)^\ell e^{2i\delta_\ell(\omega)}\,,
\end{equation}
where the real part of the phase-shift $\delta_\ell(\omega)$ captures elastic scattering and the imaginary part encodes absorptive effects. 
The phase shifts for $\ell=0,1$, obtained from the MST method~\cite{Mano:1996vt,Mano:1996gn,Mano:1996mf}, and adjusted for consistency with the reflection symmetry, Eq. \eqref{eq:crossing}, and the analytic domain derived above,  take the form
\begin{align}\nonumber
    \delta_0(\omega)& =  R_s\, \omega  \left(\frac{1}{2} \log (4R_s^2 \omega^2)+ \gamma_E -\frac{1}{2}\right)+R_s^2\omega\left(\frac{11 \pi }{12}\sqrt{\omega^2}+i\omega\right)  \\[0.3cm]&+R_s^3\omega^2\left[\left(\frac{1}{2} -\frac{1}{2}\log (4R_s^2\omega^2 )+\frac{11 }{6}\zeta_2-\frac{1}{3}\zeta_3-\gamma_E\right)\omega +i \pi\sqrt{\omega^2}\right]\label{eq:NSdelta0}\\[0.3cm]
    &+ R_s^4\omega^3\left[ \left(-\frac{\pi }{2} \left[\log (4R_s^2\,\omega^2)+2\gamma_E\right]+\frac{2299 \pi }{2160}\right)\sqrt{\omega^2}\right.\nonumber\\[0.3cm]
    &\left.\qquad\qquad+\left(\frac{11}{6}\left[\log (4R_s^2\,\omega^2)+2\gamma_E\right] +4  \zeta_2+\frac{227 }{36} \right)i\omega\right]\nonumber\\[0.3cm]
    &+R_s^5\omega^4\Bigg[\left\{\frac{11}{24} \left[\log (4R_s^2\,\omega^2)+2\gamma_E\right]^2-\left(\zeta_2+\frac{17}{36}\right) \left[\log (4R_s^2\,\omega^2)+2\gamma_E\right]-\frac{8591}{1080} \zeta_2\right.\nonumber\\[0.3cm]
    &\qquad\qquad\left.-\frac{11 }{15}\zeta_2^2-\frac{59 }{36}\zeta_3+\frac{1}{5}\zeta_5-\frac{1111}{540}\right\}\omega\nonumber\\[0.3cm]
    &\qquad\qquad+\left\{2 \pi\zeta_2+\frac{191}{36}\pi-\left(\frac{11 \pi }{6}\right) \left[\log (4R_s^2\,\omega^2)+2\gamma_E\right]\right\}i\sqrt{\omega^2}\Bigg]+\mathcal{O}(\omega^6)\,,\nonumber
\end{align}
\begin{align}\nonumber
    \delta_1(\omega)& =R_s\, \omega  \left( \frac{1}{2} \log (4R_s^2 \omega^2)+ \gamma_E -\frac{3}{2}\right)+\frac{19}{60} \pi R_s^2\omega\sqrt{\omega^2}  \\[0.3cm]& +(R_s\,\omega)^3\left( \frac{19 }{30}\zeta_2-\frac{1}{3}\zeta_3\right)+ R_s^4\omega^3\left[\frac{78037 \pi }{378000}\sqrt{\omega^2}+\frac{i}{36} \omega\right]\label{eq:NSdelta1}\\[0.3cm]&
    +R_s^5\omega^4\left[\left(\frac{251}{1800}-\frac{1}{72}\left[\log (4R_s^2\,\omega^2)+2\gamma_E\right]+\frac{78037}{189000}\zeta_2-\frac{19}{75}  \zeta_2^2+\frac{361 }{900}\zeta_3+\frac{1}{5}\zeta_5\right)\omega\right.\nonumber\\[0.3cm]&
    \qquad\qquad\left.+\frac{i \pi }{36}\sqrt{\omega^2}\right]
    +\mathcal{O}(\omega^6)\,.\nonumber
\end{align}

Let us now comment on the $\omega$-dependence of the reflection amplitude~\eqref{eq:phaseshift}. Because of analyticity~\eqref{eq:BHanaly} and reflection symmetry~\eqref{eq:crossing}, the reflection amplitude can only depend on polynomials of $i\omega$ and $\sqrt{\omega^2}$, and logarithms of $\omega^2$. Both the square-root and logarithmic singularities result in branch cuts along the full imaginary axis. To the best of our knowledge, such a branch cut in the UHP has not been mentioned before  in the literature.

The logarithms in $\delta_{0}(\omega)$ and $\delta_1(\omega)$ that appear at leading order in $(R_s \omega)$ arise from the long-range Newtonian potential $\sim 1/r$. All other logarithms capture short-distance effects and typically first appear at  $\mathcal{O}(\omega^{2\ell+3})$, with their powers increasing at higher orders in $\omega$. 
The $\sqrt{\omega^2}$ branch-cuts are due to the sub-leading long-distance corrections to the potential $\sim 1/r^n$, with $n \geq 2$.

Finally, note that the quasinormal modes discussed in Section \ref{sec:smatrixpoles} do not appear in the low-frequency expansion, as they are non-perturbative effects that cannot be captured within perturbation theory. 

\subsection{Inspection of solvable cases}\label{sec:BHsols}
Since exact solutions of the Regge-Wheeler and Zerilli equations \eqref{eq:WEBH} are not known, we will now approximate the potentials in two different ways, restricting to scalar waves ($s=0$) for simplicity. First, we consider the purely Newtonian contribution to the potential, which captures the leading long-distance behavior.
Second, we perform a high-frequency (short-wavelength) approximation of the Regge–Wheeler equation. In both cases, we will explicitly verify the analyticity domain established in Section~\ref{sec:BHinverse} and examine the analytic structure of the corresponding solutions.

\textbf{Newtonian potential.} Following the derivation in~\cite{liu_spectrum_1996,Andersson:1996cm}, in the large-$r$ limit the Regge–Wheeler potential in \eqref{eq:Vschasym} can be approximated as
\begin{equation}
V_{\text{RW}}(r) \approx \frac{\ell(\ell+1)}{r^2}\,.
\end{equation}
Under this approximation, and reverting back to the radial coordinate $r(x)$, the wave equation \eqref{eq:WEBH} reduces to
\begin{equation}\label{eq:Newtonian}
\left[\frac{d^2}{dr^2} + \omega^2 + \frac{2R_s\omega^2}{r} - \frac{\ell(\ell+1)}{r^2}\right] u(r) = 0\,,
\end{equation}
where we have also redefined the field as 
\begin{equation}
u(r) \equiv \left(1 - \frac{R_s}{r}\right)^{1/2}\! \phi(\omega,r)\,.
\end{equation}

The corresponding analyticity argument for this radial system follows closely the proof in~\cite{Taylor:1972} for non-relativistic quantum mechanics, and is analogous to the results derived previously in this paper as we will see through explicit computation. In summary, for a potential $V(r)$ that is analytic for $\text{Re}\,r>0$, the amplitude $R(\omega)$ is meromorphic everywhere in the complex-$\omega$ plane, except the imaginary axis, with poles occuring in the lower-half-plane.

The solutions of Eq.~\eqref{eq:Newtonian} are the well-known Coulomb wavefunctions.
Rather than constructing the retarded Green’s function from the Jost solutions $\phi_L(\omega,r(x))$ and $\phi_R(\omega,r(x))$, we instead employ $\widetilde\phi_L(\omega,r)$, defined as the solution that is regular and vanishes at the origin and $\widetilde\phi_R(\omega,r)\equiv \phi_R(\omega,r)$, which still represents an outgoing plane wave in the limit $x\to\infty$.
Applying these boundary conditions, the solutions are given by
\begin{equation}
    \begin{split}
        \widetilde{\phi}_L(\omega,r)&=e^{-i\omega r}\left(2i\omega r\right)^{\ell+1}\mathcal{M}\left(\ell+1+iR_s\omega,2\ell+2,2i\omega r\right)\,,\\
        \widetilde{\phi}_R(\omega,r)&=\left(-2i\omega r\right)^{\ell+1}\left(-2iR_s\omega\right)^{-iR_s\omega}e^{i\left(\omega r-\frac{\pi}{2}\ell\right)}\mathcal{U}\left(\ell+1-iR_s\omega,2\ell+2,-2i\omega r\right)\,,
    \end{split}
\end{equation}
where $\mathcal{M}$ and $\mathcal{U}$ are the confluent hypergeometric functions~\cite{abramowitz1970}.\footnote{Confluent hypergeometric functions are known to develop Stokes phenomena in their asymptotic expansions, see \cite{Olver:1981}.}

The asymptotic behavior of $\widetilde{\phi}_L(\omega,r)$ yields the connection coefficients
\begin{equation}
    \begin{split}
        A_{\textrm{in}}(\omega)&=\frac{\Gamma(2\ell+2)\left(-2iR_s\omega\right)^{-iR_s\omega}(-1)^\ell}{\Gamma\left(1+\ell-iR_s\omega\right)}\,,\\
        A_{\textrm{out}}(\omega)&=\frac{\Gamma(2\ell+2)\left(2iR_s\omega\right)^{iR_s\omega}}{\Gamma\left(1+\ell+iR_s\omega\right)}\,.
    \end{split}
\end{equation}
Following \eqref{eq:Green} the retarded Green’s function is given by
\begin{equation}
    G_R(\omega,r,r')=-\frac{\Gamma\left(1+\ell-iR_s\omega\right)}{2i\omega\, \Gamma(2\ell+2)e^{i\ell\pi/2}\left(-2iR_s\omega\right)^{-iR_s\omega}}\widetilde{\phi}_L(\omega,r)\widetilde{\phi}_R(\omega,r')\,,\qquad r<r'
\end{equation}
and the reflection amplitude is given by, via \eqref{eq:classicalLSZintro},\footnote{The amplitude \eqref{eq:Slsmallomega} exhibits a series of poles on the imaginary axis at $\omega = -in/R_s$, with $n\in \mathbb{N} $. Such poles have appeared in various contexts (see, e.g., \cite{Kabat:1992tb,Bautista:2021wfy,Bautista:2023sdf}), but they arise only within this approximation and are not present in the quasinormal-mode spectrum of the full S-matrix.}
\begin{equation}\label{eq:Slsmallomega}
    R(\omega)=(-1)^\ell\left(4R_s^2\omega^2\right)^{iR_s\omega}\frac{\Gamma\left(1+\ell-iR_s\omega\right)}{\Gamma\left(1+\ell+iR_s\omega\right)}\,.
\end{equation}

\begin{figure}[t]
    \centering
    \includegraphics[width=1\linewidth]{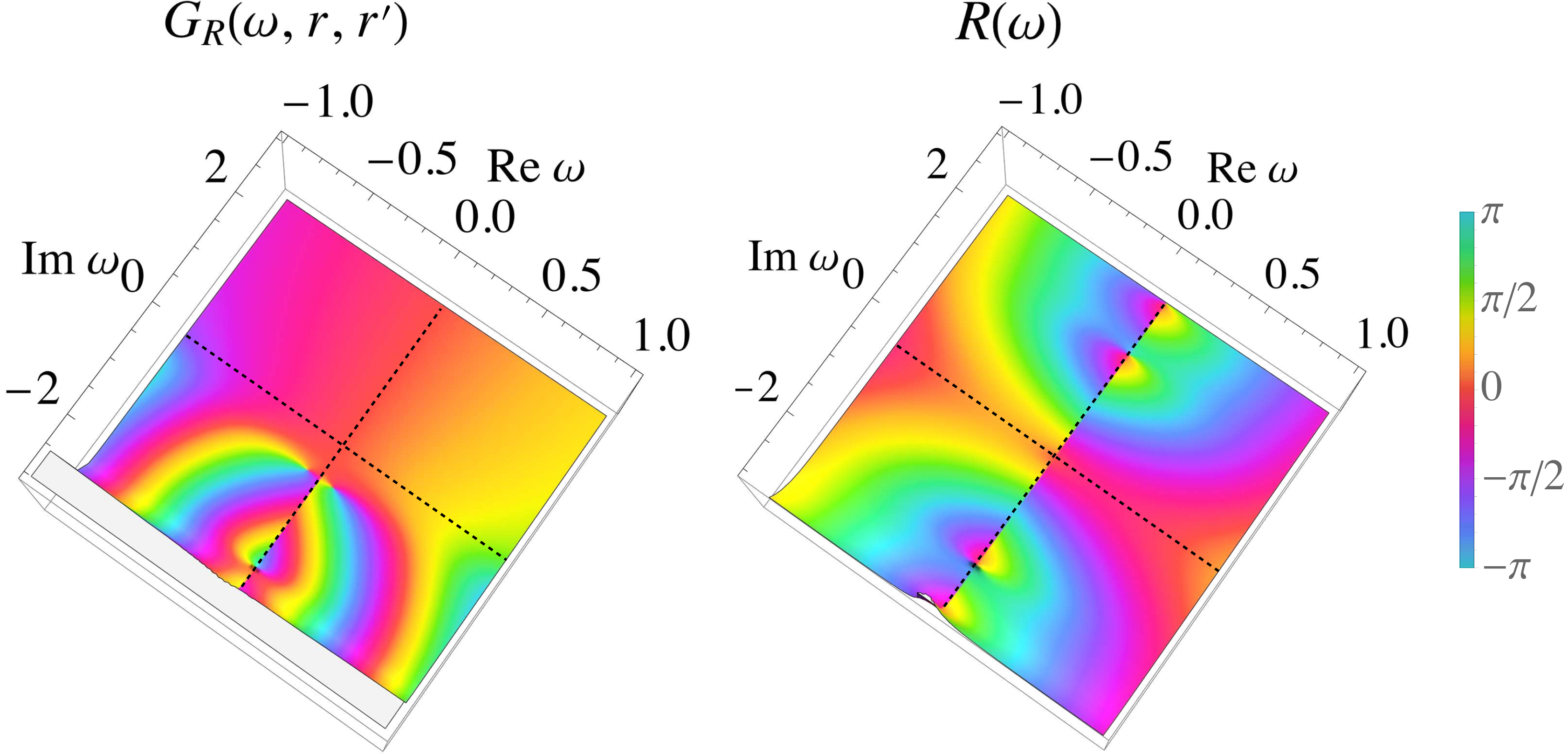}
    \caption{Solution to Newtonian potential in units $R_s=1$ for $\ell=1$. Dotted lines mark the real and imaginary axes. \textbf{Left:} $G_R(\omega, r=2,r'=2.5)$. \textbf{Right:} $R(\omega)$. The Green's function is analytic in the UHP, with QNMs (poles) appearing in the LHP, which lie on the imaginary axis. The reflection amplitude has QNMs in the LHP as well, but also has a branch cut along the entire imaginary axis arising from the Stokes phenomenon.}
    \label{fig:smallomegaGandS}
\end{figure}

The retarded Green's function is analytic in the UHP of $\omega$, since the confluent hypergeometric functions are analytic for the relevant parameter ranges. This is shown explicitly in the left panel of Figure~\ref{fig:smallomegaGandS}. Furthermore, $\widetilde{\phi}_L(\omega,r)$ is meromorphic everywhere (with poles in LHP), while $\widetilde{\phi}_R(\omega,r)$, and in turn $G_R(\omega,r,r')$, has a branch cut along the negative imaginary axis of $\omega$. The discontinuity across this branch cut contributes to the late-time power-law decay tail of the black hole ringdown~\cite{Leaver:1986gd,price_nonspherical_1972_2439,price_nonspherical_1972_2419,Casals:2012ng,Casals:2015nja}.\footnote{This can be seen by computing the inverse Fourier transform of the retarded Green's function and deforming the integration contour into the LHP. Picking up the discontinuity of the branch cut results in the power-law behavior of $G_R(t \to \infty,r,r') \sim t^{-n}$.}

Despite the fact that the retarded Green's function is analytic in the UHP, the reflection amplitude has a branch cut along the imaginary axis coming from the term
\begin{equation}
    \left(4R_s^2\omega^2\right)^{iR_s\omega}=e^{iR_s\omega\log(4R_s^2 \omega^2)}\,.
\end{equation}
The presence and logarithmic nature of the singularity are consistent with the low-frequency approximation of the previous section, where such logarithms are also present. Additionally, the location of the branch cut was anticipated from the analyticity argument of Section~\ref{sec:asymG}. We will see this same structure in the high-frequency approximation.

\textbf{High-frequency approximation.} We now approximate the Regge-Wheeler equation~\eqref{eq:WEBH} in the high-frequency limit $|R_s \omega|\gg1$. 
Again following the derivation in~\cite{liu_spectrum_1996,Andersson:1996cm}, the Regge-Wheeler equation can be written as
\begin{equation}\label{eq:DEvz}
    z\frac{d^2v(z)}{dz^2}+(1-2iR_s\omega-z)\frac{dv(z)}{dz}+f(z)v(z)=0
\end{equation}
with
\begin{equation}\label{eq:fz}
    f(z)\equiv -\frac{1}{2}+2 i R_s \omega -\frac{2 l (l+1)+1}{2(z-2 i R_s \omega )}+\frac{z}{4(z-2 i R_s \omega )^2}\,,
\end{equation}
where the field $v(r)$ is defined as
\begin{equation}
    v(r)\equiv\left(\frac{R_s}{r}\right)^{1/2}\left(\frac{r}{R_s}-1\right)^{iR_s\omega}e^{-i\omega(r-R_s)}\phi(\omega,r)\,,
\end{equation}
and the coordinate $z$ is given by
\begin{equation}
    z\equiv -2i\omega(r-R_s)\,.
\end{equation}

We expand $f(z)$ in the limit $|R_s\omega|\gg1$ and solve the differential equation order by order in $1/(R_s\omega)$.\footnote{In particular, we need to expand to order $\mathcal{O}\left(|R_s\omega|^0\right)$ so that solutions correctly asymptote to plane waves and we can define an S-matrix.} In this limit, $f(z)$ is approximated as 
\begin{equation}
    f(z)=2iR_s\omega-\frac{1}{2}+\mathcal{O}\left(|R_s\omega|^{-1}\right)\,,
\end{equation}
and the differential equation \eqref{eq:DEvz} reduces to a confluent hypergeometric differential equation
\begin{equation}
    z\frac{d^2v(z)}{dz^2}+(1-2iR_s\omega-z)\frac{dv(z)}{dz}+\left(2iR_s\omega-\frac{1}{2}\right)v(z)=0\,,
\end{equation}
whose solutions are known. The centrifugal term in the potential is subleading in this limit, so at this order the solutions are all independent of $\ell$. 
In Appendix~\ref{app:SchwHighFreqCorr} we also consider the subleading correction.

The two Jost solutions to the differential equation with the desired boundary conditions~\eqref{eq:bcLR} are given by
\begin{equation}
    \begin{split}
        \phi_L(\omega,r)&=\left(\frac{r}{R_s}\right)^{1/2}\left(\frac{r}{R_s}-1\right)^{-iR_s\omega}\\
        &\quad\times\mathcal{M}(1/2-2iR_s\omega,1-2iR_s\omega,-2i\omega(r-R_s))e^{i\omega(r-2R_s)}\,,\\
        \phi_R(\omega,r)&=(-2iR_s\omega )^{1/2-2iR_s\omega }\left(\frac{r}{R_s}\right)^{1/2}\left(\frac{r}{R_s}-1\right)^{-2iR_s\omega}\\
        &\quad\times\mathcal{U}(1/2-2iR_s\omega ,1-2iR_s\omega,-2i\omega(r-R_s)) \, e^{i\omega x}\,,
    \end{split}
\end{equation}
where $r=r(x)$ according to \eqref{eq:tortoiseSch}.
The retarded Green's function is constructed from the Jost solutions as follows,
\begin{equation}
    G_R(\omega,x,x')=-\left(\frac{-iR_s \omega}{2}\right)^{1/2}\frac{\Gamma(1/2-2iR_s \omega)}{i\omega\, \Gamma(1-iR_s\omega )}\phi_L(\omega,r(x))\phi_R(\omega,r(x'))\,,\qquad x<x'\,,
\end{equation}
and the reflection coefficient is given by,\footnote{As a consistency check, the high-frequency approximation of Eq.~(2.8) of~\cite{Bautista:2023sdf} (expressed in terms of the Nekrasov–Shatashvili functions) matches our result~\eqref{eq:S0largeomega} in the physical region $\omega>0$.
The agreement follows upon identifying 
$a=iR_s\omega$, as argued in \cite{Bautista:2023sdf}.}
\begin{equation}\label{eq:S0largeomega}
    R(\omega)=-\left(\frac{(-i\omega)^{1/2}}{(\pi i \omega)^{1/2}}\right)\,\Gamma(1/2-2iR_s\omega )\, e^{-2iR_s\omega(1-\log(2iR_s\omega)) }\,.
\end{equation}  

\begin{figure}
    \centering
    \includegraphics[width=1\linewidth]{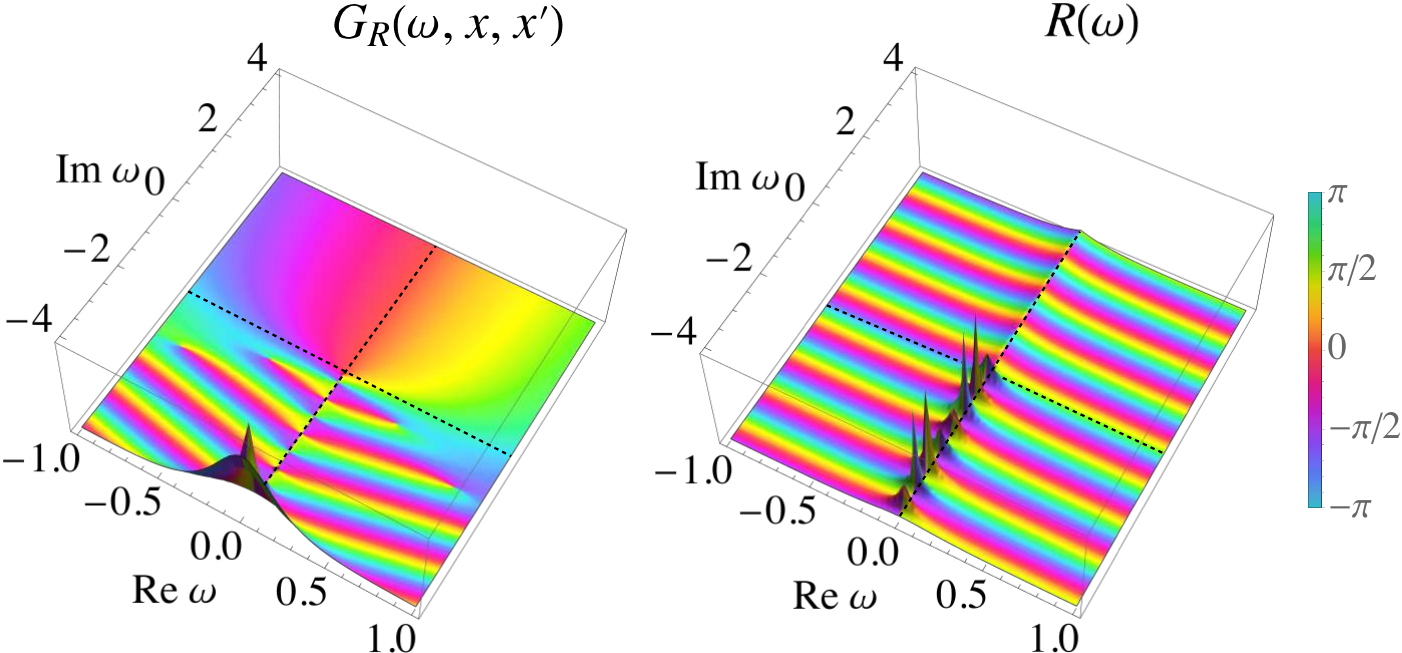}
    \caption{High-frequency solution to Regge-Wheeler potential in units $R_s=1$ for $\ell=0$. Dotted lines mark the real and imaginary axes. \textbf{Left:} $G_R(\omega, x=2,x'=3.7)$. \textbf{Right:} $R(\omega)$. The Green's function is analytic in the UHP, with QNMs (poles) appearing in the LHP. In the large $\omega$ approximation, the QNMs lie on the imaginary axis. The reflection coefficient has QNMs in the LHP as well, but also has a branch cut along the entire imaginary axis arising from the Stokes phenomenon.}
    \label{fig:SchwlargeomegaGandS}
\end{figure}

We plot the analytic structure of $G_R(\omega, x,x')$ and $R(\omega)$ in the complex-$\omega$ plane in Figure~\ref{fig:SchwlargeomegaGandS}. 
The retarded Green's function has a branch cut along the negative imaginary axis and is analytic in the UHP, which is consistent with causality. On the other hand, $R(\omega)$ has a branch cut on the positive imaginary axis, respecting the analyticity domain~\eqref{eq:BHanaly} proven in Section~\ref{sec:asymG}.

We see again the appearance of Stokes phenomena, where the reflection coefficient $R(\omega)$ has additional branch cuts along the positive imaginary axis coming from the terms $\log(2iR_s\omega)$ and $(i\omega)^{-1/2}$.

We can use this high-frequency result to analyze the behavior of $R(\omega)$ as $|\omega|\to\infty$ in different directions in the complex plane. Concretely, we find
\begin{equation}\label{eq:Romegainfty}
    \lim_{|\omega|\to\infty}R(|\omega|e^{i\theta})=i\sqrt{2}
    \begin{cases}
        e^{-2\pi R_s |\omega|}\,, & \theta\to0\\
        e^{-2i\pi R_s |\omega|}\,, & \theta\to\frac{\pi}{2}^- \quad (\text{Re}(\omega)>0)\\
        -e^{2i\pi R_s |\omega|}\,, & \theta\to\frac{\pi}{2}^+ \quad (\text{Re}(\omega)<0)\\
        -e^{-2\pi R_s |\omega|}\,, & \theta\to\pi
    \end{cases}
\end{equation}
The presence of the branch cut along the imaginary axis is evident given the difference between approaching the axis from the left ($\theta\to\frac{\pi}{2}^+$) or the right ($\theta\to\frac{\pi}{2}^-$). 
The exponential suppression $R(\omega)\sim e^{-2\pi R_s \omega}$ as $\omega\to\infty$ reflects that, at fixed angular momentum, high–frequency waves are almost entirely absorbed by the black hole. This behavior was first obtained in \cite{ben-israel_black_2017,Basha:2018bvi} via a WKB approximation. Crucially, we also show here, in Appendix~\ref{app:SchwHighFreqCorr}, that the asymptotic behavior \eqref{eq:Romegainfty} persists at subleading order in the $|R_s\omega|\gg 1$ expansion.

We note that our results differ from the behavior of the reflection amplitude at large imaginary frequency obtained in \cite{Neitzke:2003mz} that used a monodromy analysis.\footnote{Reference \cite{Neitzke:2003mz} uses an opposite sign convention for $\omega$: their limit $\omega\to-i\infty$ corresponds to our $\omega\to+i\infty$.} In that work, the author reports an asymptotic constant value $-2i$ for the reflection amplitude as $\omega\to i\infty$, with no mention of a branch cut along the positive imaginary axis. By contrast, our result \eqref{eq:Romegainfty} yields an oscillatory phase with magnitude $\sqrt{2}$, whose overall sign and phase depend on the side of the branch cut from which $\omega$ approaches.

On the other hand, our high-frequency approximation in \eqref{eq:S0largeomega} does not quantitatively recover all aspects of the black hole S-matrix; specifically, our solution \eqref{eq:S0largeomega} yields QNMs that lie exactly on the imaginary axis. They miss a non-zero real part which has been established in \cite{nollert_quasinormal_1993,andersson_asymptotic_1993}.\footnote{The absence of real part of the QNMs was also observed in the high-frequency approximation of \cite{Andersson:1996cm}.} In any case, we expect that retaining subleading terms in the high-frequency approximation can recover the missing real part, which we find evidence of in Appendix~\ref{app:SchwHighFreqCorr}.

\subsection{Analyticity for generic compact objects}\label{sec:EFT}
In the previous subsections, we have established the analytic structure of the S-matrix for the scattering of a scalar wave on a Schwarzschild black hole background and found that the S-matrix is meromorphic everywhere on the complex $\omega$ plane, except the imaginary axis, with poles (QNMs) in the lower-half-plane. It is natural to wonder if a similar analysis holds for a broader class of solutions, i.e those corresponding to scattering off a generic compact object.

In the long-wavelength limit, any compact object can be effectively treated as a point particle, with its dynamics captured by a worldline EFT, where finite-size effects are incorporated through an action of the form:\footnote{This wordline EFT is for a scalar wave, see \cite{Combaluzier--Szteinsznaider:2025eoc} for the case of gravitational perturbations.}
\begin{align}\label{eq:SLove}
  \!\!\!  S_\text{Love}&=\sum_{\ell,n} \frac{\mu^{4-d}}{\ell!}\int \mathrm{d}\tau \Big[C_{\ell,n}(\nabla_{\!\ell} \,\phi_+)\partial_\tau^n(\nabla_{\!\ell} \,\phi_-)  \Big]
\end{align}
where $\tau$ is the proper time. The Wilson coefficients 
$C_{\ell,n}$ are known as Love numbers and capture the tidal deformability properties of the compact object. In order to incorporate the absorption effects, we work in the Keldysh basis $\phi_\pm$ \cite{Keldysh:1964ud,Martin:1973zz}. The EFT expansion is then organized in terms of symmetric, traceless products of $\ell$ spatial derivatives, with  $\nabla^\mu=(g^{\mu\nu}-u^\mu u^\nu)\nabla_\nu$, where $u^\mu=(1,0,0,0)$ denotes the 4-velocity of the compact object. The dimensionful scale 
$\mu$ is introduced to render the Love numbers dimensionally consistent in 4 dimensions and also acts as a renormalization scale.

Perturbative computations of the reflection amplitude in this EFT exhibit logarithmic UV divergences \cite{Ivanov:2024sds,Caron-Huot:2025tlq}, which can be effectively regularized using dimensional regularization where $d=4-2\epsilon$. The Love numbers serve as counterterms, canceling the divergences and the $1/\epsilon$ poles are renormalized according to  a chosen subtraction scheme, with the associated logarithms governing the scale dependence of the Love numbers.

The reflection coefficient for this process was computed to high orders in \cite{Caron-Huot:2025tlq} by mapping the EFT calculation onto an effective wave equation. In this formulation, the long-distance behavior is governed by the gravitational potential, while the short-distance dynamics are encoded in a potential controlled by the Love numbers. This approach makes it possible to exploit techniques for differential equations, drastically simplifying the calculation compared to a more direct Feynman diagram approach. 

For a generic compact object described as a point particle in the worldline EFT, there is no natural notion of a horizon or a well-defined tortoise coordinate. As a result, the scattering problem cannot be formulated in the same way as we did in the previous sections for the Schwarzschild black hole, where the reflection coefficient was defined via asymptotic plane waves in tortoise coordinates. Instead, the S-matrix  is defined with respect to asymptotic free solutions in $d=4-2\epsilon$ dimensions, which effectively introduces a cutoff for the infrared logarithms (see \cite{Caron-Huot:2025tlq}). The origin of these logarithms can be traced to the fact that, in Schwarzschild coordinates, the potential decays as $1/r$, so the radial dependence is not fully absorbed into the asymptotic states as it is in tortoise coordinates, thereby introducing explicit dependence on the additional scale $\mu$.
 
For instance, the phase-shift for $\ell = 0 $ up to $\mathcal{O}(\omega^3)$ can be expressed as:
\begin{align}\label{eq:WLphase}\nonumber
\delta_{\ell=0}\,(\omega)&=R_s\,\omega\left(\frac{1}{2\epsilon}+\frac{1}{2}\log\bigg({\frac{4\omega^2}{\bar\mu_{\mathrm{IR}^2}}}\bigg) -\frac{1}{2}+\gamma_E+\frac{C_{0,0}}{4\pi R_s}\right)+R_s^2\omega\left[\left(\frac{11\pi}{12}+\frac{C_{0,0}}{4 R_s}\right)\sqrt{\omega^2}+ \frac{C_{0,1}}{4\pi R_s^2}i\omega\right]\\[0.3cm]
&+R_s^3\omega^2\Bigg[\left\{\frac{1}{4\epsilon }\left(1+\frac{11 C_{0,0}}{12 \pi  R_s}+\frac{C_{0,0}^2}{8 \pi ^2 R_s^2}-\frac{C_{0,2}^{({-1})}}{\pi  R_s^3} \right)\right.\\[0.3cm]
&\quad-\frac{3}{4} \log \left(\frac{4 \omega ^2}{\bar\mu^2}\right)\left( 1+\frac{11 C_{0,0}}{12 \pi  R_s}+\frac{C_{0,0}^2}{8 \pi ^2 R_s^2}-\frac{C_{0,2}^{(-1)}}{3 \pi  R_s^3}\right)+\frac{31}{12}+\frac{11 \pi ^2}{36}-\frac{1 }{3}\zeta_3 \nonumber\\[0.3cm]
&\quad \left.+C_{0,0}\frac{\left(89+4 \pi ^2\right)}{48 \pi  R_s}+\frac{C_{0,0}^2}{8 \pi ^2 R_s^2}-\frac{C_{0,0}^3}{192 \pi ^3 R_s^3}-\frac{C_{0,2}^{(-1)}}{2 \pi  R_s^3}-\frac{C_{0,2}^{(0)}}{4 \pi  R_s^3}\right\}\omega+\frac{ C_{0,1}}{4 R_s^2}i\sqrt{\omega^2}\Bigg]\,,\nonumber
\end{align}
where, when necessary, we expand the Love numbers as
$C_{\ell,n}=\sum_m C_{\ell,n}^{(m)}\epsilon^{m}$,
so as to isolate the divergent terms. The perturbative expansion in $\epsilon$ is then truncated at $\mathcal{O}(\epsilon^0)$. We conveniently define the scales $\bar\mu^2= \mu^2 4\pi e^{-\gamma_E}$ and $\bar \mu_{\text{IR}}^2= 4\pi \mu^2 e^{\gamma_E-1}$.

The black hole solution \eqref{eq:NSdelta0}, as a special case of a compact object, can be recovered by fixing particular values of the Love numbers. For example, the vanishing of the static Love number,
$C_{0,0}=0$, follows directly from matching the expression in equation \eqref{eq:NSdelta0} with \eqref{eq:WLphase}. Likewise, the first dissipative coefficient is fixed to $C_{0,1}=4\pi R_s^2$ \cite{Das:1996we}. In addition, the logarithmic terms determine the renormalization group (RG) flow of the Love numbers. Explicit results for the scalar black hole Love numbers, together with their RG equations up to $\mathcal{O}(G^7)$, are collected in \cite{Caron-Huot:2025tlq}.

Establishing the region of analyticity of $R(\omega)$ for a generic compact object is not possible beyond the low-frequency approximation where the worldline EFT applies. The combined requirements of reflection symmetry under $\omega \to -\omega^*$, as expressed in~\eqref{eq:crossing}, and consistency with the  phase-shift expression~\eqref{eq:WLphase} for $\omega>0$, uniquely determine the analytic structure at low frequencies. In particular, all logarithmic contributions are supported along the entire imaginary axis, in exact analogy with the black hole case.

\subsection{Effect of infrared regulators on analytic structure}\label{sec:IRfiniteS}

In this section, we show how the analyticity results of Section~\ref{sec:BHinverse} extend to the S-matrix defined with free asymptotic plane waves, which is formally an infrared-divergent object and must be regularized.\footnote{This exercise is relevant since perturbative scattering amplitude computations, including in the worldline EFT of section~\ref{sec:EFT}, are usually performed with asymptotically free particle states, where IR divergences are unavoidable.} We will consider here hard-cutoff regularization.

In hard-cutoff regularization, the potential is set to zero after some $r>R_{\text{IR}}$, where $R_{\text{IR}}$ is the hard cutoff. 
Since the potential is zero at large $r$, the solution reduces to plane waves and one can define an S-matrix ${R}^{\text{hard}}(\omega)$ in the usual way \cite{Taylor:1972}. Now, we can compare this hard-cutoff definition with the tortoise coordinate definition \eqref{eq:RT} where the potential falls-off fast enough. We find the relation
\begin{equation}\label{eq:hardcut}
{R}^{\text{hard}}(\omega)=R(\omega) \left(\frac{ R_{\text{IR}}}{R_s } \right)^{2iR_s\omega}\,.
\end{equation}
Curiously, ${R}^{\text{hard}}(\omega)$ reduces to $R(\omega)$ when the hard cutoff is set to the Schwarzschild radius, $R_{\text{IR}}=R_s$. The mapping between asymptotic states in Schwarzschild and tortoise coordinates leading to \eqref{eq:hardcut} can be interpreted as a large diffeomorphism (or large gauge) transformation acting on the asymptotic particle states \cite{He:2014laa}. Resumming the $\omega$-dependent phase generated by this transformation yields the exponential relation between the two S-matrix normalizations.\footnote{We thank Radu Roiban for bringing this point to our attention.} Equivalently, this exponentiation reproduces the Weinberg soft factor projected onto partial waves \cite{Weinberg:1965nx}. As usual for infrared effects, this additional phase has no impact on physical observables.

It is clear from the explicit form of~\eqref{eq:hardcut} that the analytic structure of ${R}^{\text{hard}}(\omega)$ is directly inherited from that of $R(\omega)$. However, the asymptotic behavior in the limit $\omega\to + i\infty$ is strongly affected by the choice of regularization. In particular, different infrared cutoffs modify the large-frequency behavior of the reflection coefficient in the complex plane, since
\begin{equation}
    \left(\frac{ R_{\text{IR}} }{R_s } \right)^{2iR_s\omega}\xrightarrow[\textrm{Im}\, \omega \to \infty]{}
    \begin{cases}
        \infty , & R_{\text{IR}}<R_s\\
        1 , & R_{\text{IR}}=R_s\\
        0 , & R_{\text{IR}}>R_s
    \end{cases}
\end{equation}
Combining this observation with the result that the tortoise coordinate reflection coefficient $|R(\omega)|$ approaches a constant at large imaginary frequencies~\eqref{eq:Romegainfty}, we conclude that for $R_{\text{IR}}<R_s$  the hard-cutoff reflection coefficient ${R}^{\text{hard}}(\omega)$ grows exponentially in the upper half of the complex $\omega$-plane.

In gapped quantum field theories, the partial wave S-matrix is expected to satisfy \textit{polynomial boundedness} in the complex frequency plane \cite{Correia:2020xtr},
\begin{equation}\label{eq:polbound}
|R (\omega\to\infty)|\leq |\omega|^N, \quad \text{Im}\,\omega>0
\end{equation}
for some finite $N$. This requirement excludes exponentially growing amplitudes and it is a key ingredient to ensure the validity of dispersion relations and causality constraints.\footnote{In QFT, unitarity on the real axis, together with analyticity in the UHP and polynomial boundedness, implies that the partial wave $S_\ell(\omega)$ remains bounded along any direction in the UHP, $|S_\ell(s)|\leq 1$ for $|s|\to\infty$. Moreover, in the presence of a mass gap, one expects $|S_\ell(s)|\to 1$ asymptotically, since the effective interaction region becomes finite. This behavior is recovered in our case by setting $R_\text{IR}=R_s$, where the S-matrix reduces to the tortoise-coordinate definition and approaches a constant in the UHP, as dictated by \eqref{eq:Romegainfty}.}
Requiring polynomial growth on the regularized reflection coefficient $R^{\text{hard}}(\omega)$ imposes a sharp lower bound on the infrared cutoff:\begin{center}
\textit{Polynomial boundedness implies a lower bound on the IR cutoff:} $R_{\text{IR}} \ge R_s$.
\end{center}
In particular, we see that this bound is saturated in tortoise coordinates, where $R_{\text{IR}}=R_s$ and the S-matrix is manifestly infrared finite without additional scales. The bound $R_{\text{IR}} \ge R_s$ is consistent with the physical expectation that the the infrared cutoff should be larger than all other scales in the problem, which in this case is only the Schwarzschild radius $R_s$. 

It is noteworthy that taking the IR cutoff $R_\text{IR}$ to be \emph{smaller} than the Schwarzschild radius $R_s$ leads to a breakdown of polynomial boundedness \eqref{eq:polbound}, which is a fundamental assumption of S-matrix theory in QFT \cite{Correia:2020xtr}. It would be interesting if such a lower bound on the IR cutoff could be derived more generally in QFT beyond the classical wave scattering setup considered here.

Typically, scattering amplitude calculations in QFT employ dimensional regularization rather than a hard cutoff, for example as in the worldline EFT computation of Section~\ref{sec:EFT}. It would be interesting to derive the corresponding relation between the reflection amplitude in $d = 4-2\epsilon$ dimensions and the reflection coefficient  \( R(\omega) \), defined via tortoise coordinates.\footnote{This would require repeating the high-frequency analysis done in Section \ref{sec:BHsols} in general $d$ dimensions which we have not attempted to pursue here.} This would clarify whether an analogue of the bound \( R_{\text{IR}} > R_s \) exists, but in terms of  \( \mu_{\text{IR}} \).

\section{Discussion and future directions}\label{sec:discussion}

In this work, we established the analytic structure of the Schwarzschild black hole S-matrix. We showed that in the lower half of the frequency plane, the partial-wave amplitudes possess poles corresponding to quasinormal modes as well as a branch cut along the imaginary axis associated with late-time tails. In the upper half-plane, which is generally protected by causality, we found that the transmission amplitude $T(\omega)$ remains analytic, while the reflection (or elastic) amplitude $R(\omega)$ develops a branch cut along the positive imaginary axis, which is again connected to late-time tails.

This breakdown of analyticity in the upper half-plane arises from Stokes phenomena, which prevent the S-matrix from inheriting the analytic structure of the retarded Green’s function. The main results are summarized in Fig. \ref{fig:GBHsketch}.

To reach these conclusions, we combined two complementary approaches:

\begin{enumerate}
\item \textbf{General proof of analyticity domain.} In Chapter \ref{sec:asymG}, using tools from inverse scattering theory we established domains of analyticity for a general class of potentials admitting analytic continuation. Compared to previous proofs of analyticity in quantum-mechanical scattering~\cite{Taylor:1972,Yafaev:1992, Reed:1979MMPIII,Newton:2013ScatteringReprint}, our results extend to long-range interactions and open systems with dissipation, as relevant for gravitational wave scattering by black holes. The proven domains of analyticity and corresponding assumptions on the potential are summarized in Figure~\ref{fig:allResults}.

\item \textbf{Verification through explicit solutions.} In Chapters~\ref{sec:poschlteller} and \ref{sec:BH}, we verified the previously proven analyticity domains in exactly solvable cases. First, we considered scattering off a Pöschl–Teller potential, which can be seen as a toy model for the gravitational potential (see Figure~\ref{fig:PTpotential}). Then, in Chapter~\ref{sec:BH}, we analyzed the Regge–Wheeler equation both at small frequencies, where the branch cuts associated with gravitational tails become manifest (see Eq. \eqref{eq:NSdelta0}), and at high frequencies, where the asymptotic quasinormal-mode spectrum in the lower-half-plane can also be observed (see Figure \ref{fig:SchwlargeomegaGandS} and Eq.~\eqref{eq:S0largeomega}). We also analyzed the exactly solvable Newtonian case where the analytic structure is manifest (see Figure \ref{fig:smallomegaGandS} and Eq. \eqref{eq:Slsmallomega}).
\end{enumerate}

\vspace{4mm}

We now make a few more detailed observations on these results and discuss some natural applications and future directions.

\textbf{Physical origin of the branch cut in the upper half-plane.}  
In QFT, singularities are typically associated with particle exchange, but this interpretation only applies partially here. 
 We can associate some of the branch-cuts seen in the low-frequency approximation of the phase-shifts in Section \ref{sec:EFT} with Feynman diagrams. In particular, the single-graviton exchange and ladder diagrams (associated with the $1/r$ Newtonian potential) generate the $\log(\omega^2)$ branch cut while triangle and higher ``fan'' diagrams, suppressed at larger distances  \cite{Duff:1973zz, Ivanov:2024sds,Caron-Huot:2025tlq}, produce square-root-type cuts $\sqrt{\omega^2}$:

\[
\vcenter{\hbox{\includegraphics[width=1.5cm]{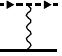}}}
\;\sim\;
\log(\omega^2),
\qquad
\vcenter{\hbox{\includegraphics[width=1.5cm]{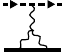}}}
\;\sim\;
\sqrt{\omega^2}\,.
\]

This indicates that the branch cut observed in the upper half-plane of the elastic amplitude $R(\omega)$ is a universal feature of massless exchange, expected to appear also in higher dimensions and in other systems governed by long-range interactions.

However, these graviton exchanges do not fully capture all contributions to the UHP branch cut of $R(\omega)$, seen explicitly in the low-frequency approximation of Section \ref{sec:EFT}. In the worldline EFT, UV divergences eventually appear and require counterterms, which correspond to the tidal Love numbers. After renormalization, the amplitude develops additional $\log^n(\omega^2)$ cuts associated with the running of dynamical Love numbers \cite{Ivanov:2024sds,Caron-Huot:2025tlq,Combaluzier--Szteinsznaider:2025eoc}, encoding the tidal response of compact objects. Whether such contributions admit a conventional QFT interpretation in terms of particle excitations remains an open question and a potential obstacle to a bootstrap program aimed at constraining Love numbers (more on this below).

\textbf{Long-range interactions and polynomial boundedness.}  
In quantum field theory, the S-matrix is defined for asymptotically free particle states. This assumption breaks down in the presence of long-range interactions in $d=4$ spacetime dimensions, leading to the well-known infrared (IR) divergence problem~\cite{Bloch:1937pw}. In the classical setup of gravitational-wave scattering considered here, this issue can be circumvented by switching from radial to tortoise coordinates, Eq.~\eqref{eq:tortoiseSch}. In tortoise coordinates, the potential decays sufficiently fast at large distances, allowing for an unambiguous and IR-finite definition of the S-matrix, without introducing additional scales or cutoffs (see Section \ref{sec:IRfiniteS}).

Beyond establishing the domain of analyticity of this IR-finite S-matrix, we also found that the partial wave amplitude remains bounded and asymptotes to a constant as $\mathrm{Im}\,\omega \to +\infty$, see Eq.~\eqref{eq:Romegainfty}. This property would be particularly relevant for bootstrap approaches, which typically rely on dispersion relations.

If instead one insists on defining the S-matrix in the standard QFT sense, in terms of asymptotically free plane waves, the result is manifestly IR divergent. Introducing a hard IR cutoff $R_{\text{IR}}$, we find that the regulated S-matrix is related to the unambiguous IR-finite one through Eq.~\eqref{eq:hardcut}. Both coincide when $R_{\text{IR}} = R_s$, indicating that the Schwarzschild radius---the only intrinsic length scale in the problem---effectively acts as an IR cutoff.  

Interestingly, decreasing the cutoff below the Schwarzschild radius, $R_{\text{IR}} < R_s$, leads to exponential growth of the regulated amplitude for $\mathrm{Im}\,\omega \to +\infty$. This violates the standard assumption of polynomial boundedness in gapped S-matrix theory \cite{Correia:2020xtr}. Requiring this assumption to hold leads to a lower bound on the IR cutoff,
\begin{equation}
R_{\text{IR}} \geq R_s.
\end{equation}
It would be interesting to see if analogous bounds hold more generally in quantum field theories with massless particles.

\textbf{S-matrix bootstrap and bounds on Love numbers.}  
Having established the analytic structure of the black hole S-matrix, it would be interesting to apply S-matrix bootstrap techniques to derive bounds on the tidal Love numbers, in analogy with existing bounds on Wilson coefficients in QFTs~\cite{Adams:2006sv,Arkani-Hamed:2020blm,deRham:2017avq,Bellazzini:2020cot,Sinha:2020win,Tolley:2020gtv,Caron-Huot:2020cmc,Beadle:2024hqg,Paulos:2017fhb,EliasMiro:2022xaa} and low-energy data in other settings \cite{Hui:2023pxc,Hui:2025aja,Creminelli:2024lhd,Creminelli:2023kze,Chowdhury:2025dlx,Chowdhury:2025qyc,DiPietro:2021sjt}. Such an approach could help constrain finite-size effects in neutron stars and, in particular, restrict different models for their equation of state. These bounds could also be compared or combined with existing observational constraints~\cite{De:2018uhw,LIGOScientific:2018cki,LIGOScientific:2018hze,Shterenberg:2024tmo,Chia:2024bwc}.  However, as discussed above, part of the branch cut in the UHP of the reflection amplitude $R(\omega)$ originates from running of the Love numbers and it appears unconstrained by unitarity in the crossed channel, unlike in  QFT, making bootstrapping $R(\omega)$ currently unfeasible. 

On the other hand, we have shown that the absorption amplitude $T(\omega)$ is fully analytic in the upper half-plane (see Chapter \ref{sec:asymG}). It satisfies unitarity, $|T(\omega)|^2 \leq 1$, and exhibits good asymptotic behavior, making it a natural candidate for a bootstrap analysis. Its low-frequency expansion is expected to contain the same Love numbers that appear in the elastic amplitude $R(\omega)$ through unitarity, Eq.~\eqref{eq:unitarity}. Although the transmission amplitude is less universal than the elastic one, being closely tied to the presence of a horizon and thus specific to black holes, it may still provide valuable insights into their properties. For instance, the vanishing of static black hole Love numbers has motivated extensive study \cite{Binnington:2009bb,Damour:2009vw,Kol:2011vg,Porto:2016zng,Charalambous:2021kcz,Hui:2021vcv,Berens:2025okm,Parra-Martinez:2025bcu}. It would be interesting to test, via bootstrap bounds, whether zero is the minimal value allowed by consistency. 
We hope to report progress in this direction soon.

Another possibility would be to consider instead the retarded Green's function $G_R(\omega,x,x')$ which is by construction also analytic in the UHP, due to causality.  Several works have already used the retarded Green's function to place bounds on low-energy coefficients in other contexts \cite{Creminelli:2024lhd,Chowdhury:2025dlx,Chowdhury:2025qyc}.

A closely related example is scattering of electromagnetic radiation on charged or polarizable materials, where no long-range interactions are present. In this case, the corresponding photon elastic amplitudes are expected to be analytic in the full upper half-plane, making them suitable for a bootstrap analysis. Such an approach could in principle constrain electromagnetic properties of compact or point-like objects, such as their electric polarizabilities \cite{Jackson:1998nia}.  In Appendix~\ref{app:bounds} we illustrate analytically how such an analysis would proceed using tools from the two-dimensional S-matrix bootstrap \cite{Paulos:2016but,EliasMiro:2019kyf,EliasMiro:2021nul}.

\textbf{Analyticity of the Kerr black hole S-matrix.}  
A natural extension of this work is to generalize the analysis to Kerr black holes and establish analyticity domains for the S-matrix. While a similar analytic structure is expected to hold, the general proof of analyticity developed in Chapter~\ref{sec:asymG} does not directly carry over. The Teukolsky equation, which governs wave scattering on a Kerr background, also admits a tortoise-type coordinate transformation~\cite{Chandrasekhar:1976zz}. However, this transformation makes the potential frequency-dependent, raising possible issues of analytic continuation. Moreover, although the low-frequency analysis can be repeated for Kerr without major difficulty \cite{Mano:1996gn,Mano:1996vt}, we are not aware of a high-frequency solution of the Teukolsky equation available in the literature.

\textbf{Analyticity in Anti-de Sitter and non-asymptotically flat backgrounds.}  
The techniques used here to establish the analyticity of the S-matrix could, in principle, be extended to non–asymptotically flat settings, such as Anti–de Sitter (AdS) space. In that case, one would similarly need to define appropriate Jost solutions in terms of the free AdS modes. However, we expect that the upper half-plane would remain analytic for all connection coefficients, even in the presence of long-range interactions in the bulk. This is because the branch cut observed in flat space arises from Stokes phenomena, which in turn originate from the presence of an irregular singular point in the differential equation—something that typically does not occur in AdS \cite{Castro:2013lba,Denef:2009yy}.

\textbf{Ruling out models with non-analytic potentials}.\footnote{We thank Clifford Cheung for this observation.}  
In Chapter~2 we showed that analyticity of the potential implies analyticity of the S-matrix. It is natural to ask whether this logic can be reversed: can demanding certain analytic properties of the S-matrix constrain the class of potentials consistent with locality and causality? For instance, all potentials arising from the virtual exchange of particles in quantum field theory admit an analytic continuation that remains bounded in the complex plane. By contrast, periodic or oscillatory potentials such as $V(x) \sim {\sin x \over x^2}$ would probably lead to additional singularities in the S-matrix, since their analytic continuation leads to exponential growth in the complex-$x$ domain.

\textbf{Tail effect obstruction and bound-to-boundary map.} A major challenge in the gravitational amplitudes program is the \emph{tail effect} appearing at $\mathcal{O}(G^4)$ in the gravitational scattering of two massive bodies~\cite{Galley:2015kus,Foffa:2011np,Kalin:2022hph,Porto:2017dgs}. This effect introduces infrared branch cuts in the S-matrix and mixes conservative and dissipative dynamics, currently obstructing the analytic continuation from the scattering regime to the bound regime of black hole binaries. As a result, existing amplitude-based results cannot yet be straightforwardly applied to this experimentally relevant domain.  

Interestingly, the physical origin of this tail obstruction in the two-body problem and the branch cut in the upper half-plane found here for the gravitational-wave amplitude are one and the same: the long-range nature of gravity. A more detailed understanding of the analytic structure of the two-body S-matrix, along the lines developed here, could provide new insights into overcoming this obstruction.

\textbf{A new path to analyticity in quantum field theory?}  
The methods developed here, based on analytic continuation of the potential, may extend beyond the present setting to establish analyticity of the S-matrix directly in quantum field theory. As shown in~\cite{Correia:2024jgr,Todorov:1970gr}, within the effective-one-body framework it is possible to define an effective potential in QFT, which is expected to exhibit good analytic properties, though with additional non-analyticities associated with particle production that are absent in the classical wave scattering case.  

It is conceivable that similar Stokes phenomena between correlators and amplitudes also occur in QFT. While the retarded correlator should remain analytic in the upper half-plane of frequency space, the amplitude itself can develop non-analyticities associated with virtual particle exchange, analogous to the branch cut observed here for the elastic partial amplitude. Investigating this connection, even perturbatively, could be highly informative.

In Appendix~\ref{sec:LSZ}, we lay some of the groundwork for such an analysis. In particular, we connect the partial-wave formalism used here to standard QFT tools, such as the LSZ reduction formula, and verify that scattering on a Yukawa potential—which naturally arises from tree-level exchange of a massive particle—also exhibits Stokes phenomena.

\vspace{2mm}

\textbf{Acknowledgments.} We are grateful to Mikhail Solon for collaboration at the early stages of this project. We thank Zvi Bern, Simon Caron-Huot, Clifford Cheung, Alba Grassi, Kelian H\"aring, Enrico Herrmann, Callum Jones, Luke Lippstreu, Joan Elias Miró, Julio Parra-Martinez, Alessandro Podo, James Ratcliffe, Radu Roiban, Borna Salehian, Marco Serone, Mikhail Solon, Pedro Vieira, and Sasha Zhiboedov for useful discussions. M.C. is supported by the
National Science and Engineering Council of Canada (NSERC) and the Canada Research Chair program, reference number CRC-2022-00421. T.G., G.I., and A.M.W. are grateful to the Mani L. Bhaumik Institute for Theoretical Physics for support. G.I. is supported by the US Department of Energy under award number DE-SC0024224 and the Sloan Foundation. A.M.W. is supported by the NSF Graduate Research Fellowship under Grant No. DGE-2034835. G.I. and A.M.W. acknowledge support by grant NSF PHY-2309135 to the Kavli Institute for Theoretical Physics (KITP).
\begin{appendix}

\section{Analyticity for compact and exponentially decaying potentials}\label{app:potentialsproof}

Following \cite{Newton:2013ScatteringReprint, Taylor:1972} for scattering in non-relativistic quantum mechanics there are two main ways of showing analyticity in the UHP, depending on the potential. As mentioned in Figure \ref{fig:allResults}, if the potential decays faster than exponential, then convergence is always guaranteed and analyticity follows. If the potential decays exactly like an exponential (like the Yukawa or P\"oschl-Teller potentials) then we can extend analyticity up to a strip in the UHP. 

\textbf{Compactly supported potentials.} A potential $V(x)$ is compactly supported if it decays faster than exponentially,
\begin{equation}
\label{eq:compactV}
    V(x) \lesssim e^{-\mu |x|}, \qquad \text{ when }  x \to \pm\infty\,,
\end{equation}
for some $\mu > 0$.

In this case we can extend the analyticity of the Jost solutions $\phi_L(\omega,x)$ and $\phi_R(\omega,x)$ to the full lower-half-plane. For concreteness, we consider the case of the left outgoing Jost solution $\phi_L(\omega,x)$, with the analysis being completely analogous for $\phi_R(\omega,x)$. 

Now, even though, the free Green's function \eqref{eq:G0} can blow up exponentially for $x \to - \infty$, 
\begin{equation}
    G_0(\omega,x \to -\infty) \sim e^{2 \,|\mathrm{Im} \,\omega| |x|} \qquad \text{ for }  \;\; \mathrm{Im}\,\omega < 0,
\end{equation}
the faster-than-exponential decay \eqref{eq:compactV} of the potential ensures that each integral $\chi_L^{(n)}(\omega,x)$ in \eqref{eq:chin} converges and the solution $\phi_L(\omega,x) = e^{i \omega x} \chi_L(\omega,x)$ is analytic in the full complex plane. The same conclusion holds for $\phi_R(\omega,x)$ if the potential decays exponentially in both directions.

For this reason, the connection coefficients $A_\text{in}(\omega)$ and $A_\text{out}(\omega)$ as the integrals  \eqref{eq:Ainchi} and \eqref{eq:Aoutchi} remain convergent. This allows us to conclude that the reflection and transmission amplitudes $R(\omega)$ and $T(\omega)$ given in \eqref{eq:RT} are meromorphic functions everywhere.

\textbf{Exponentially decaying potentials.} Let us now consider exactly exponentially decaying potentials,
\begin{equation}
\label{eq:Vexpo}
    V(x \to \pm\infty) \sim e^{- \mu_\pm \, |x|},
\end{equation}
for some pair $\mu_{\pm} > 0$. 

In this case we find that the Jost solutions cannot be extended all the way in the lower-half-plane. In particular, we see that the first iteration in \eqref{eq:chin} is given by
\begin{equation}
\chi^{(1)}_L(\omega,x) =  \int_{- \infty}^x dx'\, G_0(\omega,x-x') \,V(x')
\end{equation}
with the integrand behaving as
\begin{equation}
    G_0(\omega,x) V(x) \sim e^{(2 \,|\mathrm{Im} \, \omega| - \mu_-)\, |x| }, \qquad x \to - \infty
\end{equation}
for $\mathrm{Im} \, \omega < 0$. Thus, convergence is only guaranteed for $\mathrm{Im} \, \omega > -\mu_- / 2$, and $\chi^{(1)}(\omega,x)$ is analytic in this extended domain.

Since every iteration $\chi_L^{(n)}(\omega,x)$ with $n > 0$ is given in terms of the previous $\chi_L^{(n-1)}(\omega,x)$ via the first line of \eqref{eq:chin}, we see that in the limit $x \to - \infty$ the support of the integral vanishes and we have $\chi_L^{(n)}(\omega,x \to - \infty) \to 0$ for $n > 0$. This implies that the integration \eqref{eq:chin} over higher iterations cannot be worse converged than that of the first iteration $\chi_L^{(1)}(\omega,x)$. This implies that the first term $\chi_L^{(1)}(\omega,x)$ dictates the analyticity domain of the full solution (see \cite{Newton:2013ScatteringReprint} for the explicit case of the Yukawa potential). 

Therefore, in the case of an exponentially decaying potential \eqref{eq:Vexpo} the Jost solution
\begin{equation}
\label{eq:chiexpo}
\begin{split}
    \phi_{L}(\omega,x) \;\text{ is analytic for }& \mathrm{Im} \, \omega > - {\mu_- \over 2}\,,\\
    \phi_{R}(\omega,x) \;\text{ is analytic for } &\mathrm{Im} \, \omega > - {\mu_+ \over 2}\,,
    \end{split}
\end{equation}
where we also included the analogous conclusion for the right outgoing Jost solution $\phi_R(\omega,x)$.

Let us now discuss the analyticity of the connection coefficients $A_\text{in}(\omega)$ and $A_\text{out}(\omega)$ with the added assumption \eqref{eq:Vexpo}. We consider first $A_\text{in}(\omega)$ which we have already shown to be analytic for $\mathrm{Im} \, \omega > 0$ in section \ref{sec:jost}.  From the relation \eqref{eq:Ainchi} we see that $A_\mathrm{in}(\omega)$ will inherit the analyticity domain of $\chi_L(\omega,x)$, as long as the integral converges in this domain. The integral may only diverge if the Jost solution $\chi_L(\omega,x')$ becomes unbounded as $x' \to +\infty$. 

We can generalize the bound \eqref{eq:chibound} on $G_0(\omega,x-x')$ to $\mathrm{Im} \, \omega < 0$,
\begin{equation}
    \label{eq:G0bound2}
    |G_0(\omega,x-x')| \leq {e^{2 \, |\mathrm{Im}\, \omega|(x-x')} \over |\omega|} \qquad \text{ for }\;\; x - x' > 0 \;\; \text{ and } \;\; \mathrm{Im \; \omega < 0}.
\end{equation}
Applying this bound to $\chi^{(n)}_L(\omega,x)$ in the second line of \eqref{eq:chin} we find
   \begin{align}
    |\chi^{(n)}_L(\omega,x)| \leq {e^{2 |\mathrm{Im} \,\omega| x} \over |\omega|^n} \int_{x>x_1>\dots>x_n} |V(x_1)| \cdots |V(x_n)| \, e^{-2 |\mathrm{Im} \,\omega| x_n} \, dx_1 \dots dx_n 
\end{align}
We can write a permutation symmetric bound by noting that 
\begin{equation}
e^{- 2 |\mathrm{Im} \, \omega| \,x_n} < e^{- 2 |\mathrm{Im} \, \omega| \,x_n} \,\Theta(-x_n) + \Theta(x_n),
\end{equation}
and likewise for every other $j  < n$,
\begin{equation}
1 < e^{- 2 |\mathrm{Im} \, \omega| \,x_j} \,\Theta(-x_j) + \Theta(x_j).
\end{equation}
This leads to the bound
\begin{align}
    |\chi^{(n)}_L(\omega,x)| &\leq {e^{2 |\mathrm{Im} \,\omega| x} \over n!\, |\omega|^n} \,\left[\int_{- \infty}^0 e^{- 2 |\mathrm{Im} \, \omega| \,x'}|V(x')| \,dx' + \int_{0}^x |V(x')| \,dx'\right]^n, \;\; \mathrm{Im} \,\omega < 0.
\end{align}
The first integral is guaranteed to converge for $\mathrm{Im} \, \omega > - \mu_- / 2$ region and we find that the series \eqref{eq:serieschi} is uniformly convergent,
\begin{equation}
\label{eq:chibound2}
    |\chi_L(\omega,x)| \leq \sum_{n=0}^\infty  |\chi_L^{(n)}(\omega,x)| \leq \exp \left[ 2 \,|\mathrm{Im} \, \omega| \,x  + \int_{- \infty}^0 e^{- 2 |\mathrm{Im} \, \omega| \,x'}|V(x')| \,dx' + \int_{0}^x |V(x')| \,dx' \right],
\end{equation}
in the strip $0>\mathrm{Im} \, \omega > - \mu_-/2$, which is inside the already established analyticity domain \eqref{eq:chiexpo}.

Importantly, this result bounds the growth $|\chi_L(\omega,x \to + \infty)| \lesssim e^{2 \,|\mathrm{Im} \, \omega| \,x}$. Given the exponential decay of the potential \eqref{eq:Vexpo}, we find that the integral \eqref{eq:Ainchi} for $A_\text{in}(\omega)$ converges for $\mathrm{Im} \, \omega > - \mu_+/2$. So we conclude that
\begin{equation}
\label{eq:Ainexpo}
    A_\text{in}(\omega) \;\; \text{ is analytic for } \;\; \mathrm{Im} \,\omega > - {\min(\mu_-,\mu_+) \over 2}.
\end{equation}
In other words, $A_\text{in}(\omega)$ is analytic in the intersection of the domains of analyticity of each Jost solution $\chi_L(\omega,x)$ and $\chi_R(\omega,x)$, given in \eqref{eq:chiexpo}. This could have also been derived by noting that $A_\text{in}(\omega)$ is the Wronskian between the two Jost solutions. 

Let us now discuss the case of $A_\text{out}(\omega)$. Taking $\mathrm{Im} \, \omega > 0$ we see that the integrand in \eqref{eq:Aoutchi} may only blow-up in the limit $x' \to + \infty$ where it behaves as $\sim e^{(2 \,|\mathrm{Im} \, \omega| - \mu_+) \,x'}$ (recall that $\chi_L(\omega,x)$  is bounded by \eqref{eq:chibound} for $\mathrm{Im} \, \omega > 0$). Therefore, for $0<\mathrm{Im} \, \omega < \mu_+/2$ we will have convergence and $A_\text{out}(\omega)$ will be analytic in this region.

There is also analyticity in the lower half-plane. For $\mathrm{Im} \, \omega < 0$ the integrand of \eqref{eq:Aoutchi} instead may now diverge for $x' \to - \infty$ where it behaves as $\sim e^{(2 \,|\mathrm{Im} \, \omega| - \mu_-) \,|x'|}$ (recall that $\chi_L(\omega,x\to - \infty) \to 1$ according to the boundary condition \eqref{eq:bcLR} where $\chi_L(\omega,x) = e^{i \omega x} \phi_L(\omega,x)$). This implies that $|\mathrm{Im}\,\omega| < \mu_-/2$, or $\mathrm{Im}\,\omega >- \mu_-/2$, in order to have convergence.\footnote{Note that the opposite limit $x' \to + \infty$ is safe: the exponential growth of $\chi_L(\omega,x)$ from \eqref{eq:chiexpo} is canceled by the pre-factor $e^{-2 i \omega x'} \sim e^{- 2 |\mathrm{Im} \,\omega| x'}$ in the integrand.} We thus establish that
\begin{equation}
\label{eq:Aoutexpo}
    A_\text{out}(\omega) \; \text{ is analytic for } \; {\mu_+ \over 2} >\mathrm{Im} \,\omega > - {\mu_- \over 2}.
\end{equation}

Putting everything together we find the following domains of analyticity for the S-matrix. The transmission amplitude $T(\omega)= 1/A_\text{in}(\omega)$ will be meromorphic in the domain where $A_\text{in}(\omega)$ is analytic \eqref{eq:Ainexpo},
\begin{equation}
\label{eq:Texpo}
    T(\omega) \; \text{ is meromorphic for } \;\; \mathrm{Im} \,\omega > - {\min(\mu_-,\mu_+) \over 2}.
\end{equation}
The reflection amplitude $R(\omega) = - {A_\text{out}(\omega) / A_\text{in}(\omega) }$ on the other hand will be meromorphic in the intersection of the domains \eqref{eq:Ainexpo} and \eqref{eq:Aoutexpo},
\begin{equation}
\label{eq:Rexpo}
    R(\omega) \; \text{ is meromorphic for } \; {\mu_+ \over 2} >\mathrm{Im} \,\omega > - {\min(\mu_-,\mu_+) \over 2}.
\end{equation}

These analyticity domains are represented in Figure \ref{fig:allResults}.

\section{Bound-state singularities, causality and the Laplace transform}\label{app:boundstates}

For potentials that are negative in some part of its domain, there can exist bound state poles in the UHP. An example is the P\"oschl-Teller potential, in Eq. \eqref{eq:PTpot}, for $\lambda>1$ \cite{Cevik:2016mnr}. This might naively lead one to believe that existence of bound states violate causality. 

However, the use of the Fourier transform of the wavefunction, in Eq. \eqref{eq:FT}, is invalidated by the presence of a bound state pole: the time-domain wavefunction for a bound state grows exponentially with time (see Section \ref{sec:smatrixpoles}).\footnote{Note that in quantum mechanics, the time dependence of a bound-state wavefunction is given $e^{-i E t}=e^{i k^2 t}$, so this tension does not arise.}
\begin{equation}
\label{eq:expgrow}
    \phi(t,x)\sim e^{-i\omega t}=e^{k t}\to\infty\text{\,\, as \,\,} t\to\infty\,,
\end{equation}
where $\omega=ik$ with $k>0$.

This means that in order to discuss the implications of causality in frequency space, we need to ensure that our frequency-space wavefunctions and Green's functions are well defined. A good definition is found by using the Laplace transform instead of the Fourier transform, which effectively shifts the contour above the bound state pole, ensuring convergence.

The Laplace transform reads
\begin{equation}
    \phi(\omega,x) = \int_{0}^{\infty} dt \, e^{i \omega t} \phi(t,x) \,,
\end{equation}
which only converges for $\text{Im}\,\omega > k$, due to the exponential growth \eqref{eq:expgrow}. In turn, the inverse Laplace transform is given by
\begin{equation}
    \phi(t,x)= \int_{-\infty+i\beta}^{\infty+i\beta}\frac{d\omega}{2\pi} \, e^{-i \omega t}  \phi(\omega,x) \,,
\end{equation}
for any $\beta>k$, ensuring that the integral remains convergent throughout the contour.

With this careful relationship between $\phi(t,x)$ and $\phi(\omega,x)$ (and the corresponding one for $G_R(t,x,x')$ and $G_R(\omega,x,x')$), the time-domain wavefunction and retarded Green's function vanish for $t<0$ thus ensuring consistency with causality when bound states are present. 

 \section{Sub-leading corrections to high-frequency limit}\label{app:SchwHighFreqCorr}

We consider the subleading (NLO) correction to $f(z)$, given in Eq.~\eqref{eq:DEvz}, in the limit $|R_s\omega|\gg1$:
\begin{equation}
    f(z)=2iR_s\omega-\frac{1}{2}-\frac{i\left(1+2\ell(\ell+1)\right)}{4R_s\omega}+\mathcal{O}\left(|R_s\omega|^{-2}\right)\,.
\end{equation}
At the order $\mathcal{O}\left(|R_s\omega|^{-2}\right)$, the solutions to~\eqref{eq:DEvz} in this limit asymptote to plane waves. 
We solve for the Jost solution at this order, denoted by $\phi_L^\text{NLO}(\omega,r)$, obtaining
\begin{equation}
    \phi_L^\text{NLO}(r)=e^{i \omega  (r-2  R_s )} \left(\frac{r}{ R_s }\right)^{\frac{1}{2}} \left(\frac{r}{ R_s }-1\right)^{-i  R_s  \omega } \mathcal{M}\left(i\beta-2 i  R_s  \omega +\frac{1}{2},1-2 i  R_s  \omega ,-2 i \omega(r- R_s ) \right)
\end{equation}
with $\beta\equiv \frac{ 1+2\ell(\ell+1)}{4  R_s  \omega }$. From the asymptotic limit of $\phi_L^\text{NLO}(\omega,r \to \infty)$ we obtain the reflection coefficient, via Eq. \eqref{eq:BCplus},
\begin{equation}
    R^\text{NLO}(\omega)=-\frac{e^{-2 i  R_s  \omega } (2i  R_s  \omega )^{-1+2 i  R_s  \omega }  (2 R_s) ^{1-2i\beta} \left(\omega ^2\right)^{\frac{1}{2}-i\beta} \Gamma \left(i\beta-2 i  R_s  \omega +\frac{1}{2}\right)}{\Gamma \left(\frac{1}{2}-i\beta\right)}
\end{equation}
which reduces to Eq.~\eqref{eq:S0largeomega} when $\beta=0$.
The behavior of $R^\text{NLO}(\omega)$ as $|\omega|\to\infty$ in different parts of the complex plane exactly matches the leading-order behavior in Eq.~\eqref{eq:Romegainfty}. The high-frequency approximation therefore converges for $|R_s\omega|\gg 1$. 

The leading-order reflection coefficient~\eqref{eq:S0largeomega} has approximate QNMs that lie on the negative imaginary axis and have no real part, unlike the actual QNMs \cite{nollert_quasinormal_1993,andersson_asymptotic_1993}. However, from $R^\text{NLO}(\omega)$ we see a small nonzero real part of the first several QNMs (depending on $\ell$), which are solutions to the equation
\begin{equation}
    i\beta-2 i  R_s  \omega +\frac{1}{2}=\frac{i(1+2\ell(\ell+1)}{4R_s\omega}-2 i  R_s  \omega +\frac{1}{2}=-n\,,\qquad n=0,1,2,...
\end{equation}
This equation yields the QNMs:
\begin{equation}
    \omega_n^\pm=\frac{-i}{8R_s}\left(1+2n\pm \sqrt{-7-16\ell(\ell+1)+4n(n+1)}\right)
\end{equation}
which have nonzero real parts when 
\begin{equation}
    -7-16\ell(\ell+1)+4n(n+1)<0\,.
\end{equation}
Therefore this approximation gives pairs of QNMs with non-zero real part symmetric about the negative imaginary axis for $n\leq 2\ell+1$.
We presume from this analysis that keeping more terms in the large frequency expansion produces better approximations for the QNMs.

\section{Bootstrap bounds on Love numbers without gravity}\label{app:bounds}
As teased in the introduction and conclusions of the paper, bounding Love numbers of compact objects using the S-matrix bootstrap program is an exciting possibility, and in this appendix, we discuss how one would go about doing so. As a first attempt, we look at theories without gravity and potentials that have good fall-off behavior as stated in Section \ref{sec:asymG}. Love numbers in such theories could, for example, relate to the electric and magnetic polarizabilities of nucleons \cite{Drechsel:2002ar,Griesshammer:2012we}. Along with assumptions of unitarity and analyticity of the S-matrix, we use the  maximum modulus principle to bound Wilson coefficients for a point-particle EFT.

The maximum modulus principle \cite{ConwayF1CV2e} in complex analysis states that if a function $f(z)$ is holomorphic in a compact domain, then the modulus $|f(z)|$ cannot exhibit a strict maximum that is within the domain of $f(z)$. This principle has been used before in various S-matrix and conformal bootstrap works such as \cite{Hogervorst:2013sma,Paulos:2017fhb,He:2018uxa,Guerrieri:2018uew}. We will be following the analysis in \cite{EliasMiro:2019kyf}, where this theorem is used to bound the EFT coefficients of the $d=3$ QCD flux tube.

We assume that the potential is such that the reflection coefficient $R(\omega)$ is analytic and that $|R(\omega)|$ is bounded in the UHP. Additionally, unitarity implies that $|R(\omega)|\leq1$ for $\omega\in\mathbb{R}$. Hence, by the Phragmén-Lindel\"of principle \cite{ahlfors1973}, $|R(\omega)|\leq 1$ for all values of $\omega$ in the UHP. 

We match this setup to a wave scattering against a generic compact object. The compact object is described by a static potential which is derived from a worldline EFT \cite{Ivanov:2024sds,Caron-Huot:2025tlq}. The Wilson coefficients of the EFT are called Love numbers, and encode information about the size and internal structure of the object. 

The reflection coefficient is identified with the $\ell$-th partial wave amplitude in the spherically-symmetric system,
which is obtained from the action~\eqref{eq:SLove} (using the conventions of~\cite{Caron-Huot:2025tlq})
\begin{equation}
        R(\omega)=\frac{1+i\omega\,\frac{ F_\ell(\omega)}{4\pi}}{1-i\omega\,\frac{ F_\ell(\omega)}{4\pi}}
\end{equation}
with
 \begin{equation} F_\ell(\omega)=\sum_{n=0}^\infty (i\omega)^n C_{\ell,n}\,.
\end{equation}
$C_{\ell,n}$ are
the tidal Love numbers. The low-frequency expansion of the amplitude is given by
\begin{equation}
    R(\omega) = 1 + i \omega \, \frac{C_{\ell,0}}{2\pi}-\omega^2\,\frac{C_{\ell,0}^2+4\pi C_{\ell,1}}{8\pi^2}+\mathcal{O}(\omega^3)
\end{equation}

Armed with an expression for the reflection coefficient in terms of the Love numbers, we use the assumptions of unitarity and analyticity to constrain the allowed values. 
From the maximum modulus principle, we know that $|R(\omega)| \leq 1$ for all values of $\omega$ in the UHP. 

We expand at low frequencies in a direction $\theta$ via the relation $\omega=\epsilon e^{i\theta}$,  keeping only the leading term in $\epsilon$. We find
\begin{equation}
    R(\epsilon e^{i\theta}) = 1 + i A \epsilon e^{i \theta} - B\epsilon^2 e^{2i \theta} + \mathcal{O}(\epsilon^3)
\end{equation}
with
\begin{equation}
    \begin{split}
        A &= \frac{C_{\ell,0}}{2\pi}\,,\qquad B = \frac{C_{\ell,0}^2+4\pi C_{\ell,1}}{8\pi^2}\,.
    \end{split}
\end{equation}
We then apply the constraint
\begin{equation}
    |R(\epsilon e^{i\theta})|^2 = 1 - 2A\epsilon \sin(\theta) - 2B \epsilon^2 \cos(2 \theta) +A^2 \epsilon^2 + \mathcal{O}(\epsilon^3)\leq 1\,.
\end{equation}
Approaching the origin with $\theta=\pi/2$, we bound the leading term
\begin{equation}
    -2A \leq 0 \Rightarrow C_{\ell,0}\geq 0\,.
\end{equation}
If we instead set $\theta = 0$, then the $O(\epsilon)$ term vanishes and we get
\begin{equation}
    A^2 -2B \leq 0 \Rightarrow C_{\ell,1} \geq 0 \,.
\end{equation}
By assuming the reflection coefficient is analytic and bounded, we have constrained the first two Love numbers in our worldline EFT. Higher-order coefficients could be bounded using the procedure set out in 
\cite{EliasMiro:2019kyf,EliasMiro:2021nul}.

\section{Connection with Quantum Field Theory}\label{app:QFT}

In this appendix, we reformulate the framework of the main text in a language closer to quantum field theory. We first review how the retarded Green’s function can be expressed in terms of two-point correlators within the Schwinger–Keldysh formalism. We then verify that the relation between the asymptotic behavior of 
$G_R$ and the S-matrix \eqref{eq:classicalLSZintro}, while derived classically, is consistent with the LSZ reduction procedure. Finally, we work out an explicit QFT example, the tree-level Yukawa interaction, and show that the Stokes phenomenon also arises in $2\to2$ scattering, when considering the retarded Green’s function of the corresponding effective one-body Schrödinger equation.

\subsection*{Schwinger-Keldysh formalism and the retarded Green's function}\label{sec:LSZ}
A widely used approach to capture the dynamics of an open system is the in-in (Schwinger-Keldysh) formalism \cite{Keldysh:1964ud,Martin:1973zz}, which amounts to evolving the system along a time contour from $t=-\infty$ to $t=+\infty$ (forward branch) and then all the way back to $-\infty$ (backward branch). This contour evolution ensures that we compute expectation values of operators in a given state, rather than transition amplitudes as in the in-out formalism. 

The path integral representation of this contour evolution is implemented by doubling the fields in the action:
\begin{equation}\label{eq:Ssk}
S_{\mathrm{SK}}[\phi_1, \phi_2]=S[\phi_1]-S[\phi_2]\,.
\end{equation}
Here, the index $1$ or $2$ labels the branch of the contour on which the field is evaluated, and the minus sign reflects the reversed time orientation of the backward branch.

To obtain correlation functions with a more transparent physical interpretation, it is convenient to work in the Keldysh basis \cite{Keldysh:1964ud}: $\phi_+=\frac{1}{2}(\phi_1+\phi_2)$ and $\phi_-=\phi_1-\phi_2$.
In this basis, the field $\phi_+$ has the interpretation of the classical expectation value of the field, and its equations of motion follows from varying the in-in action $\delta S_{\mathrm{SK}}/\delta \phi_-$. Equivalently, its mean field configuration can be obtained from the generating function $W[J_+,J_-]$ as\,\footnote{Notice that the source term in the action takes the form $J_+ \phi_- + J_- \phi_+$, meaning that insertions of $\phi_+$ are obtained by applying $\delta/\delta J_-$ and vice versa.}
\begin{equation}
\langle \phi_+(x)\rangle =i\frac{\delta W[J_+,J_-]}{\delta J_-(x)}\bigg\rvert_{J_-=J_+=0}\,.
\end{equation}
Hence, we can identify the classical field satisfying the equation of motion \eqref{eq:waveeq} with $\langle \phi_+(x)\rangle$.
Notice that all correlators here are understood as evaluated on the given background or initial state, not necessarily in the vacuum.

The two point correlation functions in this basis are given by
\begin{equation}\label{eq:2ptG}
\begin{pmatrix}
\langle \phi_+(x) \phi_+(y)\rangle & \langle \phi_+(x) \phi_-(y)\rangle \\
\langle \phi_- (x)\phi_+(y)\rangle & \langle \phi_-(x) \phi_-(y)\rangle 
\end{pmatrix}
=
i\begin{pmatrix}
 G_K(x,y) &  G_R(x,y) \\
 G_A(x,y) & 0 
\end{pmatrix}\,,
\end{equation}
where $G_R$ and $G_A $ is the retarded and advanced Green's function, the Keldysh Green's function $G_K = \frac{1}{2}\langle \{\phi(x),\phi(y)\} \rangle$ encodes fluctuation effects and statistical/thermal properties of the state and $\langle \phi_-\phi_-\rangle=0$ follows from the largest time equation \cite{tHooft1974}.

In particular, the retarded Green's function is given by $G_R(x,y)= - \delta \langle \phi_+(x)\rangle/\delta J_+(y)$, matching the standard result from linear response theory
\begin{equation}
\langle \phi_+(x)\rangle = - \int d^4 y \, G_R(x,y)J_+(y)\,.
\end{equation}

\vspace{0.3cm}

The on-shell retarded amplitude is obtained from the retarded two point function $ \langle \phi_+(x) \phi_-(y)\rangle$ by a generalization of the LSZ procedure \cite{Caron-Huot:2023vxl}.  For massless scalars, the amputated retarded two point function reads
\begin{equation}\label{eq:LSZ}
 \int d^4 x \,e^{i p'\cdot x}\int d^4 y \,e^{-i p\cdot y}\,(-i\partial_x^2)\,(-i\partial_y^2)\, \langle \phi_+(x)\phi_-(y)\rangle\,,
\end{equation}
with $p=(\omega,\mathbf{p})$ the incoming momentum and $p'=(\omega',\mathbf{p}')$ the outgoing momentum.

Following \cite{Caron-Huot:2023vxl}, we define the asymptotic creation ($a_p^{\mathrm{in}}$) and annihilation ($a_p^{\mathrm{out}}$) operators as the on-shell limit of the current 
\begin{equation}
j(p)=\int d^4x \,e^{ip\cdot x}(-i\partial_x^2)\phi(x)\,,
\end{equation}
which reduces to
\begin{align}\label{eq:rulej}
\mathrm{incoming}:\quad &j(p)\rightarrow a_p^{\dagger \,\mathrm{in}}-a_p^{\dagger \,\mathrm{out}}\\\nonumber
\mathrm{outogoing}:\quad &j(p)\rightarrow a_p^{\mathrm{out}}-a_p^{\mathrm{in}}
\end{align}

Applying this amputation procedure to \eqref{eq:LSZ}, the on-shell limit yields the S-matrix:
\begin{equation}\label{eq:2ptainaout}
\langle[a_{p'}^{\mathrm{out}},a_p^{\dagger \,\mathrm{in}}] \rangle =2\pi \delta(\omega'-\omega)\,2 \omega \,S(\mathbf{p},\mathbf{p}') 
\end{equation}
where the expectation value is again taken in the chosen background state. Replacing the two point function in \eqref{eq:LSZ} with the retarded Green's function, we obtain the LSZ reduction formula
\begin{equation}\label{eq:LSZ2}
2\pi \delta(\omega'-\omega)2 \omega \,S(\mathbf{p},\mathbf{p}') = i \int d^4 x \,e^{i p'\cdot x}\int d^4 y \,e^{-i p\cdot y}\,(-i\partial_x^2)\,(-i\partial_y^2)\,G_R(x,y)\,.
\end{equation}

\subsection*{Consistency of the asymptotic behavior with LSZ procedure}
In this section, we show that the LSZ reduction \eqref{eq:LSZ2} is consistent with the asymptotic behavior 
\begin{align}\label{eq:classicalLSZ} 
G_R^\ell(\omega,r,r')\xrightarrow[r,r'\to\infty]{} \frac{1}{2i\omega}\left[e^{i\omega\left( r'-r\right)}-(-1)^\ell\,S_\ell(\omega)\,e^{i\omega(r+r')} \right]\,
\end{align}
where $S_\ell(\omega)$ is the partial wave amplitude.
 
  To this end, we start by manipulating the right hand side (rhs) of \eqref{eq:LSZ2}. The integrand can be expressed as a total derivative:
\begin{equation}
\!\!\!\!\!\text{(\ref{eq:LSZ2} rhs)}=i \int d^4 x \int d^4 y \frac{\partial}{\partial x^\mu} \frac{\partial}{\partial y^\nu}\left[ \,e^{i p'\cdot x}\,e^{-i p\cdot y}\,(-i\partial_x^\mu+p'^\mu)\,(-i\partial_y^\nu-p^\nu)\, G_R(x,y)\right].
\end{equation}
By Stoke's theorem, the volume integrals reduce to surface integrals at infinity. Assuming asymptotically flat spacetimes, it is convenient to parametrize the 3D hypersurfaces as
\begin{equation} d\Sigma_\mu = \lim_{|\mathbf{x}|\to \infty }|\mathbf{x}|^2\,\hat x_\mu\, d\Omega\,dt\,,
\end{equation}
with $\hat x_\mu = (0,\hat x)$ the unit normal vector to the surface at spatial infinity at fixed time. 

Time translation invariance $G_R(x_0-y_0,\mathbf{x},\mathbf{y})$ allows us to handle the time integrals trivially using the Fourier representation in \eqref{eq:FTGR}, and the retarded Green's function in the frequency domain can be expanded in spherical harmonics $Y_{\ell m}$:
\begin{equation}\label{eq:GRGl}
G_R(\omega,\mathbf{x},\mathbf{y}) =\frac{1}{rr'} \sum_{\ell m }Y_{\ell m}(\hat x)Y^*_{\ell m}(\hat y)\,G_R^\ell(\omega,r,r')\, ,
\end{equation}
with $|\mathbf{x}|=r$ and $|\mathbf{y}|=r'$.
Since the integral is evaluated at spatial infinity, we can safely replace $G_R(\omega,r,r')$ with its asymptotic form \eqref{eq:classicalLSZ}.

After the smoke clears, we find
\begin{align}
\!\!\!\!\!\text{(\ref{eq:LSZ2} rhs)}&=\!\lim_{r,r'\to \infty } 2\pi i \delta(\omega-\omega')\sum_{\ell m} \,\frac{r r' }{2i\omega}\int  d\Omega_x \, e^{-i \mathbf{p}'\cdot \mathbf{x}} \,Y_{\ell m}(\hat x)\int d\Omega_y\, e^{i \mathbf{p}\cdot \mathbf{y}}\,Y_{\ell m}^*(\hat y)\\\nonumber
&\times (-\omega+\hat y\cdot \mathbf{p})\left[(-\omega+\hat x\cdot \mathbf{p'} )e^{i\omega(r'-r)}-(-1)^\ell\,S_\ell(\omega)\,(-\omega-\hat x\cdot \mathbf{p'} )e^{i\omega(r+r')}\right]\label{eq:rhs3}\,.
\end{align}

The remaining angular integrals are highly oscillating everywhere except when the momentum direction $\mathbf{p}$ (or $\mathbf{p}'$) is aligned either with   the source ($r$) or observer ($r'$) locations. We can therefore perform a saddle point approximation around $\hat x \sim  \mathbf{p}$ and $\hat y \sim - \mathbf{p}'$ obtaining
\begin{align}
\!\!\!\!\!\text{(\ref{eq:LSZ2} rhs)}&=-\frac{4 \pi^3 \delta(\omega-\omega')}{\omega|\mathbf{p}||\mathbf{p}'|}  \sum_{\ell m}Y_{\ell m}(\hat p')Y_{\ell m}^*(-\hat p)\left[\,e^{-i |\mathbf{p}'| r}\,e^{-i |\mathbf{p}| r'}\right]\\\nonumber
&\times (-\omega- | \mathbf{p}|)\left[(-\omega+| \mathbf{p'}| )e^{i\omega(r'-r)}-(-1)^\ell\,S_\ell(\omega)(-\omega-| \mathbf{p'} |)e^{i\omega(r+r')}\right]\,.
\end{align}
Lastly, we take the on-shell limit $|\mathbf{p}|=\omega$, $|\mathbf{p}'|=\omega$, and the above expression collapses to the partial wave expansion of the S-matrix,\footnote{Where we used $Y_{\ell m }(-\hat p) = (-1)^\ell Y_{\ell m }(\hat p) $ and $\sum_{m=-\ell}^{\ell} Y_{\ell m}(\hat x)Y^*_{\ell m}(\hat y) = \frac{2\ell+1}{4\pi} P_\ell(\hat x \cdot \hat y)$.}, finding
 \begin{equation}
\!\!\!\!\!\text{(\ref{eq:LSZ2} rhs)}=\pi\delta(\omega-\omega')2\omega\left[\frac{\pi}{\omega^2 }\sum (2\ell+1)P_\ell(\hat p\cdot \hat p') S_\ell(\omega)\right]\,,
\end{equation}
which matches the expected result of \eqref{eq:LSZ2}.\footnote{Notice that in our convention $S(\mathbf{p},\mathbf{p}') = \frac{\pi}{\omega^2}\sum_\ell (2\ell+1)S_\ell(\omega) P_\ell(\hat p \cdot \hat p')$.}

The two pieces in  the second line of Eq. \eqref{eq:rhs3} have a clear physical meaning. The first term, proportional to $e^{i \omega (r'-r)}$, corresponds to the ``incoming'' wave propagating from the source to the observer; in the on-shell limit it vanishes, reflecting that the free incoming field does not contribute to the amputated amplitude. 

The second term, proportional to $S_\ell(\omega)$ and $e^{i \omega (r+r')}$, encodes the outgoing scattered wave at spatial infinity.  
Thus, the asymptotic classical relation \eqref{eq:classicalLSZ} is captured by the more general LSZ reduction formula.

\subsection*{Stokes phenomena in the Yukawa potential at tree-level}\label{sec:yukawa}

In this section we illustrate Stokes phenomena for the tree-level Yukawa interaction. Namely, the scattering of two scalar particles via exchange of another scalar of mass $\mu$. This example is convenient because the solutions involve elementary special functions with a well-understood analytic structure, making the Stokes phenomenon particularly transparent.

The effective one-body equation \cite{Correia:2024jgr} for this process is
\begin{equation}\label{eq:YukEOB}
\left[\frac{d^2}{d r^2}-\frac{\ell(\ell+1)}{r^2} +|\mathbf{p}|^2\right]\psi_\ell(r) = V(r)\psi_\ell(r)
\end{equation}
where $\mathbf{p}$ is the momentum in the center-of-mass frame. The Yukawa potential is given by\footnote{This follows from the  Fourier transform of the tree-level expression
$\alpha/(|\mathbf{q}|^2+\mu^2)$, where 
$\alpha$ is the product of the cubic couplings of the external scalars to the exchanged field of mass $\mu$, and $\mathbf{q}$ is the momentum transfer.}
\begin{equation}\label{eq:Yukpot}
V(r) =-\alpha\frac{e^{-\mu r}}{ r}\,.
\end{equation}
Here $\psi_\ell(r)$ is the radial wavefunction of the effective one-body problem that encodes the two-body dynamics, rather than a wave on a fixed background as in the main text.

Since a closed-form solution of \eqref{eq:YukEOB} with the Yukawa potential is not known we will work perturbatively at leading order in the Born series. 

We first compute the retarded Green’s function
\begin{equation}\label{eq:GreenWE}
\left[\frac{\partial^2}{\partial r^2}-\frac{\ell(\ell+1)}{r^2} +|\mathbf{p}|^2-V(r)\right]G_R^\ell(|\mathbf{p}|,r,r') = \delta(r-r')
\end{equation}
at leading order. We do not know the analytic properties of this effective one-body Green’s function, since it effectively encodes the information of a four-point correlator. We will now get some insight onto these properties by analyzing the explicit example of the Yukawa potential at leading order.

 Notice first that in the case of free propagation, $V(r)=0$, the retarded Green's function simplifies to
\begin{equation}\label{eq:freeG}
\begin{split}
G_0(r,r')&\equiv G_R^\ell(|\mathbf{p}|,r,r')\bigg\rvert_{V=0}\\
&=-\frac{1}{|\mathbf{p}|}\left[\theta(r'-r) j_\ell(|\mathbf{p}|r)h_\ell^+(|\mathbf{p}|r')+\theta(r-r') j_\ell(|\mathbf{p}|r')h_\ell^+(|\mathbf{p}|r)\right]\,,
\end{split}
\end{equation}
where $j_\ell(z)$ is a spherical Bessel function of the first kind, related to the Riccati-Hankel function by $j_\ell(z)=\frac{1}{2i}\left(h_\ell^+(z)-h_\ell^- (z)\right)$. 
 
The two independent homogenous solutions at leading order in $\alpha$ can be constructed from the free solutions $j_\ell(|\mathbf{p}| r)$ and $h^+_\ell(|\mathbf{p}| r)$ as follows:
\begin{align}\label{eq:psi1Born}
\bullet\;\,\,\psi^\text{in}_\ell(r)&=-2i\left[ j_\ell(|\mathbf{p}| r)+i\int_0^\infty dr' \, G_0(r,r') \, V(r') \,j_\ell(|\mathbf{p}| r')\right]\\[0.8ex]\nonumber
&\xrightarrow[r\to\infty]{}e^{-i\left(|\mathbf{p}| r-\frac{\pi}{2}\ell\right)}-\Bigg(\underbrace{1-\frac{2i}{|\mathbf{p}| }\int_0^\infty j_\ell(|\mathbf{p}| r')\,V(r') \,j_\ell(|\mathbf{p}| r')}_{S_\ell(|\mathbf{p}|)}\Bigg)\,e^{i\left(|\mathbf{p}| r-\frac{\pi}{2}\ell\right)}\,,\\[1.5ex]\label{eq:psi2Born}
\bullet\;\,\,\psi^\mathrm{up}_\ell(r)&=h_\ell^+(|\mathbf{p}| r)+\int_0^\infty dr' \, G_0(r,r') \, V(r') \,h_\ell^+(|\mathbf{p}| r')\\[0.8ex]\nonumber
&\xrightarrow[r\to\infty]{}\Bigg(\underbrace{1-\frac{1}{|\mathbf{p}|}\int_0^\infty j_\ell(|\mathbf{p}| r')\,V(r')\, h_\ell^+(|\mathbf{p}| r')dr' }_{B_{\mathrm{out}}(|\mathbf{p}|)}\Bigg)\,e^{i\left(|\mathbf{p}| r-\frac{\pi}{2}\ell\right)}\,.
\end{align}
By construction, their asymptotic behavior matches the expected boundary conditions, allowing one to extract the S-matrix $S_\ell(\omega)$ and the normalization coefficient $B_\mathrm{out}(\omega)$. Using standard Bessel function integrals\footnote{For instance identity 6.612.3 of Gradshteyn and Ryzhik \cite{GradshteynRyzhik} where $\int_0^\infty e^{-\alpha x}\, J_{\nu}(bx)\, J_{\nu}(\gamma x) \,dx =\frac{1}{\pi \sqrt{\gamma b}}Q_{\nu-\frac{1}{2}}\left(\frac{\alpha^2+b^2+\gamma^2}{2b\gamma} \right)$. } we obtain 
\begin{align}\label{eq:SYuk}
S_\ell(|\mathbf{p}|) &=
\displaystyle
1+\frac{i\alpha}{|\mathbf{p}|}\,
Q_\ell\!\left(1+\frac{\mu^2}{2|\mathbf{p}|^2}\right)\,,
\\[1.5ex]
B_{\mathrm{out}}(|\mathbf{p}|) &=
\displaystyle
1+\frac{\alpha}{2|\mathbf{p}|}\Bigg[
   i\,Q_\ell\!\left(1+\frac{\mu^2}{2|\mathbf{p}|^2}\right)
   +P_\ell\!\left(1+\frac{\mu^2}{2|\mathbf{p}|^2}\right)
      \arctan\!\left(\frac{2|\mathbf{p}|}{\mu}\right)
   -\sum_{m=0}^{\ell}\gamma_m\!\left(\frac{\mu}{|\mathbf{p}|}\right)^{2m+1}
\Bigg]\,,\notag
\end{align}
where $\gamma_m$ are numerical coefficients taking different values depending on the partial wave, whose exact form will not play a role in this discussion, and $Q_\ell$ represents Legendre polynomials of second kind. 

The wavefunctions \eqref{eq:psi1Born} and \eqref{eq:psi2Born} at finite $r$ can be expressed as
\begin{align}\nonumber
\psi^\mathrm{in}_\ell(r)=&-\frac{h_\ell^-(|\mathbf{p}| r)}{2i}\Bigg[ 1-\frac{i\alpha}{2|\mathbf{p}|}P_\ell\left(1+\frac{\mu^2}{2|\mathbf{p}|^2}\right)\bigg( \Gamma(0,\mu r-2i|\mathbf{p}| r)-\Gamma(0,\mu r)\bigg)\\\label{eq:psi1sol}
&+e^{-\mu r}\sum_{m=1}^{2\ell}\frac{ e^{2i|\mathbf{p}| r}a_m(|\mathbf{p}|)-b_m(|\mathbf{p}|)}{r^m}\Bigg]\\\nonumber
&+\frac{h_\ell^+(|\mathbf{p}| r)}{2i}\Bigg[S_\ell(|\mathbf{p}|)+\frac{i\alpha}{2|\mathbf{p}|}P_\ell\left(1+\frac{\mu^2}{2|\mathbf{p}|^2}\right)\bigg( \Gamma(0,\mu r+2i|\mathbf{p}| r)-\Gamma(0,\mu r)\bigg)\\\nonumber
&-e^{-\mu r}\sum_{m=1}^{2\ell}\frac{ e^{-2i|\mathbf{p}| r}a_m^*(|\mathbf{p}|)-b_m(|\mathbf{p}|)}{r^m}\Bigg]\,,
\end{align}
\begin{align}\nonumber
\psi^\mathrm{up}_\ell(r)&=\, h_\ell^+(|\mathbf{p}| r)\,\Bigg[ B_\mathrm{out}(|\mathbf{p}|)+\frac{\alpha}{2i|\mathbf{p}|}P_\ell\left( 1+\frac{\mu^2}{2|\mathbf{p}|^2}\right)\Gamma(0,\mu r) +\frac{\alpha}{2i|\mathbf{p}|}e^{-\mu r }\sum_{m=1}^{2\ell}\frac{b_m(|\mathbf{p}|)}{r^m}\Bigg]\\ 
&-h_\ell^-(|\mathbf{p}| r)\frac{\alpha}{2i|\mathbf{p}| }\left[ P_\ell\left( 1+ \frac{\mu^2}{2|\mathbf{p}|^2}\right)\Gamma(0,\mu r -2i|\mathbf{p}| r)+e^{2i|\mathbf{p}| r-\mu r}\sum_{m=1}^{2\ell}\frac{a_m(|\mathbf{p}|)}{r^m}\right]\,,\label{eq:psi2sol}
\end{align}
where $a_m(|\mathbf{p}|)$ and $b_m(|\mathbf{p}|)$ are polynomials in $|\mathbf{p}|$ and $\mu$ of degree and $\Gamma(s,z)$ is the  incomplete gamma function
\begin{equation}
\Gamma(s,z) = \int_{z}^\infty z^{s-1}e^{-t}dt\, .
\end{equation}
Since $\Gamma(0,r)$ is exponentially suppressed at large $r$, it is straightforward to verify that these expressions reduce to $S_\ell(|\mathbf{p}|)$ and $B_\mathrm{out}(|\mathbf{p}|)$ in the appropriate limits \eqref{eq:psi1Born} and \eqref{eq:psi2Born}. Given the exact solutions \eqref{eq:psi1sol} and \eqref{eq:psi2sol}, the full retarded Green's function at tree level can then be constructed using \begin{equation}\label{eq:Gpsi1psi2}
G_R^\ell(|\mathbf{p}|,r,r') =-\frac{1}{W(\psi_\ell^{\mathrm{in}},\psi_\ell^{\mathrm{up}})}\left[\theta(r'-r)\,\psi_\ell^{\mathrm{in}}(r)\,\psi_\ell^{\mathrm{up}}(r')+\theta(r-r')\,\psi_\ell^{\mathrm{in}}(r')\,\psi_\ell^{\mathrm{up}}(r) \right]\,,
\end{equation}
where $W(\psi_\ell^{\mathrm{in}},\psi_\ell^{\mathrm{up}})$ is the Wronskian.

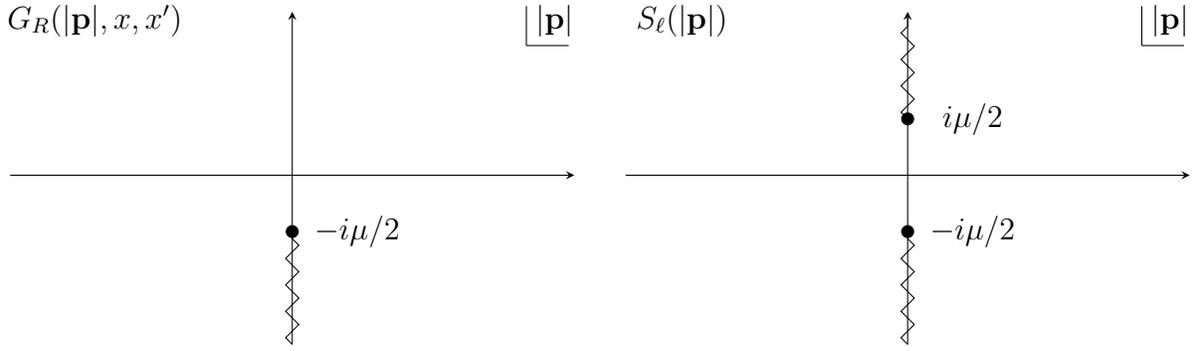
\begin{figure}[ht]
  \centering
  \begin{tikzpicture}[>=stealth, scale=.75]
    \node at (-3.5,5.25) {$G_R(|\mathbf{p}|,x,x')$};

    \draw[black, ->] (-5,2.5) -- (5,2.5);

    \draw[black, decorate, decoration=zigzag] 
           (0,-.5) -- (0,1.5);

    \draw[->] (0,-0.5) -- (0,5.4);

    \node at (4.90-.25,5.25) {$|\mathbf{p}|$};
    \draw (4.6-.45,5.5) -- (4.6-.45,4.8);
    \draw (4.6-.45,4.8) -- (5.15-.25,4.8);

\filldraw[black] (0,1.5) circle (3pt);

\node at (1.15,1.5) {$-i\mu/2$};

  \end{tikzpicture}
  \quad
  \begin{tikzpicture}[>=stealth, scale=.75]
    \node at (-4,5.25) {$S_\ell(|\mathbf{p}|)$};
    
    \draw[black, ->] (-5,2.5) -- (5,2.5);

  \draw[black, decorate, decoration=zigzag] 
           (0,3.5) -- (0,5.3);

\draw[black, decorate, decoration=zigzag] 
           (0,-.5) -- (0,1.5);

\filldraw[black] (0,1.5) circle (3pt);

\filldraw[black] (0,3.5) circle (3pt);

\node at (1.15,1.5) {$-i\mu/2$};
\node at (1.15,3.5) {$i\mu/2$};

    \draw[->] (0,-0.5) -- (0,5.4);

      \node at (4.90-.25,5.25) {$|\mathbf{p}|$};
    \draw (4.6-.45,5.5) -- (4.6-.45,4.8);
    \draw (4.6-.45,4.8) -- (5.15-.25,4.8);
    

  \end{tikzpicture}
  \caption{Analytic structure of $G_R(|\mathbf{p}|,x,x')$ (left) and $S_\ell(|\mathbf{p}|)$ (right) for the Yukawa potential. }
  \label{fig:Yuk}
\end{figure}

The analytic structure of the retarded Green's function and the S-matrix in the $|\mathbf{p}|$-plane is illustrated in Fig.~\ref{fig:Yuk}, and can be understood as follows. All functions appearing in the explicit wavefunctions develop branch cuts along the imaginary axis, in particular for $|\mathbf{p}| \ge i\mu/2$ and $|\mathbf{p}| \le -i\mu/2$. 

Focusing on the upper half-plane (UHP), we define the discontinuity
$\mathrm{Disc}_U S_\ell(|\mathbf{p}|) = S_\ell(i |\mathbf{p}|+\epsilon) - S_\ell(i |\mathbf{p}|-\epsilon)$, with $|\mathbf{p}| > \frac{\mu}{2}$,  and $ \epsilon > 0$,
corresponding to crossing from the first to the second quadrant of the $|\mathbf{p}|$-plane. 
The discontinuity of the S-matrix is controlled by the Legendre function of the second kind, \(Q_\ell(z)\), which develops a branch cut on the interval $z \in [-1,1]$, leading to:
\begin{equation}
\mathrm{Disc}_U\, S_\ell(|\mathbf{p}|) = -\frac{\pi\alpha}{|\mathbf{p}|}\, P_\ell\left(1 + \frac{\mu^2}{2|\mathbf{p}|^2}\right)\,.
\end{equation}
In contrast, the combination \eqref{eq:SYuk} that defines $B_{\mathrm{out}}(|\mathbf{p}|)$ is free of branch cuts: the discontinuity from $Q_\ell$ is exactly compensated by that of the arctangent, leaving $B_{\mathrm{out}}(|\mathbf{p}|)$ analytic in the UHP.

An analogous mechanism occurs for $\psi_\ell^{\mathrm{in}}$: the incomplete gamma function $\Gamma(0,z)$ develops a constant discontinuity for $z<0$, which cancels precisely against the one inherited from $S_\ell(|\mathbf{p}|)$.
Finally, $\psi_\ell^{\mathrm{up}}$ only has branch cuts in the lower half-plane.

We thus conclude, at leading order in the case of a Yukawa potential,
\begin{equation}
\text{Disc}_U \,G_R^\ell(|\mathbf{p}| ,r,r') = 0\,,
\end{equation}
so the retarded Green's function is analytic in the UHP, as expected from causality.

This example highlights that the emergence of Stokes phenomena is a universal aspect of asymptotic analysis, appearing also in $2\to2$ scattering amplitudes in quantum field theory.

\end{appendix}

\bibliography{biblio}

\end{document}